\documentclass{article}
\usepackage{amsmath,amsfonts,amsthm}
\usepackage{amstext}
\usepackage{amsgen}
\usepackage{amsbsy}
\usepackage{amsopn}
\usepackage{amssymb}
\usepackage{graphicx}
\usepackage{epsfig}

\newtheorem{thm}{Theorem}[section]

\newtheorem{prop}[thm]{Proposition}
\theoremstyle{definition}

\theoremstyle{remark}
\newtheorem{rem}[thm]{Remark}
\numberwithin{equation}{section}
\newtheorem{theorem}{Theorem}

\newtheorem{algorithm}[theorem]{Algorithm}

\newtheorem{remark}[theorem]{Remark}

\newcommand{\Real}{\mathbb R}

\newcommand{\Natural}{\mathbb N}

\newcommand{\bigO}{\mathbf{O}}

\def\IMAGESPATH{.}

\begin{document}

\title{Complexity and Stop Conditions for NP as General Assignment Problems,
the Travel Salesman Problem in $\mathbb{R}^2$, Knight Tour Problem
and Boolean Satisfiability Problem}
\author{Carlos Barr\'{o}n Romero \\
cbarron@correo.azc.uam.mx\\
\\
Universidad Aut\'onoma Metropolitana, Unidad Azcapotzalco \\
Av. San Pablo No. 180, Col. Reynosa Tamaulipas, C.P. 02200, \\
MEXICO }
\date{2011}
\maketitle

\begin{abstract}
This paper presents stop conditions for solving General Assignment
Problems (GAP), in particular for Travel Salesman Problem in an
Euclidian 2D space the well known condition Jordan's simple curve
and opposite condition for the Knight Tour Problem. The Jordan's
simple curve condition means that a optimal trajectory must be
simple curve, i.e., without crossing but for Knight Tour Problem
we use the contrary, the feasible trajectory must have crossing in
all cities of the tour. The paper presents the algorithms,
examples and some results come from Concorde's Home page. Several
problem are studied to depict their properties. A classical
decision problem SAT is studied in detail.

\end{abstract}





\section{Introduction}


Building software is similar to prove theorems, i.e., to study,
think, plan, revise, correct, cut, change, understand in order to
pursuing an esthetic beauty for an efficiency procedure for
solving questions or problems. I prefer constructivism on
mathematical model, properties, theories over heuristic or
recipes. The NP problems~\cite{pnp:page} and the 2D Euclidian
Traveller Salesman (hereafter, TSP) have been studied abroad. The
Combinatorics and Optimization Department of the University of
Waterloo has made an splendid research and offer free software for
TSP (the Concorde software~\cite{concorde:page}).

This paper complement my previous papers~\cite{arXiv:Barron2010,
arXiv:Barron2010b} in the study of relationship between the class
NP and the class P.

 Here, I present a carefully study of the properties of similar problems TSP,
Knight tours problem (KTP), polygonals (as TSP type), and Boolean
Satisfiability Problem (SAT).

For TSP and KTP, there are abroad articles and publications, I
mentions two for my preference, of course other are quite
important: ~\cite{Applegate:2007:TSP:1374811, knuth:2011,
Sandifer:www}.

The property of a Jordan's simple curve is  for the first time
used for developing an efficient algorithm. The Jordan's simple
curve come from~\cite{Applegate:2007:TSP:1374811} figures 1.21 and
1.22 (pag. 24) as uncrossing the tour segments. To my knowledge,
what makes different my novel approach (using Jordan's simple
curve) is the combination of visualization data techniques with a
very simple optimization algorithms using greedy and uniform
search strategies and the study of mathematical properties for
understanding stop conditions and when they are necessary or
sufficient. This means that optimization program are polynomial
time when the search space for TSP is convex but yes-no program
for GAP$_n$in general could have a large research space without an
appropriate data organization, therefore there is not polynomial
time algorithm for answering them.

In general, for the TSP under euclidian distance (or where the coseno law
states) Jordan's simple curve is a necessary condition on a plane but
sufficient. This is similar to solve by the necessary condition $f^{\prime
}(x)=0$, an global optimization problem with objective function $f$, which
is derivable. This is the well known first order necessary condition in
optimization. It does not imply the optimality. To be sufficient it is
needed, by example, that $f$ must be quadratic or convex. For the TSP,
Jordan's simple curve condition provides a stop condition (necessary
condition, see Prop.~\ref{prop:JscNecCond}) to build an approximation to the
solution and with other geometrical properties as convexity, it becomes a
necessary and sufficient condition (see Prop.~\ref{prop:JscNecSufCond}).

I include some parts to make this paper self-explicatory from my
papers~\cite{arXiv:Barron2010, arXiv:Barron2010b}.


\section{Travel Salesman Problem, and Knight Tour Problem like General
Assign Problem}~\label{sc:TSPKTPGAP}

The well known Travel Salesman Problem (TSP) and Knight Tour Problem (KTP)
are particular cases of the General Assign Problem(GAP), some properties and
characteristics are ~\cite{arXiv:Barron2010}.

The GAP$_n$ consists a complete graph, a function which assigns a value for
each edge, and objective function, $\text{GAP} \_n$ $=$ $(G_n,c,f)$, where $%
G_n =\left( V_n, A\right) $, $V_n$ $\subset$ $\mathbb{N}$ $A$ $=$ $\left\{
\left( i,j\right) \, | \, i,j \in V\right\}$, $c : V \times V \rightarrow
\overline{\mathbb{R}}$, and $f$ is a real function for the evaluation of any
path of vertices.

In particular, without lost of generality, I focus in minimization to
solving a GAP. It means to look for a minimum value of $f$ over all
hamiltonian cycles of $G_n$. Also, it is trivial to see that any GAP$_n$ has
a solution.

The main proposition in ~\cite{arXiv:Barron2010}, it depicts
contradictory properties and characteristics for solving
GAP$_{n}$:

\noindent \textbf{Proposition 6.12.} An arbitrary and large
GAP$_n$ of the NP-Class has not a polynomial algorithm for
checking their solution.

In this article, SAT$_{n\times m}$ is used to state:

\noindent \textbf{Proposition~\ref{prop:SATnxm}.} SAT$_{n\times
m}$ has not property or heuristic to build an efficient algorithm.

\noindent \textbf{Proposition~\ref{prop:NPnP}.} NP has not
property or heuristic to build an efficient algorithm.

In my previous articles, I did not give an explication of why I
selected graph as the main core of my study. From
the~\cite{ams:borndofer}: "Euler called this new mathematics
without numbers, in which only the structure of the configuration
plays a role, but not size and form, Geometria situs, geometry of
position (he took the concept from a letter of Leibniz from the
year 1679)." Methods for solving the NP look for a common, and
global property but using graphs, they depict local properties,
i.e., properties related to immediately vicinity of the nodes.
Therefore, what is valid in a situs could be invalid in other
positions. Here, on the other hand, a simple tipo of SAT$_{n}$ is
studied to support that there is not a property to build an
efficient algorithm to avoid an exhaustive searching of the search
space or to let to build knowledge from a polinomial population of
cases.

The travel salesman problem with cities as points in a $m$-dimensional
euclidian space (TSP$_{n}$) is a special case of a GAP$_{n}$ where the
edge's cost is given by the matrix $\mathcal{C}=\left[ c(i,j)\right]
,c\left( i,i\right) =\infty ,i=1,\ldots ,n,j=1,\ldots ,n.$ $c(i,j)=%
\hbox{euclidian distance}(i,j)$ $\forall i\neq j\in V_{n}$, and the vertex $%
i $ is a point in $\mathbb{R}^{m},m \geq 2$. The elements of the diagonal of
$\mathcal{C}$ are equals to $\infty $, this is to avoid getting stuck$.$
TSP2D (Hereafter, TSP) is a problem with cities as points in $\mathbb{R}^2$
and edges's values come from their euclidian distances between cities.

In similar way, the Knight tour problem (hereafter, KTP) is an
special special case of a GAP$_{n}$ where the vertices are fixed
in each square's center of a chessboard $n \times m$. The
squares'coordinate are given by their integer coordinates of the
chessboard's squares. The edge's cost is given by the matrix
$\textsl{C}$ $=\left[ c(i,j)\right] ,c\left( i,i\right) =\infty
,\hbox{Euler distance}(i,j)$ $\forall i\neq j$. To honor the great
Mathematician Leonard Euler, I called it as Euler's distance (see ~\cite%
{Sandifer:www}).

$\hbox{Euler's distance}(i,j)=\left\{
\begin{array}{ll}
c_{1} & \text{if }\left( i-i^{\prime }\right) ^{2}+\left( j-j^{\prime
}\right) ^{2}=5 \\
\infty & i=j \\
\sqrt{\left( \left( i-i^{\prime }\right) ^{2}+\left( j-j^{\prime }\right)
^{2}\right)} +4 & \text{otherwise}%
\end{array}%
.\right. $

My first approach was select $c_{1}=0.01$. However, in order to privilege
the knight moves, $c_1$ needs a more carefully tuning (see section~\ref%
{sc:AlgKTP}). The problem consists to look for a hamiltonian cycle in a
chessboard of $8\times 8$ with cost less than 4.

The ~\cite{knuth:2011} depicts in chapters 40 through 42 results about
uncrossed, celtic, long and skinny knight's tours.

The following propositions are known, I include for clarity.

\begin{prop}
Any TSP is soluble, i.e. there is a minimum cost hamiltonian cycle.
\begin{proof}
It obvious, a TSP has n cities, and its a complete graph. Then by
comparing the all (n-1)! hamiltonian cycles the solution is found.
\end{proof}
\end{prop}

\begin{prop}
~\label{prop:ktpeq5} A graph knight is a graph $m$ $\times $ $n$ squares if
only if $\forall mn$ vertices $(i,j)$, $(i^{\prime },j^{\prime })$ with $%
1\leq i<i^{\prime }\leq m,$ and $1\leq j<j^{\prime }\leq n$, the square
distance, $\left( i-i^{\prime }\right) ^{2}+\left( j-j^{\prime }\right)
^{2}=5$.
\begin{proof}
See~\cite{knuth:2011}. A knight walks in a chessboard in L form
from a white square to a black square. In order, for a knight
moving around all vertices, they need to be adjacent vertices to
allowable the motion of the knight.
\end{proof}
\end{prop}

\begin{rem}
For the TSP and KTP we are interesting to find a minimum hamiltonian cycle.

However, the previous proposition does not considere exactly KTP, a
hamiltonian cycle with only knight%
\'{}%
s move but hamiltonian path.

The idea of the Euler distances function come from the previous proposition
to provide space and to favor the moves of a knight. Some results from ~\cite%
{knuth:2011} can be verified by doing exhaustive search. I create a a simple
Matlab programs to verify some knight's tour in small chessboards (if you
want a copy, email me).

From the last two propositions, I define what type solutions to focus in
this paper:

\begin{enumerate}
\item for the TSP, a minimum cost hamiltonian cycle, and

\item for the KTP a crossing hamiltonian cycle with only knight%
\'{}%
s move.
\end{enumerate}
\end{rem}

The TSP correspond to calculate and prove the optimality and for the KTP is
to find a special hamiltonian cycle.

They are quite different. Nevertheless, for any objective function $f$ of a
GAP$_{n}$, hamiltonian's path or cycle are computable problems and the
minimum selection procedure could be used to find the solution (see
proposition 3.2 in ~\cite{arXiv:Barron2010}). The minimum selection
procedure in mathematical notation is:
\begin{equation}
y^{\ast }=\arg \min_{y=(v_{1},v_{2},\ldots ,v_{n},v_{1}),v_{1}\in
V_{n},v_{i}\in V_{n},v_{i}\neq v_{j},1\leq i<j\leq n}\left\{ f(y)\,|\,%
\hbox{GAP}_{n}=(G_{n},c,f),G_{n}=(V_{n},A)\right\}  \label{eq:MinSelProc}
\end{equation}

It is clear that both TSP and KTP are GAP$_{n}^{\prime }$s type with a
similar type of an objective function, $f,$ which is the summation of the
edges' cost over the consecutive pairs of vertices of a path. However, TSP
is a hard NP problem, similar to a global optimization problem. On the other
hand, KTP is a decisi\'{o}n problem to look for a crossing hamiltonian cycle.

The Research Space of GAP$_n$ is finite and numerable, and it has $(n-1)!$
elements (see Prop.3.4 ~\cite{arXiv:Barron2010}). It is totally impractical,
to solve eq.~\ref{eq:MinSelProc} by an exhaustive searching procedure.


\section{A greedy algorithm for Eq.~\protect\ref{eq:MinSelProc}}
~\label{sc:greedy}

Algorithms for solving global optimization problems have abundant
literature. For arbitrary objective functions on a bounded subset, $B\subset
\mathbb{R}^{m}$, uniform searching combined with local optimization can be
used to estimate the global optimal solution (see Theorems of Convergence:
Global and Local Search in~\cite{MOS:SolisWets:1981}). The convergence to
the solution has not guaranty of polynomial time complexity and the solution
can be found at the cost of thoroughly searching on $B$. This motivate to
create heuristic methods, by example, I worked in heuristic methods:

\begin{enumerate}
\item Classical and Exponential Tunnelling method~\cite{immas:Barron1991,
immas:Gomez1991, acago:gomez:2001}. These methods do not guaranty to find
the global solution but a descend of the objective function's value. The
user defines and decides the number of iterations or the time's execution.

\item Genetic Algorithms for Global Optimization for the LJ Problem~\cite%
{aml:Barron1996, aml:Barron1999, arXiv:Barron2005}.
\end{enumerate}

The following heuristic greedy algorithm execute an uniform search on the
vertices, or select the closed vertex, and repeat this procedure $K$ times ($%
K>>0$):

\begin{algorithm}
~\label{alg:GreedyGAP} \textbf{Input:} GAP$_{n}$ with $\mathcal{C}=\left[
c(i,j)\right] ,$ matrix of the edge's cost, a hamiltonian cycle, $%
V_{o}=(v_{1},v_{2},\ldots ,v_{n},$ $v_{1})$, with minimum cost
$c_{o}$.

\textbf{Output:} $v_{o}=(v_{1},v_{2},\ldots ,v_{n},v_{1})$ a hamiltonian
cycle with minimum cost $c_{o}.$

\begin{enumerate}
\item \textbf{For }$r=1$ to $K$

\item \hspace{0.5cm} \textbf{Select randomly an initial vertex by column or
row }$i_{v}$, or $j_{v}$ from $1\leq i_{v},j_{v}\leq n$.

\item \hspace{0.5cm} \hspace{0.5cm} $j_{v}=\arg min_{1\leq j\leq n}\left[
c(i_{v},j)\right] $ ~ \textbf{or} $i_{v}=\arg min_{1\leq i\leq n}\left[
c(i,j_{v})\right] ;$

\item \hspace{0.5cm} \hspace{0.5cm} $v_{1}$ = $i_{v}$ or $j_{v}$;

\item \hspace{0.5cm} \textbf{end select}

\item \hspace{0.5cm} $i=1;$

\item \hspace{0.5cm} $V_{a}=(v_{1})$.

\item \hspace{0.5cm} \textbf{While} $V_{a}$ \textbf{is not a hamiltonian path%
}

\item ~\label{inAlgGreedy:rndDet} \hspace{0.5cm} \hspace{0.5cm} $v_{i+1}=$
\textbf{Select randomly }$\left\{
\begin{array}{c}
\arg min_{1\leq j\leq n}\left[ c(v_{i},j)\right] \\
\mathbf{Select randomly\ } v_{j}\notin v_{a}%
\end{array}%
\right. ;$

\item \hspace{0.5cm} \hspace{0.5cm} \textbf{if} $v_{i+1}$ $\notin$ $V_{a}$ ~
\textbf{then}

\item \hspace{0.5cm} \hspace{0.5cm} \hspace{0.5cm} \textbf{add} $v_{i+1}$
\textbf{to} $V_{a};$

\item \hspace{0.5cm} \hspace{0.5cm} \textbf{else}

\item \hspace{0.5cm} \hspace{0.5cm} \hspace{0.5cm} $i=mod(i+1,n)+1;$

\item \hspace{0.5cm} \textbf{end if}

\item \hspace{0.5cm} \textbf{end while}

\item \hspace{0.5cm} $V_{a}=(v_{1},v_{2},\ldots ,v_{n},v_{1});$

\item \hspace{0.5cm} $c_{a}=f(V_{a});$

\item \hspace{0.5cm} \textbf{if} $c_{a}<c_{o}$ \textbf{then}

\item \hspace{0.5cm} \hspace{0.5cm} $V_{o}=V_{a};$

\item \hspace{0.5cm} \hspace{0.5cm} $c_{o}=c_{a};$

\item \hspace{0.5cm} \textbf{end if}

\item \textbf{end for }$r$
\end{enumerate}
\end{algorithm}

\begin{rem}
~\label{rem:AlgGreRnd} A putative minimum hamiltonian cycle can be obtained
by the previous algorithm. Or it keeps the current putative minimum
hamiltonian cycle. The value of $K$ is $200$.

To behave only greedy the step~\ref{inAlgGreedy:rndDet} could be changed to $%
v_{i+1}=$ $\arg min_{1\leq j\leq n}\left[ c(v_{i},j)\right]$. And
reciprocally, to set on uniform search this step could be changed to $%
v_{i+1}=$ $\mathbf{Select\ randomly\ } v_{j}\notin v_{a}$
\end{rem}


\section{Algorithm for TSP}

~\label{sc:AlgTSP}

Prop. 6.11 in ~\cite{arXiv:Barron2010} depicts that for any quadrilateral,
its sides are lower than its diagonal. This section depicts algorithms for
solving TSP using this property.

Algorithm 5 in ~\cite{arXiv:Barron2010} is used to order the vertices
according to a given hamiltonian cycle. Hereafter, we assume that the
current hamiltonian cycle is in ascending order of the consecutive cities.
It is not necessary, but assuming that the vertices of the hamiltonian cycle
are in order facilitate the description of the following algorithms.

\begin{figure}[tbp]
\centerline{\psfig{figure=\IMAGESPATH/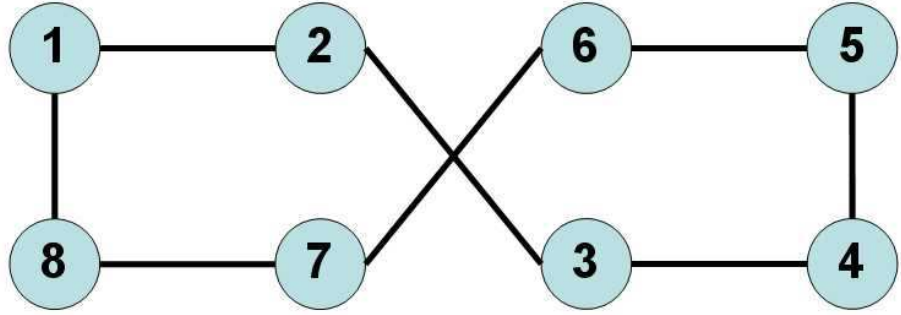,height=50mm}}
 ~\caption{Hamiltonian cycle with a crossing in the cities: $%
v_{2},v_{3},v_{6},v_{7}.$}
\label{fig:CrossNC_v0}
\end{figure}

Let's assume that with algorithm~\ref{alg:GreedyGAP}, there is a hamiltonian
cycle for a given TSP. An image of the current solution can be obtained by a
graphic interface as in Concorde or using the graphic's tool of Matlab or
Octave. By example see fig.~\ref{fig:CrossNC_v0}.

\begin{algorithm}
~\label{alg:Detcross} \textbf{Input:} An black and white image of the
current hamiltonian cycle, $v=(v_{1},v_{2},\ldots ,v_{n},v_{1})$ where the
cities are in order consecutive.

\textbf{Output:} Stop, the image of the hamiltonian cycle can be colored by
two colors. Otherwise, the hamiltonian cycle has a crossing at the cities $%
v_{i},v_{i+1},v_{j},v_{j+1}$.

\begin{enumerate}
\item \textbf{Using }$v_1$ detects its location in the input image.

\item \textbf{Using }$v_1$ detects its frontier in the image for selecting
two points, one inside and one outside of the $v_1$.

\item \textbf{Using} a flood or a paint graphic tool with the point outside
of the $v_1$, colored the image on green.

\item \textbf{Using} a flood or a paint graphic tool with the point inside
of the $v_1$, colored the image of red.

\item two\_color="yes"\textbf{;}

\item \textbf{for} $i:=1$ \textbf{to} $n$

\item \hspace{0.5cm} \textbf{Using }$v_i$ detects its location in the
colored input image$;$

\item \hspace{0.5cm} \textbf{if} the vicinity of the $v_i$ has red, green
and black \textbf{then}

\item \hspace{1cm} \textbf{continue}$;$

\item \hspace{0.5cm} \textbf{else}

\item \hspace{1cm} two\_color="no";

\item \hspace{1cm} \textbf{mark }$v_i;$

\item \textbf{end\ for} $i;$

\item \textbf{if} \textbf{\ }two\_color is "yes" \textbf{then}

\item ~\hspace{0.5cm} \textbf{Stop. }"The image of the hamiltonian cycle can
be colored by two colors,$v=(v_{1},v_{2},\ldots ,v_{n},v_{1})$ is a
hamiltonian cycle as Jordan's simple curve, i.e., without crossing.";

\item \textbf{select }marked vertex, let assume it is $v_i$

\item \textbf{using }$v_i$ detects its location in the colored input image;

\item \textbf{using }$v_i$ detects its four closed marked neighbors;

\item \textbf{Stop. }There is crossing at the cities: $%
v_{i},v_{i+1},v_{j},v_{j+1}$.
\end{enumerate}
\end{algorithm}

\begin{figure}[tbp]
\centerline{\psfig{figure=\IMAGESPATH/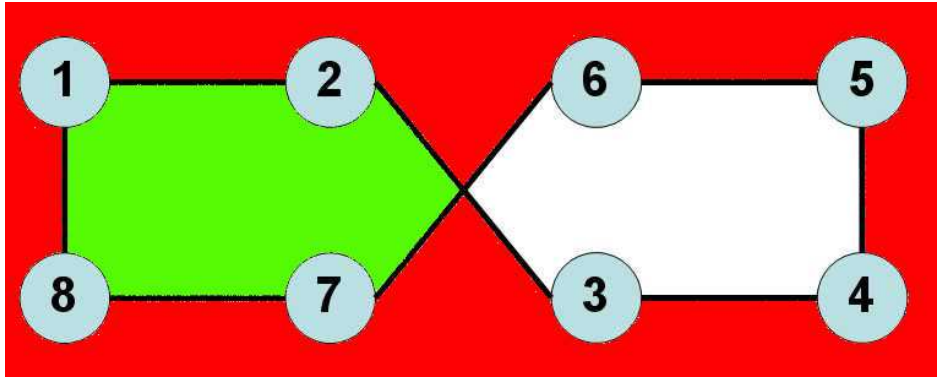,height=50mm}} ~
\caption{Colored image of a hamiltonian cycle with a crossing in the cities:
$v_{2},v_{3},v_{6},v_{7}.$}
\label{fig:CrCol_v0}
\end{figure}

\begin{rem}
Flood or paint graphic tool algorithm has the complexity of the image's size
is $\mathbf{O}(kn^{2})$ where $kn$ is related to the resolution of the
vicinity of the $n$ vertices. The value of factor $k$ depends of the minimum
distance of the vertices. $k$ must be allow to have a black-white image
where the lines of hamiltonian cycle are clearly distinguish. Figure ~\ref%
{fig:CrCol_v0} depicts when there is a crossing.
\end{rem}

\begin{figure}[tbp]
\centerline{\psfig{figure=\IMAGESPATH/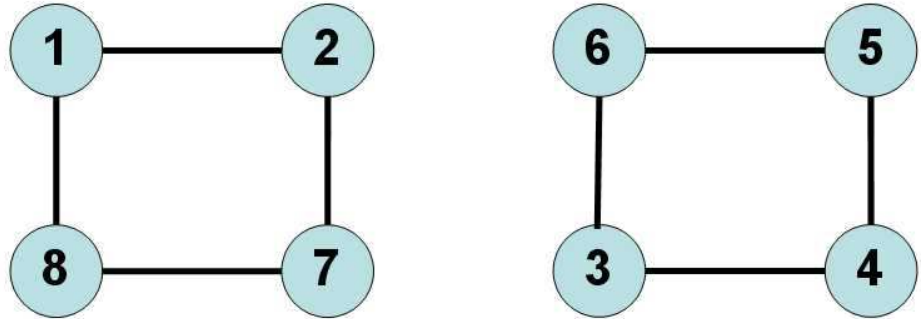,height=50mm}} ~
\caption{Wrong connection of crossing in the cities: $%
v_{2},v_{3},v_{6},v_{7}.$}
\label{fig:WrCon_v0}
\end{figure}

\begin{figure}[tbp]
\centerline{\psfig{figure=\IMAGESPATH/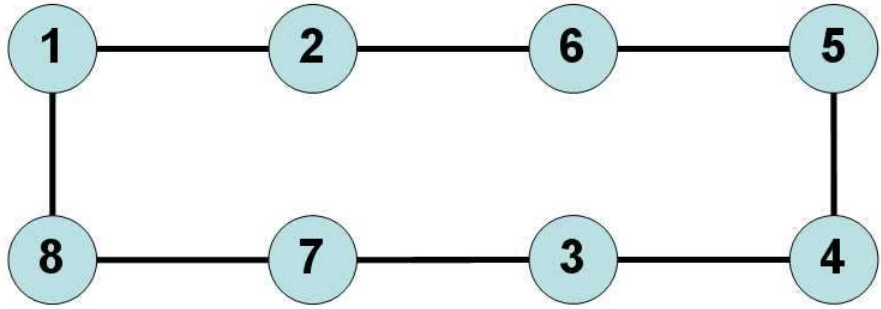,height=50mm}} ~
\caption{Hamiltonian cycle without the crossing in the cities: $%
v_{2},v_{3},v_{6},v_{7}.$}
\label{fig:JCSol_v0}
\end{figure}

Assuming that vertices of the hamiltonian cycle are in order, the only one
solution for figure~\ref{fig:CrCol_v0} is depicted in figure~\ref%
{fig:JCSol_v0}. Figure~\ref{fig:WrCon_v0} depicts the wrong connection.
There are two possibilities when a crossing is found, but only one is the
correct on the current hamiltonian cycle. Nevertheless, the cost of the
cycle always decreases. Therefore, if there is crossing at the cities $%
v_{i},v_{i+1},v_{j},v_{j+1}$ the hamiltonian cycle must connects $v_{i}$ to $%
v_{j}$ and $v_{i+1}$ to $v_{j+1}$ and there is always a guaranty for
decreasing cycle's cost from Prop. 6.11 in ~\cite{arXiv:Barron2010}.

Without loss of generality, let's assume that $i<j$.

\begin{algorithm}
~\label{alg:Uncross} \textbf{Input:} $V=(v_{1},v_{2},\ldots ,v_{n},v_{1})$ a
hamiltonian cycle, where the cities are in order consecutive and with a
crossing in four cities: $v_{i},v_{i+1},v_{j},v_{j+1}.$

\textbf{Output:} $V=(v_{1},v_{2},\ldots ,v_{n})$ a hamiltonian cycle without
the crossing at the cities $v_{i},v_{i+1},v_{j},v_{j+1}.$

\begin{enumerate}
\item $V^{\prime } = V;$

\item $l=j$

\item \textbf{for} k:= i+1 \textbf{to} j

\item \hspace{0.5cm} $V(k)=V^{\prime }(l)$

\item \hspace{0.5cm} $l=l-1;$

\item \textbf{end for} $k;$

\item \textbf{stop.} The hamiltonian cycle without the crossing is $%
V=(v_{1},v_{2},\ldots ,v_{i},v_{j},v_{j-1},\ldots ,v_{i+1},v_{j+1},\ldots
,v_{n},v_{1})$.
\end{enumerate}
\end{algorithm}

\begin{rem}
The complexity of the previous algorithm is $\mathbf{O}(n)$. The next steps
are for ordering the cities of the hamiltonian cycle, perhaps to repeat a
greedy algorithm for looking a less hamiltonian cycle's cost around the
current solution. The next paragraph depicts the complete algorithm for
solving the TSP, using the previous algorithms to get a simple Jordan's
simple curve. This is a necessary condition to stop, because the crossing of
the cities is a visual property for detecting where the hamiltonian cycle's
cost decreases.
\end{rem}

\begin{algorithm}
~\label{alg:CompTSP} \textbf{Input:} TSP.

\textbf{Output:} $V=(v_{1},v_{2},\ldots ,v_{n},v_{1})$ a hamiltonian cycle
as Jordan's simple curve, i.e., without any crossing on the path of cities.

\begin{enumerate}
\item \textbf{Repeat}

\item \hspace{0.5cm} \textbf{execute algorithm}~\ref{alg:GreedyGAP};

\item \hspace{0.5cm} \textbf{execute algorithm 5} in ~\cite{arXiv:Barron2010}
to order the hamiltonian cycle by their consecutive cities;

\item \hspace{0.5cm} \textbf{create a black-white image};

\item \hspace{0.5cm} \textbf{execute algorithm}~\ref{alg:Detcross}

\item \textbf{until get a Jordan's simple curve from the current hamiltonian
cycle.}

\item \textbf{Stop.} The hamiltonian cycle is the putative solution of the
TSP.
\end{enumerate}
\end{algorithm}

\begin{prop}
~\label{prop:JscNecCond} Given a TSP$_{n}$, where cities are points in $%
\mathbb{R}^2$, and the cost matrix correspond to euclidian distances between
cities. The Jordan's simple curve is a necessary condition for the minimum
hamiltonian cycle's cost.
\begin{proof}
With out loss of generality, if the putative optimal hamiltonian
cycle has a crossing then for Prop. 6.11 in
~\cite{arXiv:Barron2010}, the hamiltonian cycle obtained by
algorithm~\ref{alg:Uncross} has lower value than the putative
optimal hamiltonian cycle.
\end{proof}
\end{prop}

\begin{prop}
~\label{prop:concur}~\label{prop:JscNecSufCond} Given a TSP$_{n}$, if cities
are located around a closed convex curve, the algorithm ~\ref{alg:CompTSP}
finds the hamiltonian cycle of minimum cost.
\begin{proof}
The optimality of the  hamiltonian cycle comes from the
vertices'configuration around of convex curve, points on a circle,
elipse, n-poligonal. The algorithm~\ref{alg:GreedyGAP} by
construction selects the closed next vertex. The hamiltonian cycle
corresponds to the solution of the variational problem of the the
minimum length´s curve (Jordan's simple curve) with the maximum
convex area.
\end{proof}
\end{prop}

\begin{figure}[tbp]
\centerline{ \psfig{figure=\IMAGESPATH/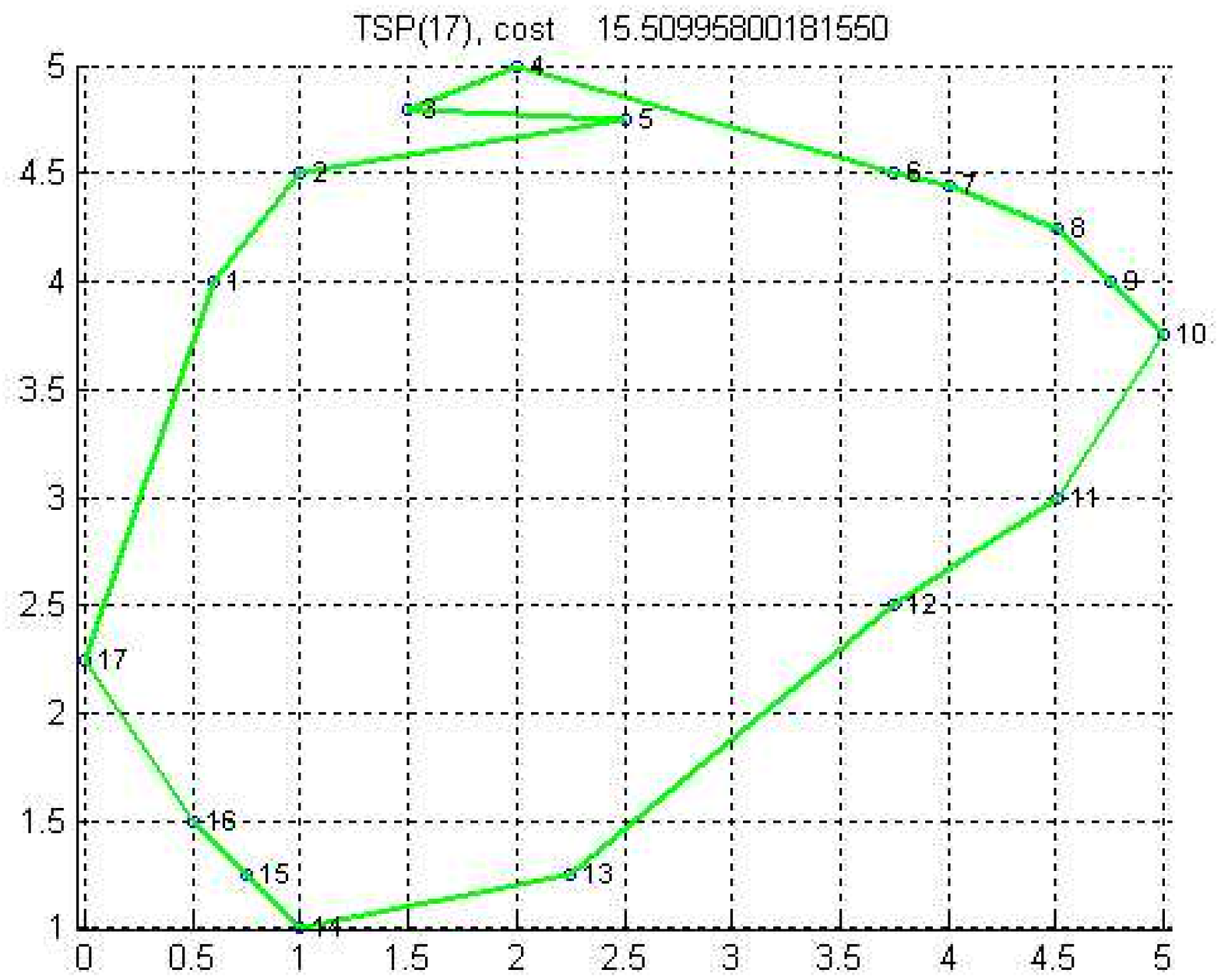,
height=50mm} \psfig{figure=\IMAGESPATH/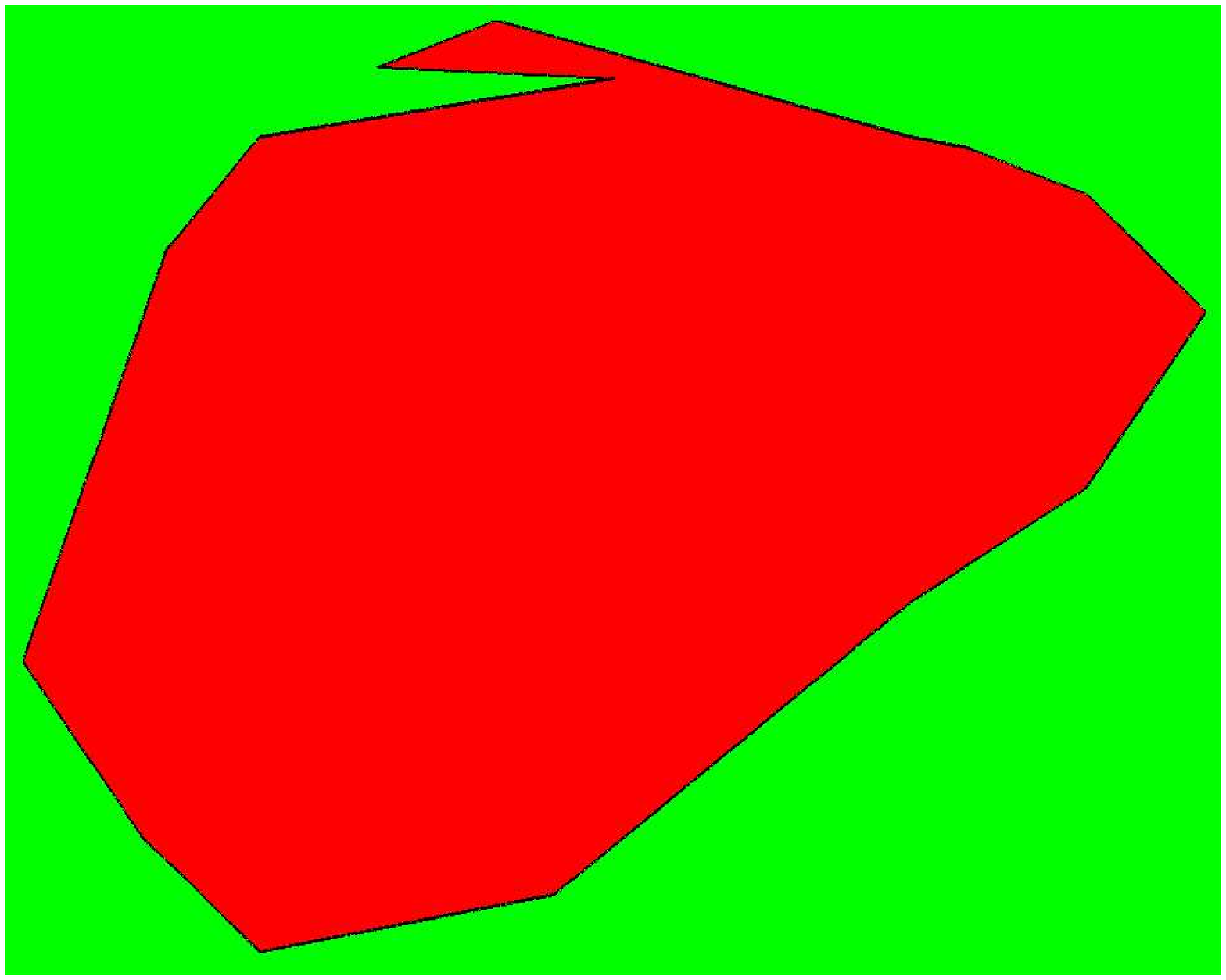,
height=50mm} }
\centerline{ \makebox[2.2in][c]{ a)
}\makebox[2.2in][c]{ b) } } ~
\caption{TSP$_{17}$. a) Not optimal hamiltonian cycle with Jordan's simple
curve, and b) its two colored image.}
\label{fig:tsp17Conv_nOpt}
\end{figure}

\begin{figure}[tbp]
\centerline{ \psfig{figure=\IMAGESPATH/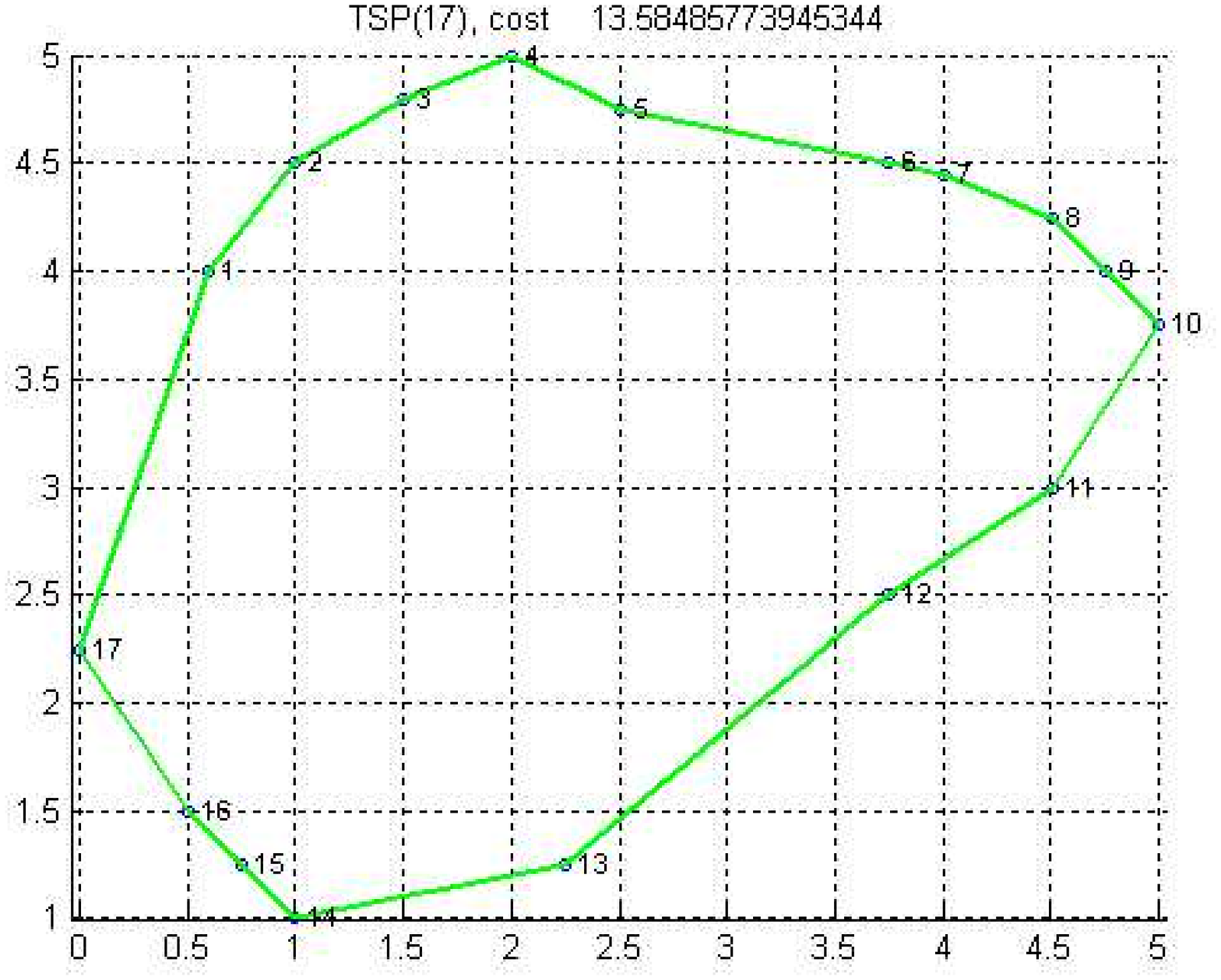,
height=50mm} \psfig{figure=\IMAGESPATH/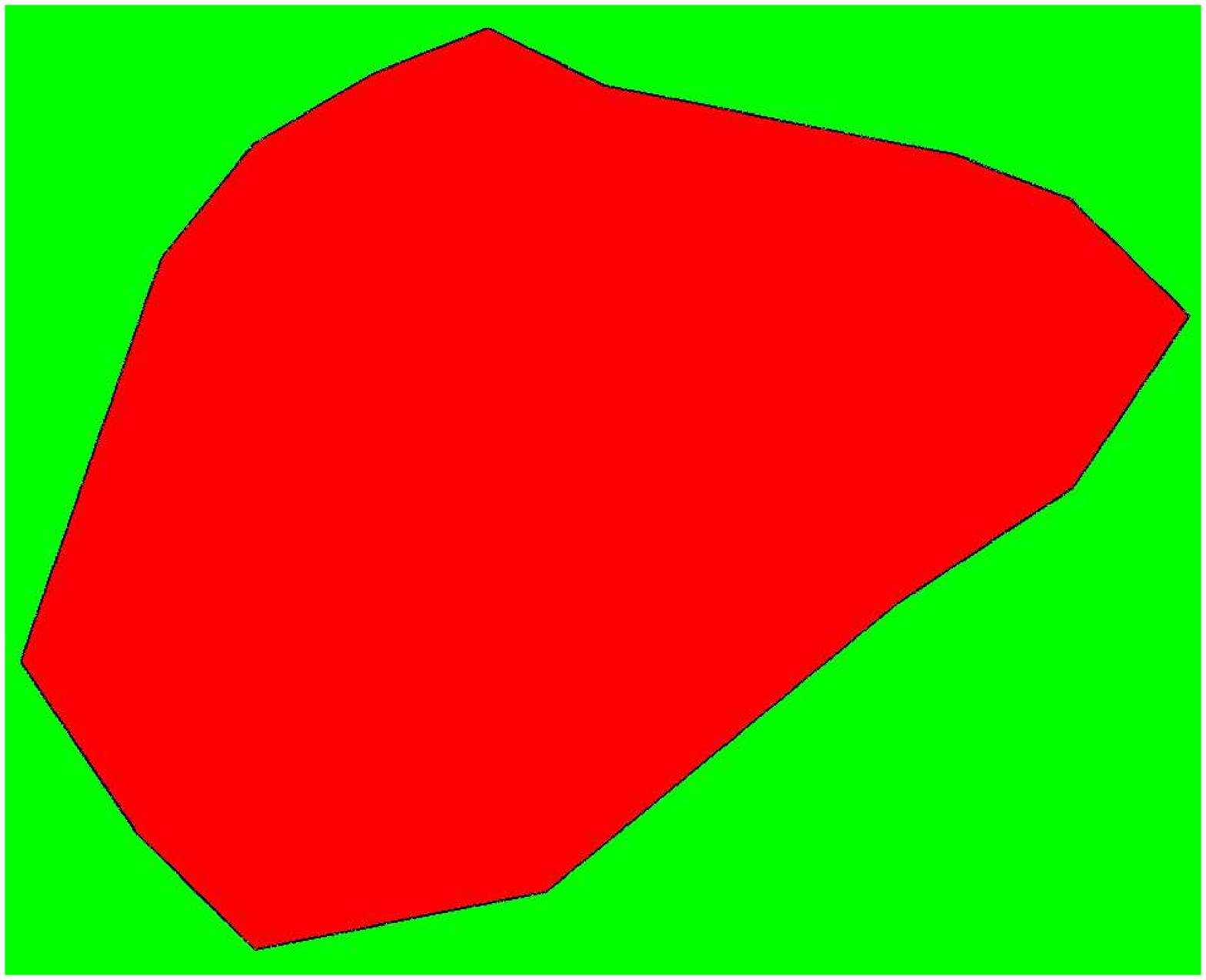,
height=50mm} }
\centerline{ \makebox[2.2in][c]{ a)
}\makebox[2.2in][c]{ b) } } ~
\caption{TSP$_{17}$. a) Optimal hamiltonian cycle with Jordan's simple
curve, and b) its two colored image.}
\label{fig:tsp17Conv_Opt}
\end{figure}

\begin{figure}[tbp]
\centerline{
\psfig{figure=\IMAGESPATH/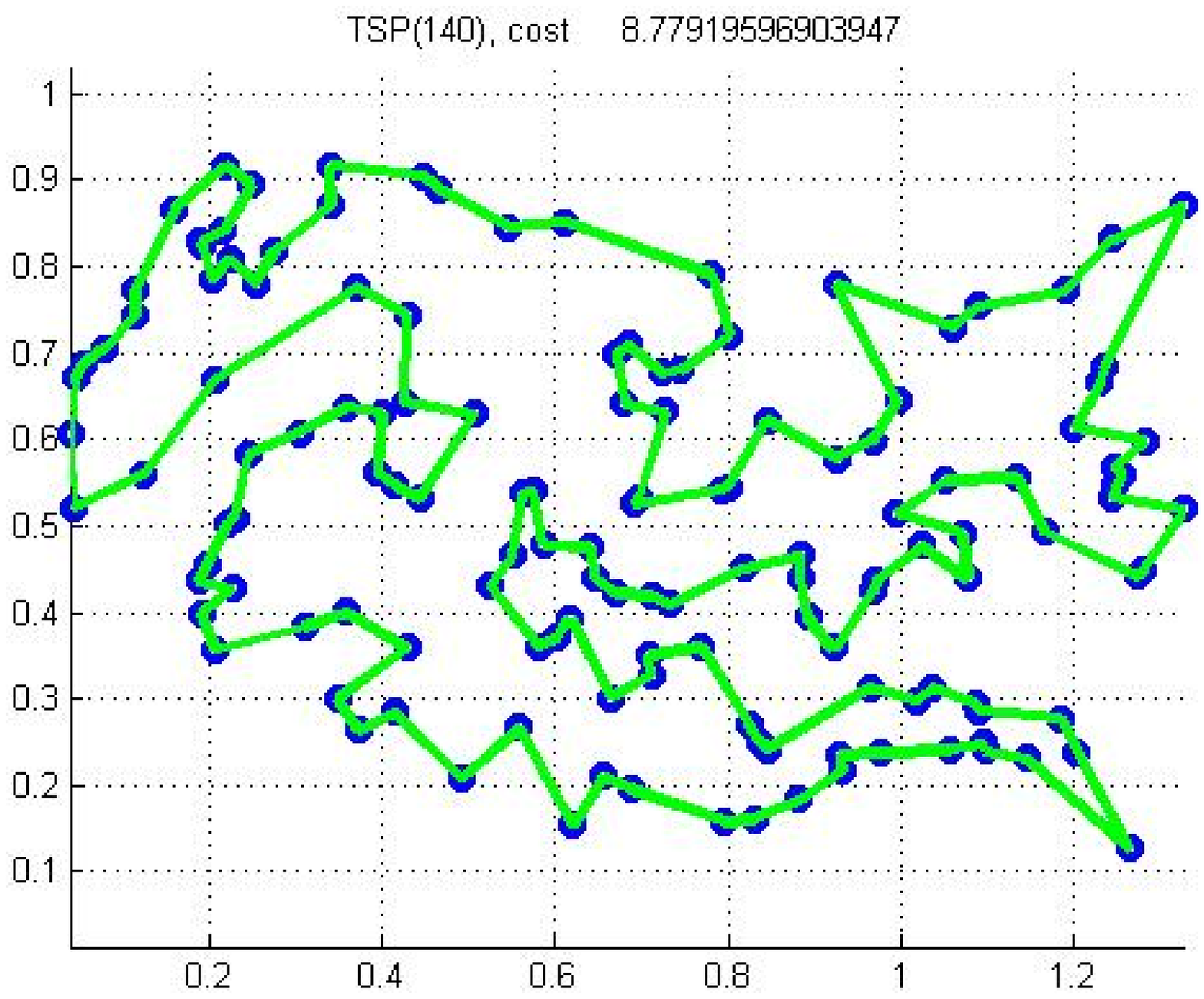,
height=50mm}
\psfig{figure=\IMAGESPATH/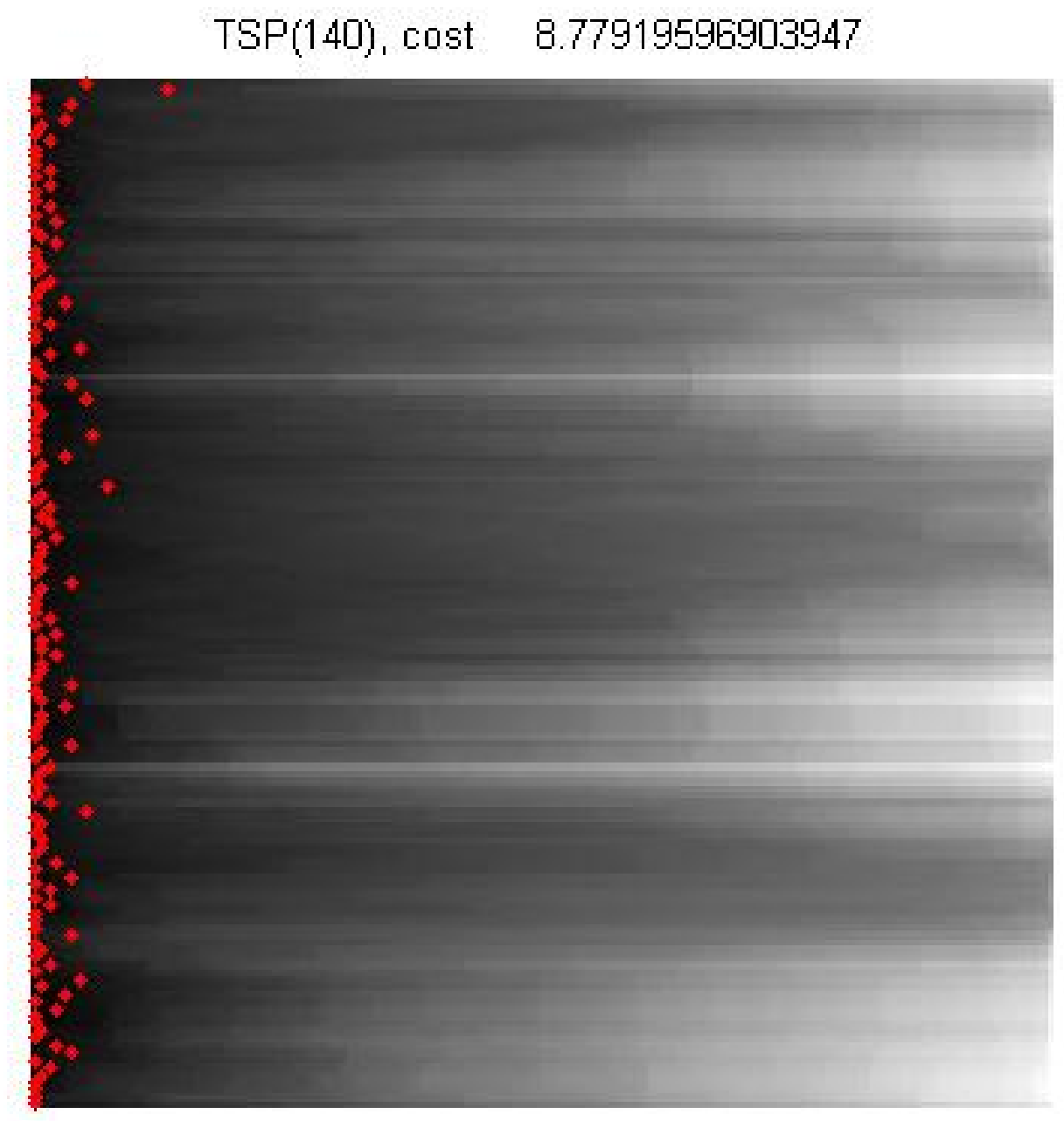,
height=50mm} }
\centerline{ \makebox[2.2in][c]{ a)
}\makebox[2.2in][c]{ b) } } ~
\caption{TSP$_{140}$. a) Hamiltonian cycle with the minimum cost, and b)
sorted cost matrix $\mathcal{M}$. The red dots on the left side implies that
there are not many alternatives hamiltonian cycles.}
\label{fig:ex_140_cities}
\end{figure}

\begin{rem}
Figure~\ref{fig:tsp17Conv_nOpt} depicts a case where the algorithm~\ref%
{alg:Detcross} ~ fails to detect the zone where it is possible to descend
the cost. However, there are many algorithms where this not happen. I hope,
that the algorithms of the Concorde~\cite{concorde:page} for TSP2D get
advantage of the Jordan's simple curve property in the future.

Unfortunately, for TSP where the objective is to get a minimum hamiltonian
cycle's cost related to money, time travel or arbitrary values then Jordan's
simple curve property is not useful. This includes the max distance giving
by: $d_m(x,y) = \max \{ |x_1-y_1|,|x_2-y_2\ \},$ where $x=(x_1,x_2),$ and $%
y=(y_1,y_2) \in \mathbb{R}^2$. It is easy to verify that a quadrilateral
under $d_m$ has diagonals not larger than its sides.

Figures~\ref{fig:tsp17Conv_nOpt} and~\ref{fig:tsp17Conv_Opt} depict a case
where the previous proposition is verified for a small TSP$_{17}$. Red area
of the former figure has $415,570$ pixels with cost = $15.5100$, and red
area of next figure has $417,719$ pixels with cost = $13.5849$. Also, these
figures shows why the Jordan's simple curve is a necessary property.

Using algorithm~\ref{alg:CompTSP}, big TSP can be solved to a putative
optimal hamiltonian cycle. Figure~\ref{fig:ex_140_cities} taken from ~\cite%
{arXiv:Barron2010} (Figure 16) depicts a Jordan's simple curve. The red dots
depicts the minimum hamiltonian cycle, it corresponds to vertices'
enumeration closed to the left side. In this case the complexity of the
algorithm~\ref{alg:CompTSP} until getting a Jordan's simple curve is
polinomial time bounded by $\mathbf{O}(n^{k})$ with $k$ a small positive
integer, and the time depends of the many crossings that the greedy
algorithm~\ref{alg:GreedyGAP} can to avoid.
\end{rem}


\section{Algorithm for KTP}
~\label{sc:AlgKTP}

The Euler's distance of the KTP does not connect the squares of a chessboard
as usual, but it privilege the knight's moves. It is clear that Prop. 6.11
in ~\cite{arXiv:Barron2010} (sides versus diagonals of quadrilaterals) can
not be used for solving KTP, i.e., the algorithm for the KTP is looking for
the contrary of a simple Jordan's simple curve as in TSP, in fact, the
property is a hamiltonian cycle with many crossing as possible.

The algorithm~\ref{alg:GreedyGAP} is used without a modification. But, here
the goal is to stop when the cost of the hamiltonian cycle is less than $4.$

To my knowledge there is not a proposition or theorem for proving that
exists a complete crossing knight tour for any chessboard's size. In~\cite%
{aggoun:www} there is a table of solutions for some chessboard's size. I
tune the Euler's distance function using the description of the article~\cite%
{Sandifer:www}. The knight's motion has three different behaviors each
quadrant of size $4\times 4$:

\begin{enumerate}
\item Two romboides with direction $(-1,-1)$ to $(1,1)$;

\item Two romboides with direction $(-1,1)$ to $(1,-1)$;

\item Two squares.
\end{enumerate}

\begin{algorithm}
~\label{alg:CompKTP} \textbf{Input:} KTP for a chessboard of $8\times 8$.

\textbf{Output:} $V=(v_{1},v_{2},\ldots ,v_{n},v_{1})$ a hamiltonian cycle
with cost less than 4.

\begin{enumerate}
\item \textbf{Execute algorithm}~\ref{alg:GreedyGAP}.

\item \textbf{while current hamiltonian cycle's cost $\geq $ 4 do}

\item \hspace{0.5cm} \textbf{execute algorithm 5} in ~\cite{arXiv:Barron2010}
to order the hamiltonian cycle by their consecutive vertices;

\item \hspace{0.5cm} \textbf{execute algorithm}~\ref{alg:GreedyGAP};

\item \textbf{end while.}

\item \textbf{Stop.} The hamiltonian cycle is the solution of the KTP.
\end{enumerate}
\end{algorithm}

There is not guaranty that the previous algorithm is computable, i.e., it
could not to stop for an arbitrary chessboard's size. I mean that the tuning
of $c_1$ works for a chessboard of $8\times8$. For arbitrary chessboard's
size, it could not provide a hamiltonian path because the corresponding GAP
could not have a hamiltonian cycle with cost less than $4$.

The selection of the value $4$ is in the hope that the greedy part of the
algorithm~\ref{alg:GreedyGAP}'s behaves greedy when it searches for minimum
hamiltonian cycle's cost. It will prefer to choice edge's cost between
vertices that corresponds knight's paths. By construction the total distance
of a tour with only knight's motion is less than $4.$ For any pair of
squares that not correspond to knight's paths the distance is giving by sqrt$%
\left( \left( i-i^{\prime }\right) ^{2}+\left( j-j^{\prime }\right)
^{2}\right) +4$, .i.e. when $\left( i-i^{\prime }\right) ^{2}+\left(
j-j^{\prime }\right) ^{2}\neq 5$ where $(i,j)$ and $(i^{\prime },j^{\prime
}) $ are the integer coordinates of the squares.

For a knight's paths the value of $c_{1}$ of the Euler's distance is $0.04,$
but in each quadrant of size $4\times 4$ the value of $c_{1}$ is:

\begin{enumerate}
\item $0.01$ for the two romboides with direction $(-1,-1)$ to $(1,1);$

\item $0.03$ for the two romboides with direction $(-1,1)$ to $(1,-1);$

\item $0.02$ for the two squares$.$
\end{enumerate}

\begin{figure}[tbp]
\centerline{ \psfig{figure=\IMAGESPATH/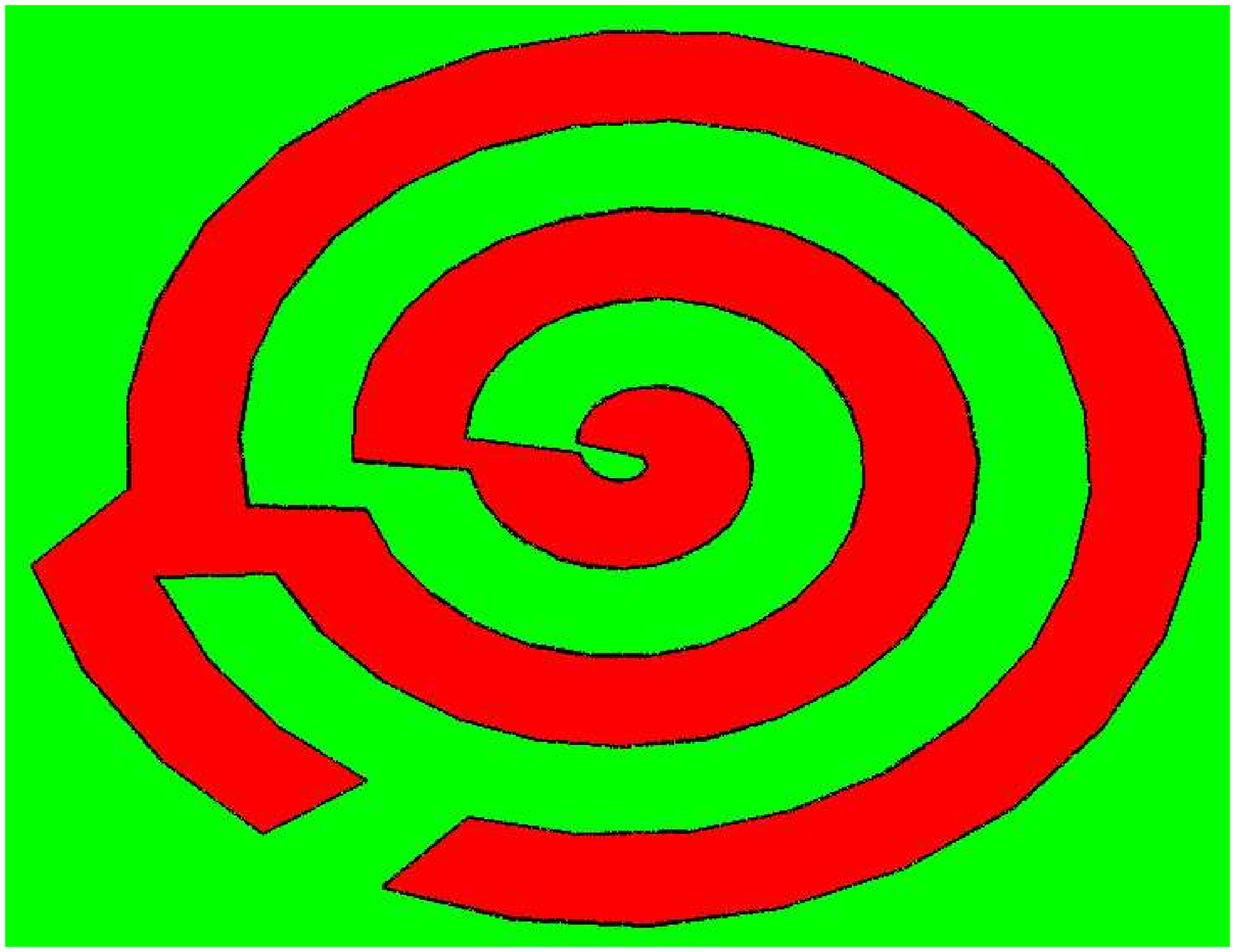,
height=50mm} \psfig{figure=\IMAGESPATH/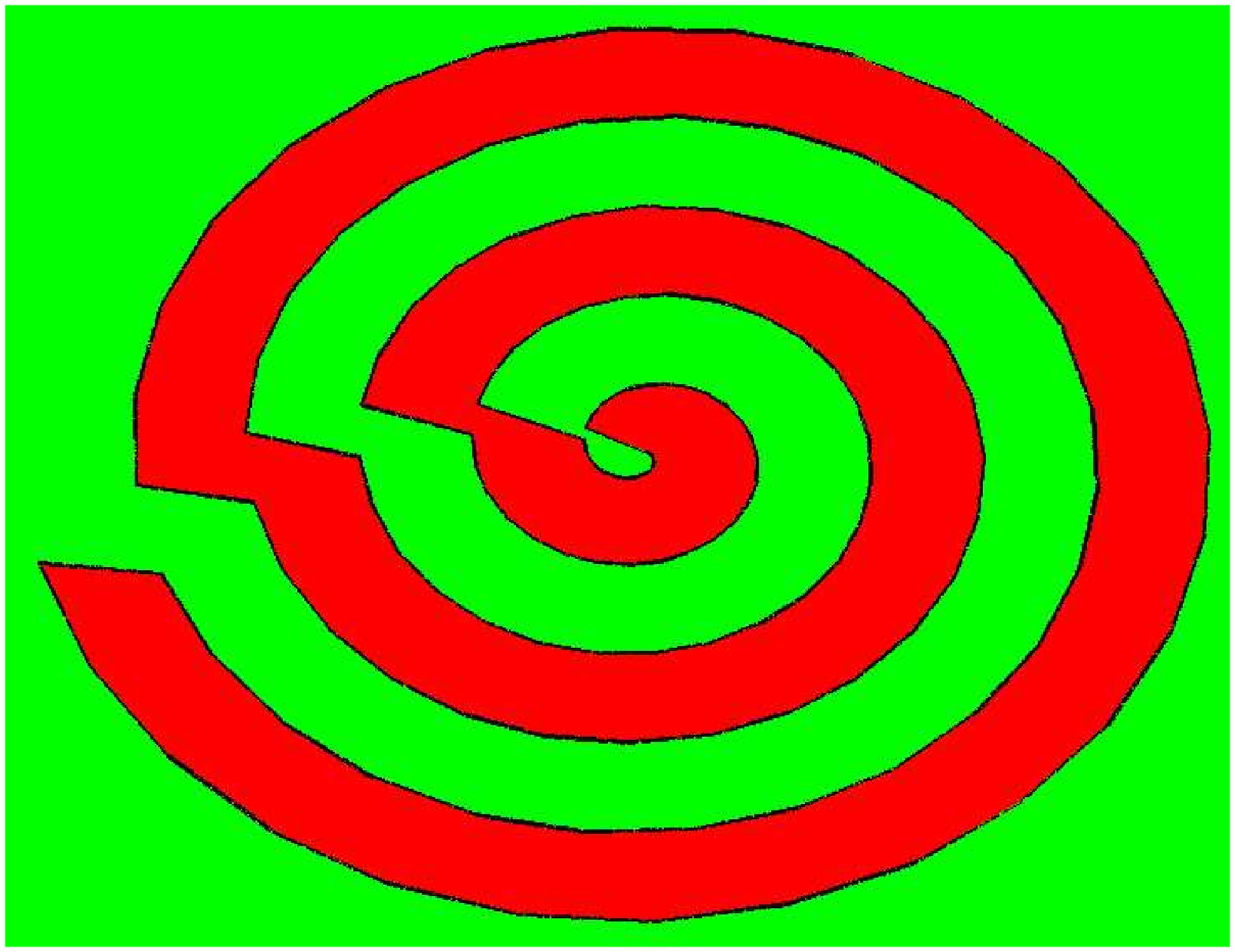,
height=50mm} }
\centerline{ \makebox[2.2in][c]{ a)
}\makebox[2.2in][c]{ b) } } ~
\caption{TSP$_{152}$ on a spiral. Two colored Jordan's simple curve: a) not
optimal hamiltonian cycle $21.1661$, and b) optimal hamiltonian cycle $%
21.1451$.}
\label{fig:tsp152_espiral}
\end{figure}

\begin{figure}[tbp]
\centerline{ \psfig{figure=\IMAGESPATH/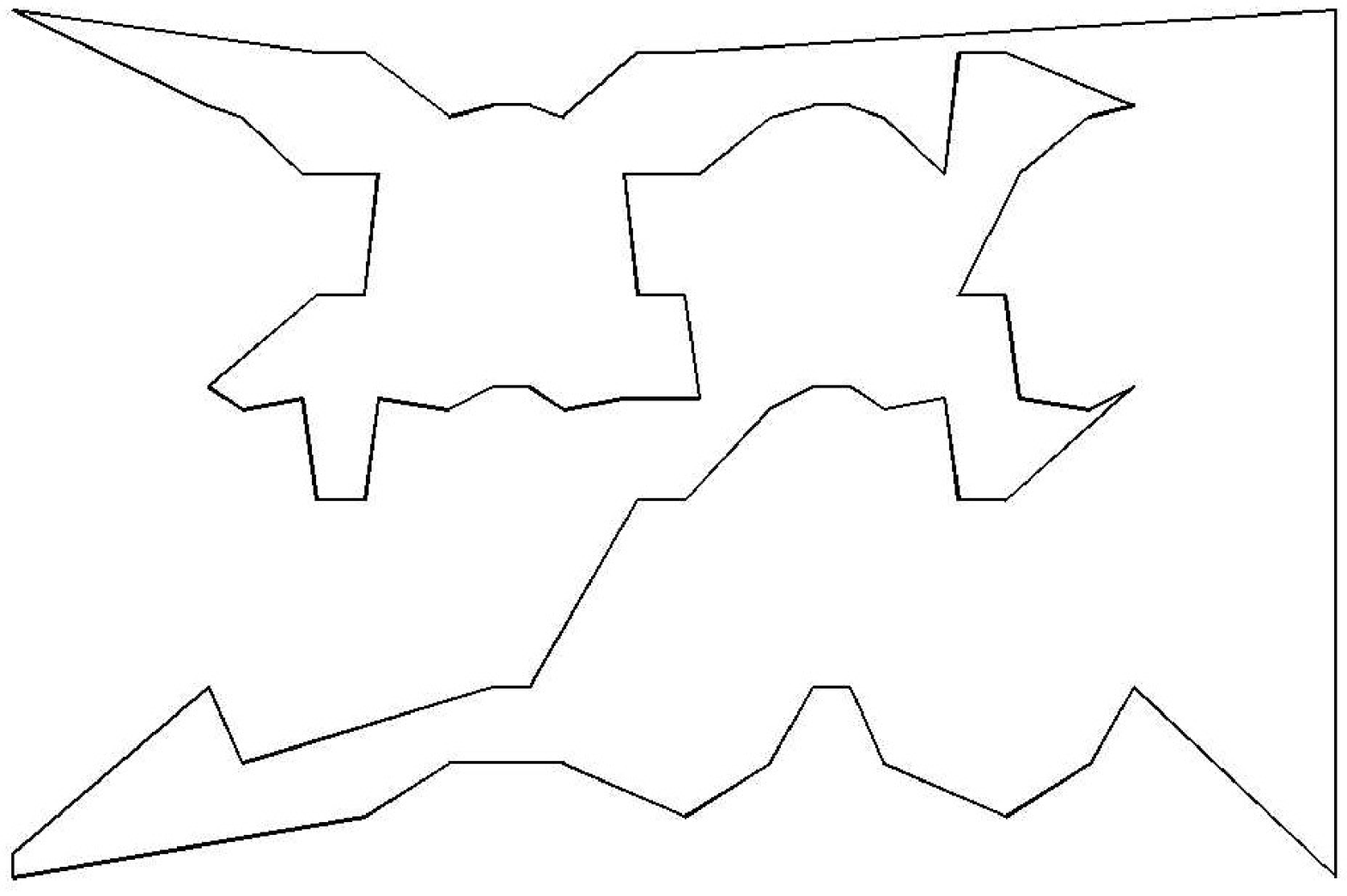, height=50mm}
\psfig{figure=\IMAGESPATH/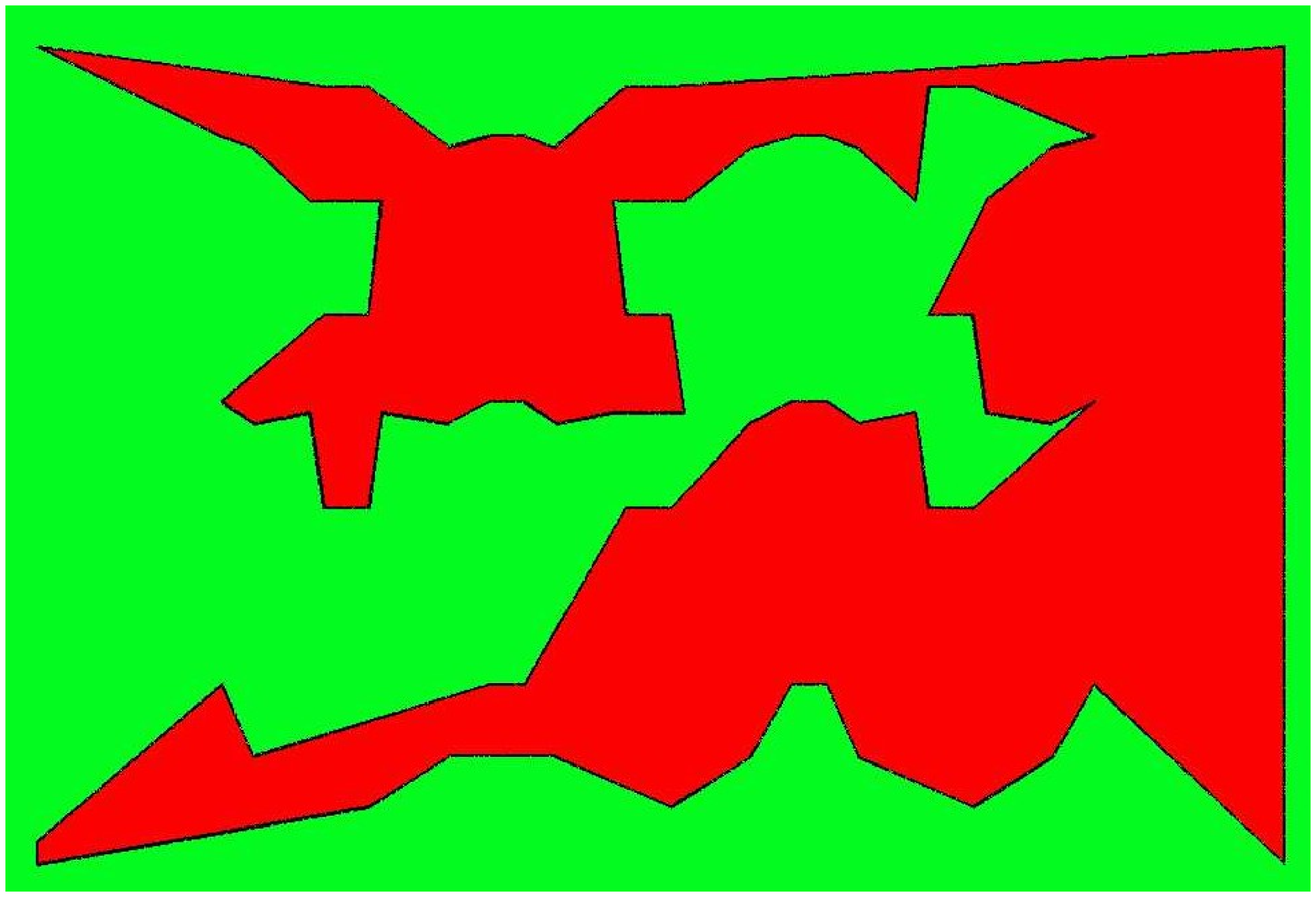, height=50mm} }
\centerline{ \makebox[2.2in][c]{ a) }\makebox[2.2in][c]{ b) } } ~
\caption{TSP$_{76}$(file pr76.tsp). a) final hamiltonian cycle
with Jordan's simple curve, and b) its two colored image.}
\label{fig:tsppr76}
\end{figure}

\begin{figure}[tbp]
\centerline{ \psfig{figure=\IMAGESPATH/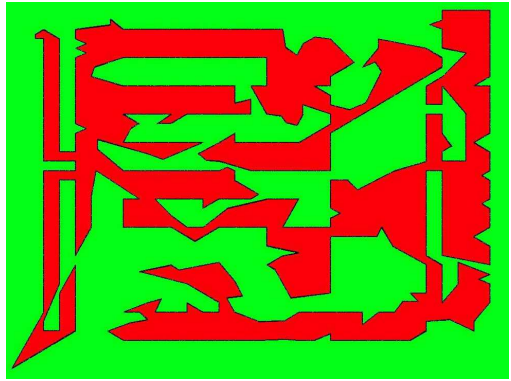,
height=50mm} } ~ \caption{TSP$_{442}$(file pcb442.tsp). Final two
colored hamiltonian cycle with cost = $56,601.1738.$}
\label{fig:tsppcb442}
\end{figure}

\begin{rem}
The heuristic idea to assign $c_1$ values come from the article~\cite%
{Sandifer:www}, and without any formal proof, it works for KTP $8\times 8$
using the GAP$\_n$'s algorithm~\ref{alg:GreedyGAP}.
\end{rem}

\begin{figure}[tbp]
\centerline{
\psfig{figure=\IMAGESPATH/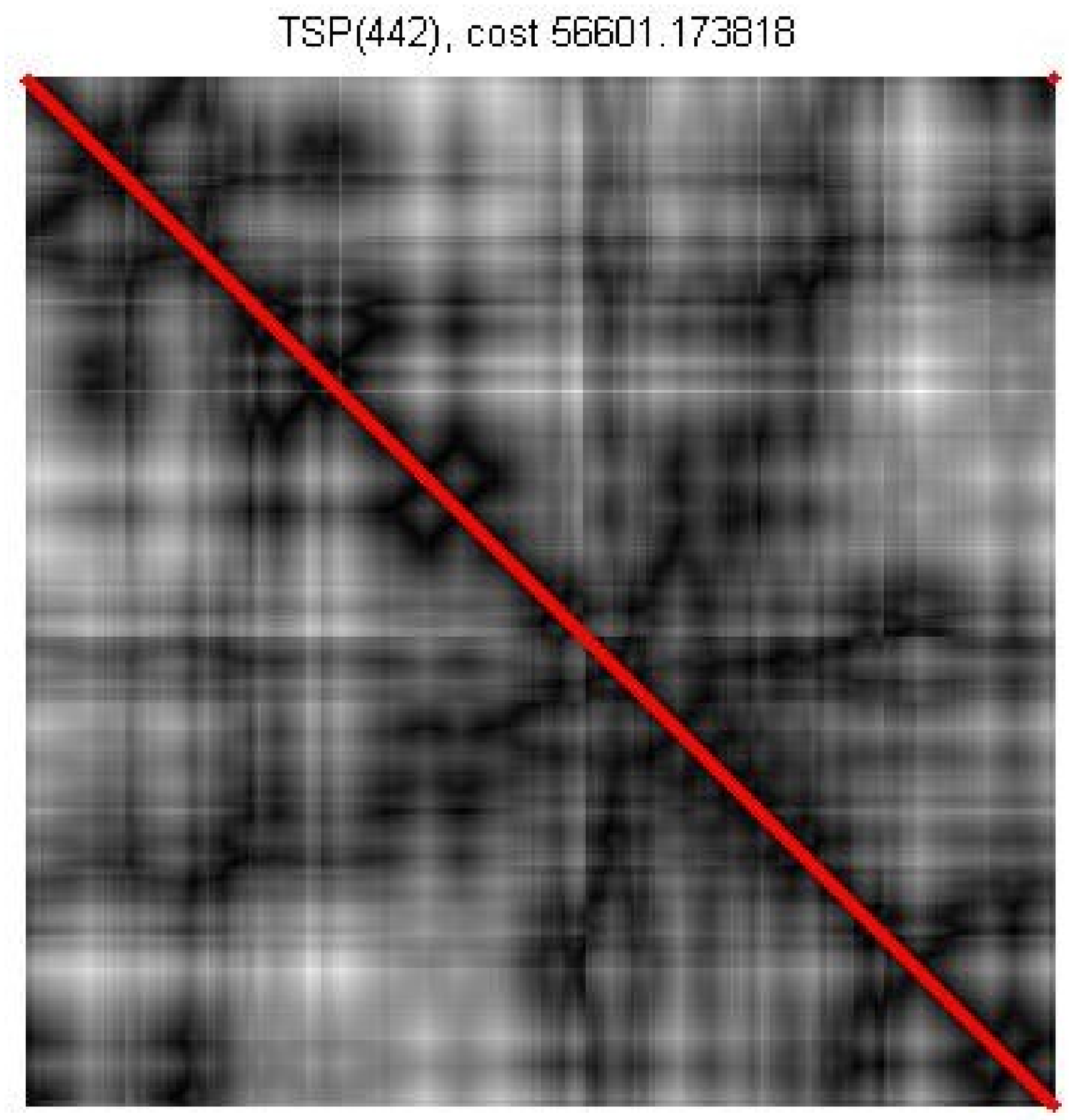, height=50mm}
\psfig{figure=\IMAGESPATH/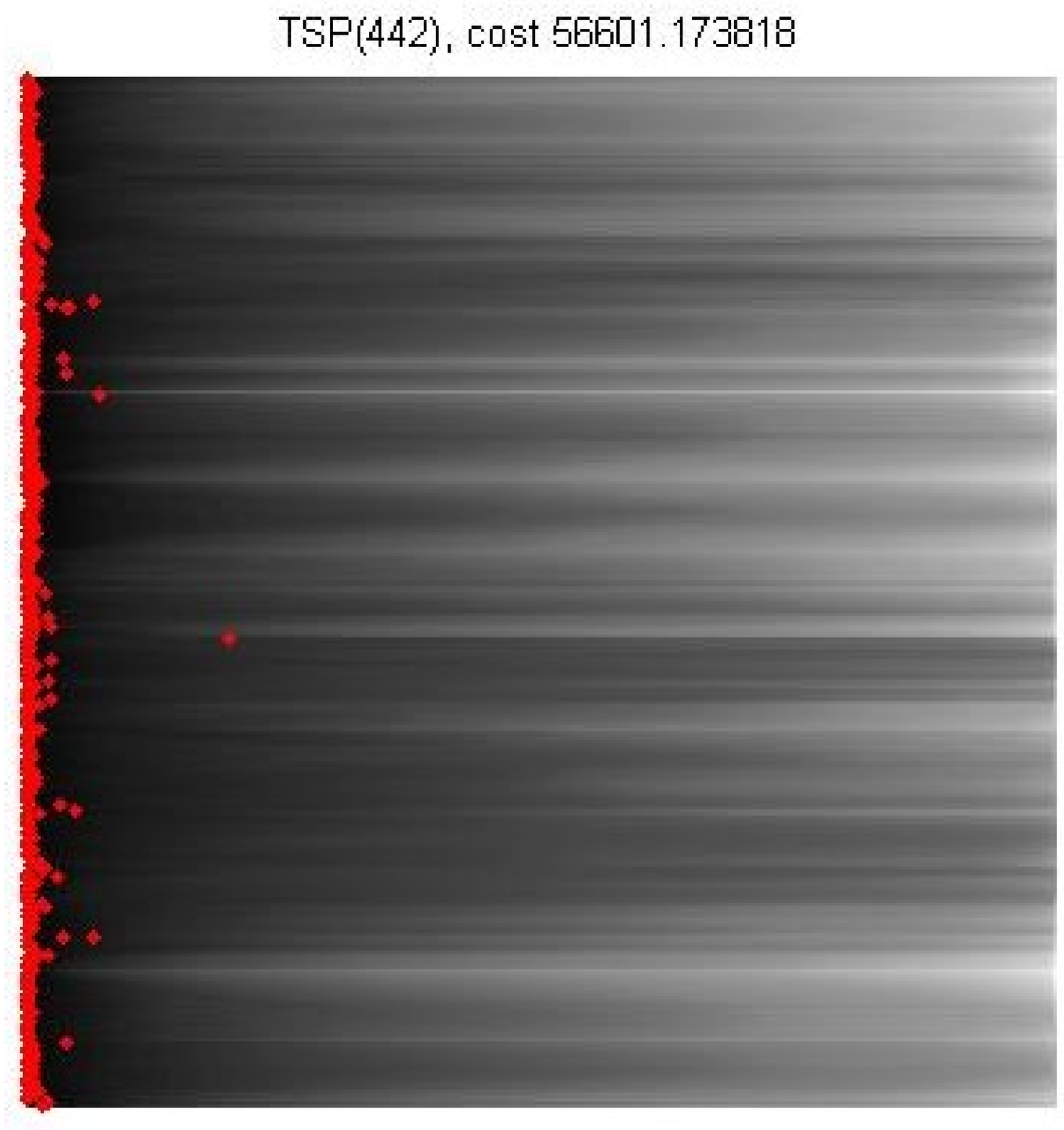,
height=50mm} }
\centerline{ \makebox[2.2in][c]{ a)
}\makebox[2.2in][c]{ b) } } ~
\caption{TSP$_{442}$ final hamiltonian cycle. a) Ordered cost matrix, and b)
sorted cost matrix $\mathcal{M}$.}
\label{fig:tsp442DgMFT}
\end{figure}


\section{Numerical Experiments for TSP and KTP}

~\label{sc:ExpTSPKTP}

The first numerical experiment is for TSP with 152 cities located on a
spiral curve. TPS$_{152}$. Figure~\ref{fig:tsp152_espiral} depicts two very
closed hamiltonian cycles. On a) is the not optimal, its red area has $%
161,279$ pixels, and on b) is the putative optimal, its red area has $%
158,813 $ pixels. The cost's difference between the two
hamiltonian cycles is only $0.021$. Also, the cities locations are
not fulfil the hypothesis of proposition~\ref{prop:concur},
therefore their red areas'size are not as in figures~\ref{fig:tsp17Conv_nOpt} and~\ref%
{fig:tsp17Conv_Opt}.

From~\cite{concorde:page} the files pr76.tsp and pcb442.tsp
provide the results depicted on figures~\ref{fig:tsppr76}
and~\ref{fig:tsppcb442}. The putative optimal complex hamiltonian
cycle of the TSP$_{442}$ is not the global optimal.
Figure~\ref{fig:tsp442DgMFT} depicts on b) many red dots far away
of the left side, as I explains in~\cite{arXiv:Barron2010} this
means that exists yet many alternatives hamiltonian cycles to
explore. A rough estimation of this left research space's size is
$5.625417\times 10^{39}$ cycles. With the simple
algorithm~\ref{alg:CompTSP}, I did not expect to solve this
problem but to get a hamiltonian cycle less than initial cost of
the $221435.5555$. The reduction of the cost is $164,834.3817$.
Because, algorithm~\ref{alg:Detcross} is done by hand, the total
time of execution can not be estimated.

\begin{figure}[tbp]
\centerline{ \psfig{figure=\IMAGESPATH/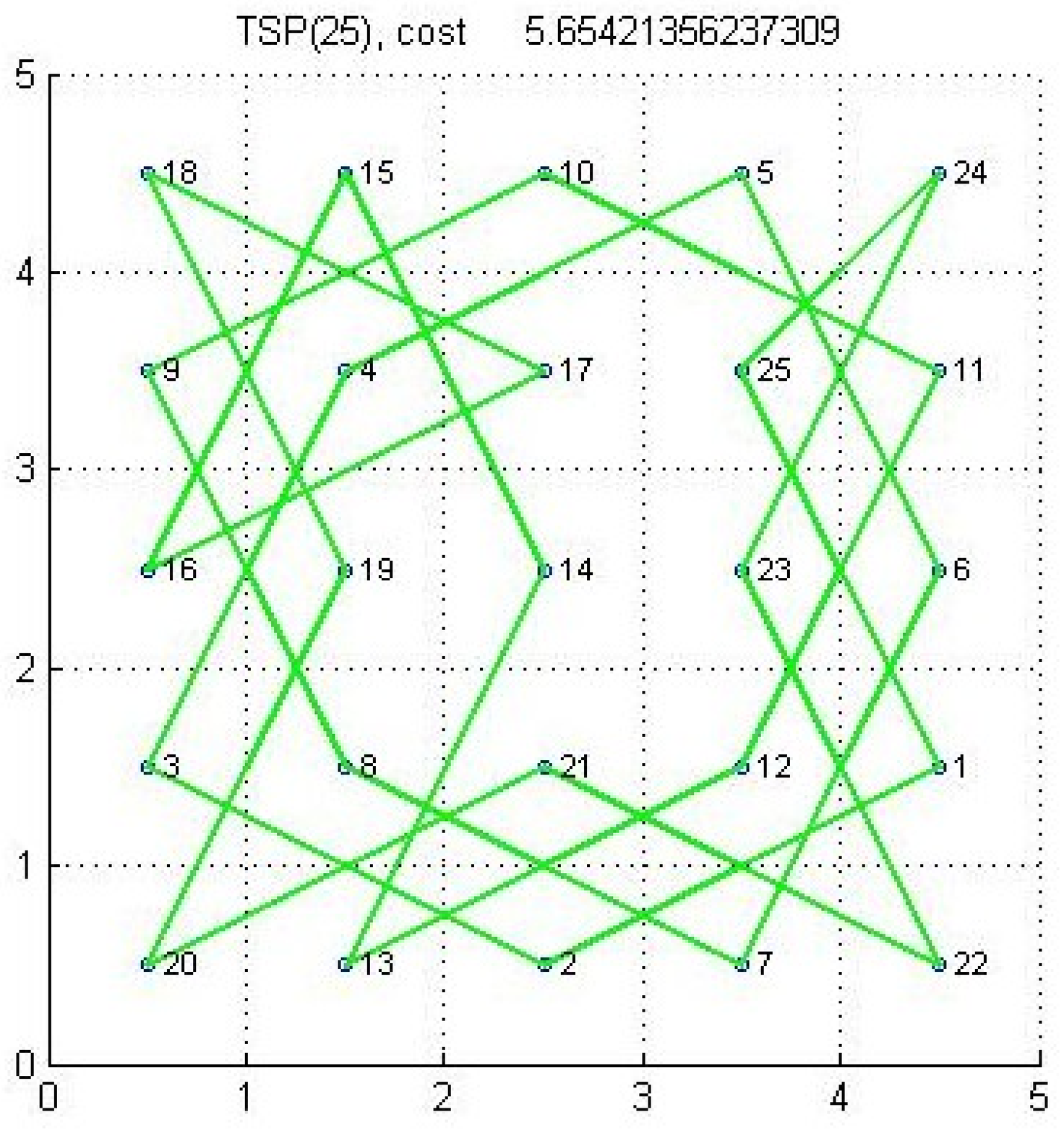,
height=50mm}
\psfig{figure=\IMAGESPATH/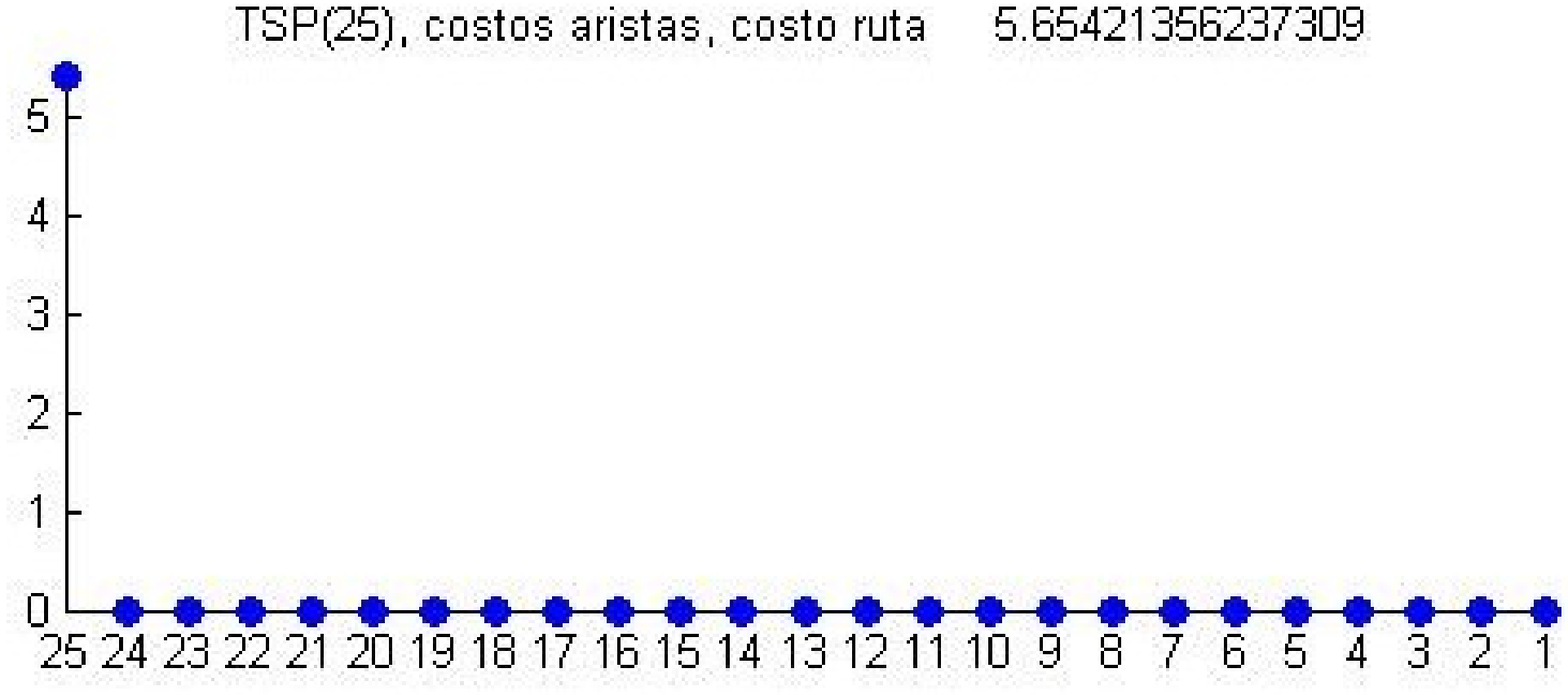,
height=50mm} }
\centerline{ \makebox[2.2in][c]{ a)
}\makebox[2.2in][c]{ b) } } ~
\caption{TKP$_{5\times 5}$. a) final tour, and b) edges'cost.}
\label{fig:5x5FTCT}
\end{figure}

Figures~\ref{fig:5x5FTCT}, and ~\ref{fig:7x7FTCT} depict solutions
where all moves are knight's path but one is not. The b) edge's
cost point out where there is not a knight's move.

\begin{prop}
~\label{prop:evenchessbord} A KTP exists in any chessboard of $m$ $\times $ $%
n$ squares if only if $nm$ is even.
\begin{proof}
A knight jumps from a white square to a black square. Therefore,
in order to complete a hamiltonian cycle with only knight's path,
it is necessary to have equal number of white squares and black
squares.
\end{proof}
\end{prop}

\begin{rem}
Proposition~\ref{prop:ktpeq5} states a very strong property but the
verification is exhaustive. The previos proposition is easy to verify and
together with prop.~\ref{prop:ktpeq5} it is immediately that for KTP in $%
n\times n$ chessboard, $n$ must be greater or equal 4, and even to guaranty
the solution of KTP. A version of the previous proposition with black and
white domino tiles, also requires $nm$ even in order to be tiling a $n
\times m$ chessboard.
\end{rem}

\begin{figure}[tbp]
\centerline{
\psfig{figure=\IMAGESPATH/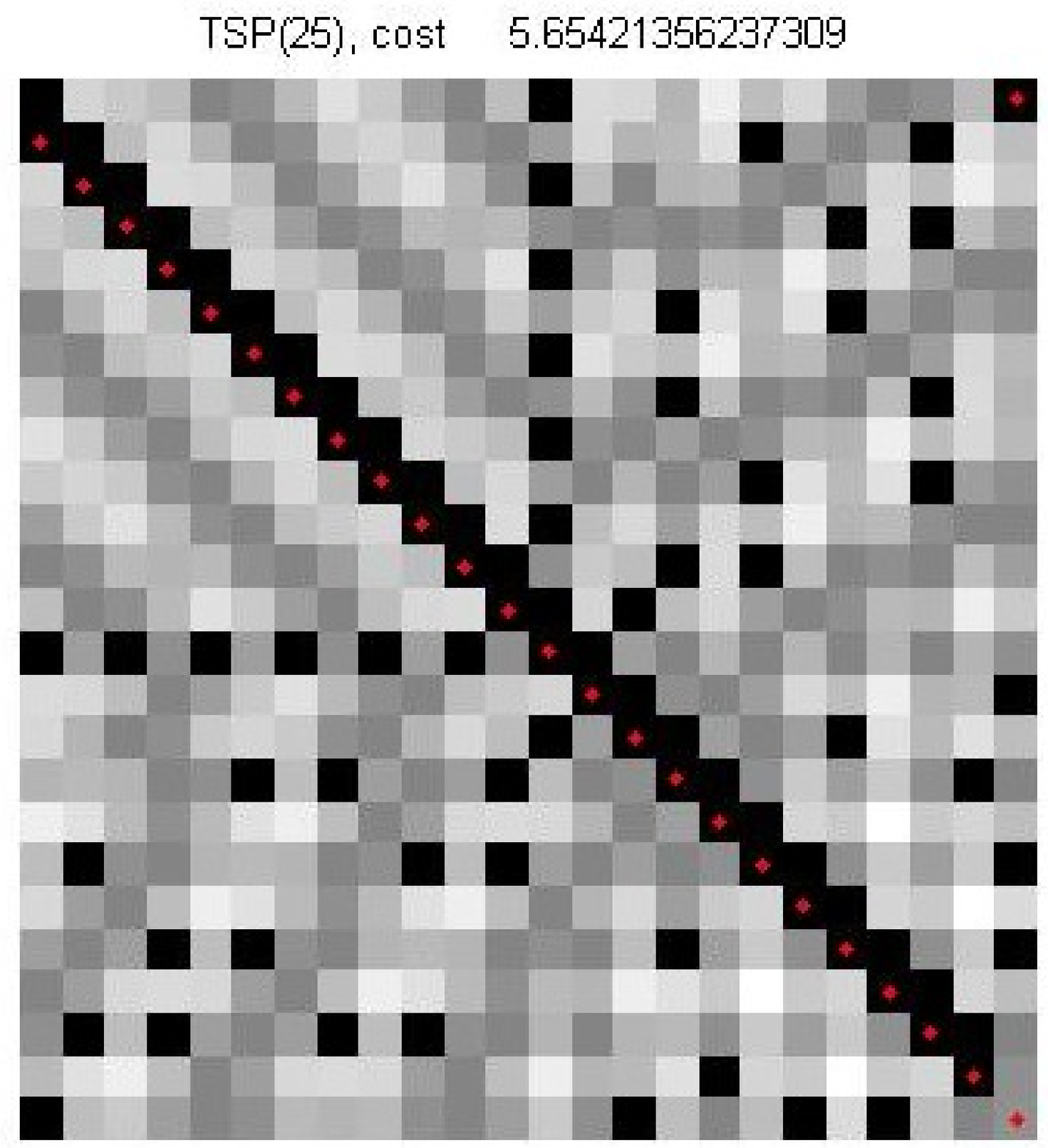,
height=50mm}
\psfig{figure=\IMAGESPATH/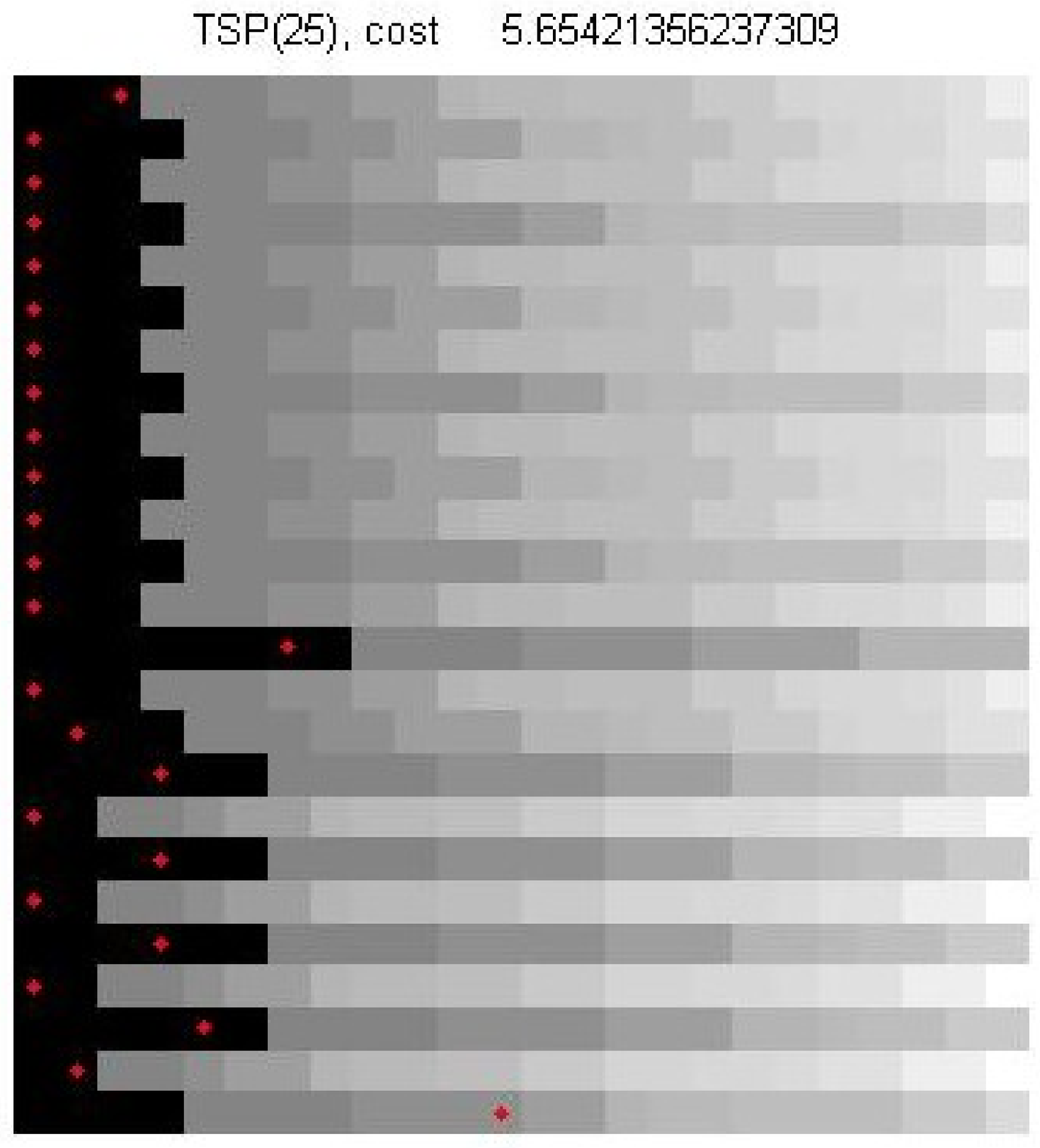,
height=50mm} }
\centerline{ \makebox[2.2in][c]{ a)
}\makebox[2.2in][c]{ b) } } ~
\caption{TKP$_{5\times 5}$ of the final tour. a) Ordered cost matrix, and b)
sorted cost matrix $\mathcal{M}$.}
\label{fig:5x5DgMFT}
\end{figure}

Therefore, for $5\times5$ and $7\times 7$ chessboards there are not
hamiltonian cycles with only knight's path. Algorithm~\ref{alg:CompKTP} is
computable because it was designed for a chessboard $m$ $\times $ $n$
squares with $nm$, even. For clarity, prop.~\ref{prop:ktpeq5} states that
hamiltonian trajectories exist in any knight graph. The prop.~\ref%
{prop:ktpeq5} and ~\ref{prop:evenchessbord} are the kind of
details that frequently happen in the design of efficient methods
for instances of NP like problems. Previous knowledge, formal
propositions, or heuristic ideas work very well for a instance
problem or case, but, they do not work for similar and related
problem. In order to solve, $5\times5$ and $7\times 7$ chessboards
repeat-end control structure for 200 times replaces the while-end
in lines 2 through 5. The sorted cost matrix $\mathcal{m}$ in
figures~\ref{fig:5x5DgMFT}, and ~\ref{fig:7x7DgMFT} b) depict that
there are many similar solutions because not all red dots are on
the left side. The value of the final cost in
fig.~\ref{fig:5x5FTCT}, and ~\ref{fig:7x7FTCT} b) depict a
hamiltonian trajectory omitting one edge. The cost values or the
edge's cost graphic are indicators to allow us the corroboration
of the solution without additional computational cost.


\begin{figure}[tbp]
\centerline{ \psfig{figure=\IMAGESPATH/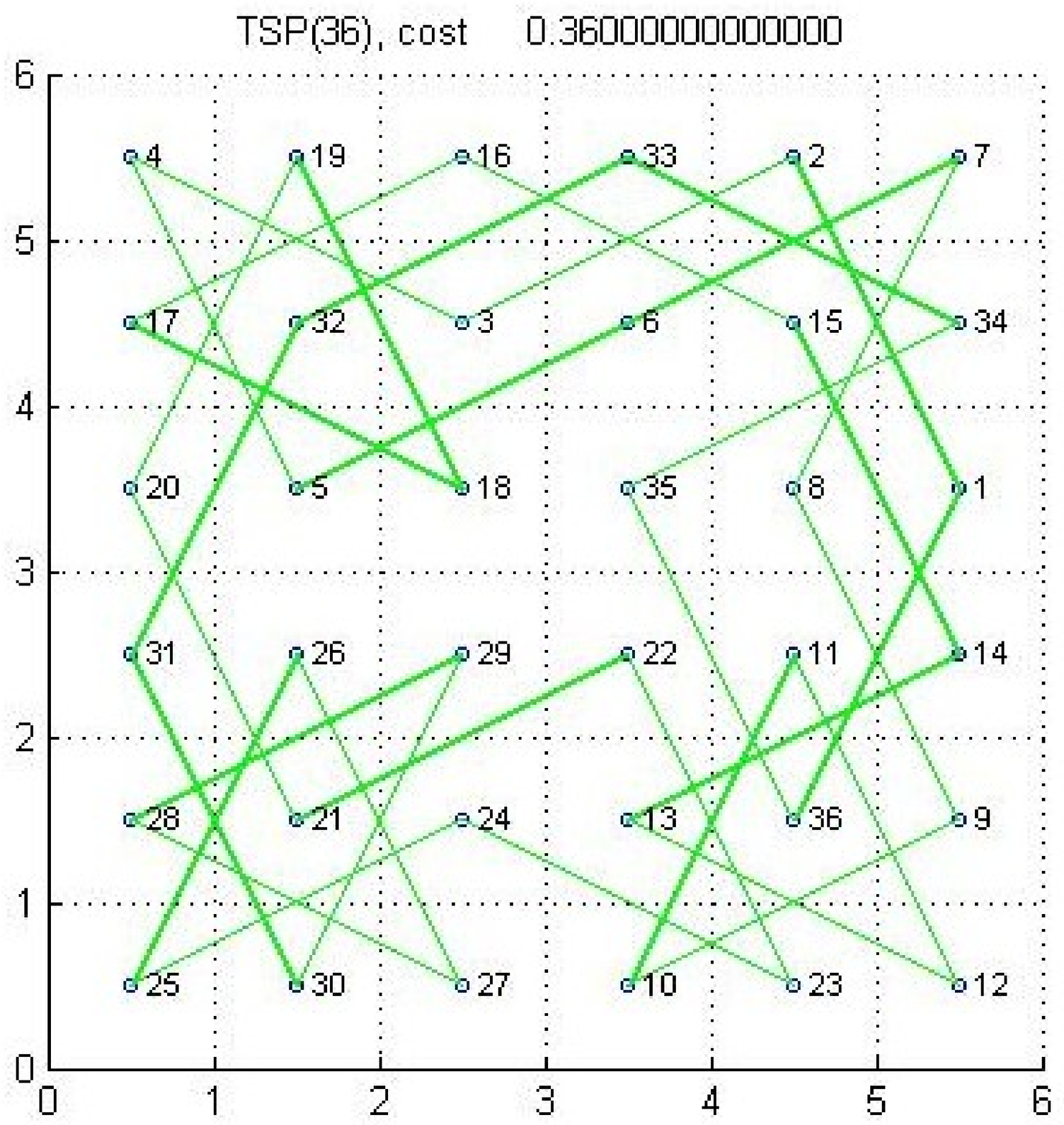,
height=50mm}
\psfig{figure=\IMAGESPATH/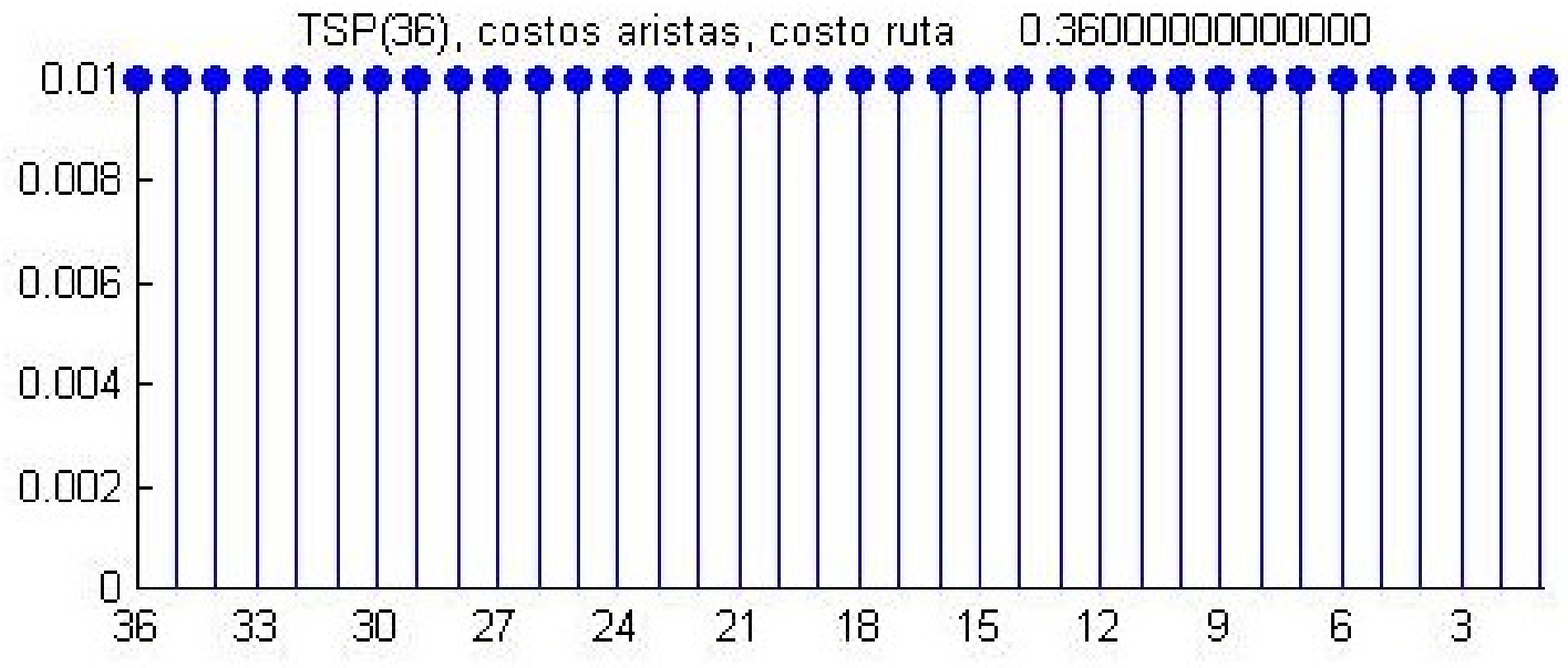,
height=50mm} }
\centerline{ \makebox[2.2in][c]{ a)
}\makebox[2.2in][c]{ b) } } ~
\caption{TKP$_{6\times 6}$. a) final tour, and b) edges'cost.}
\label{fig:6x6FTCT}
\end{figure}

\begin{figure}[tbp]
\centerline{
\psfig{figure=\IMAGESPATH/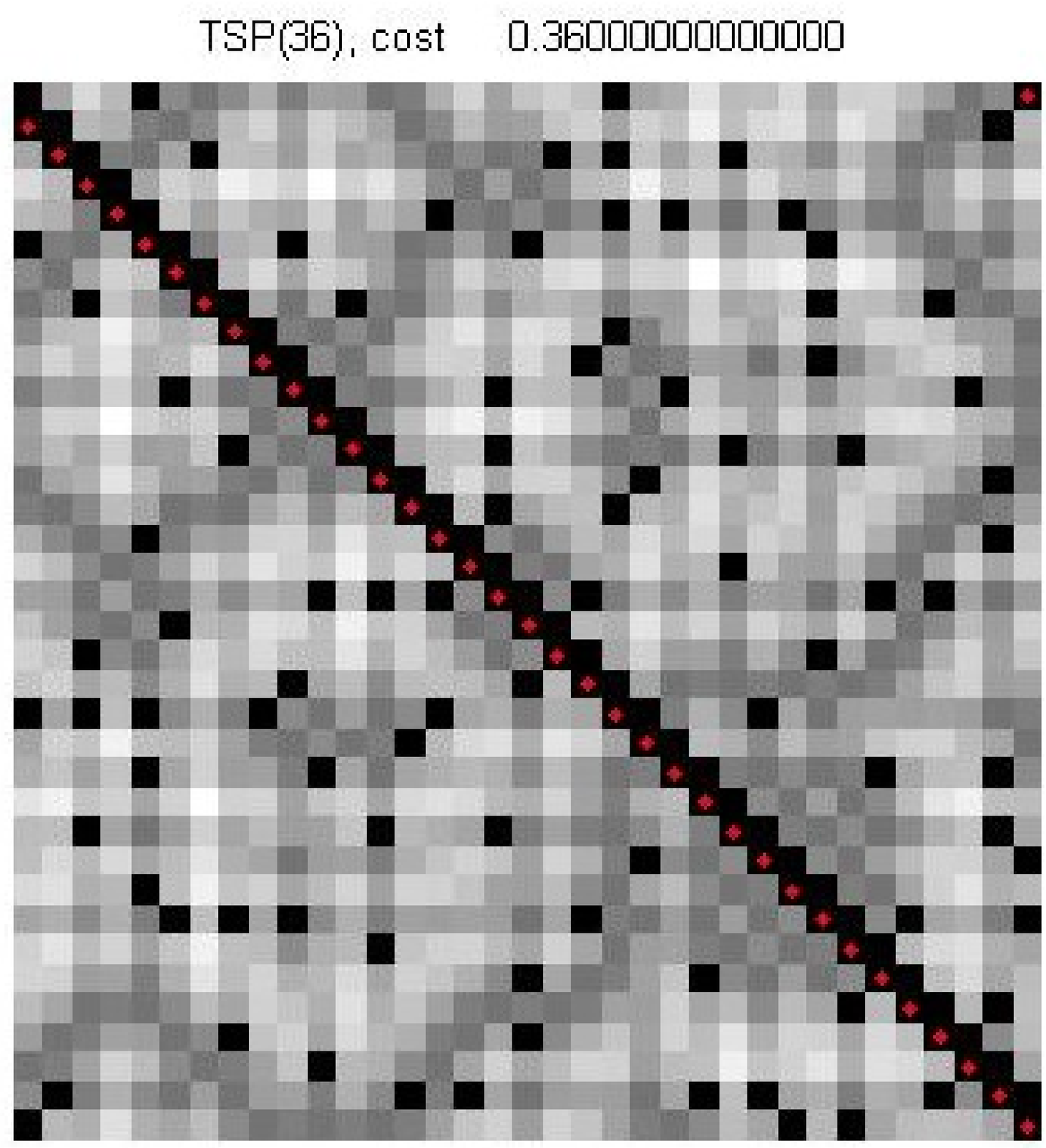,
height=50mm}
\psfig{figure=\IMAGESPATH/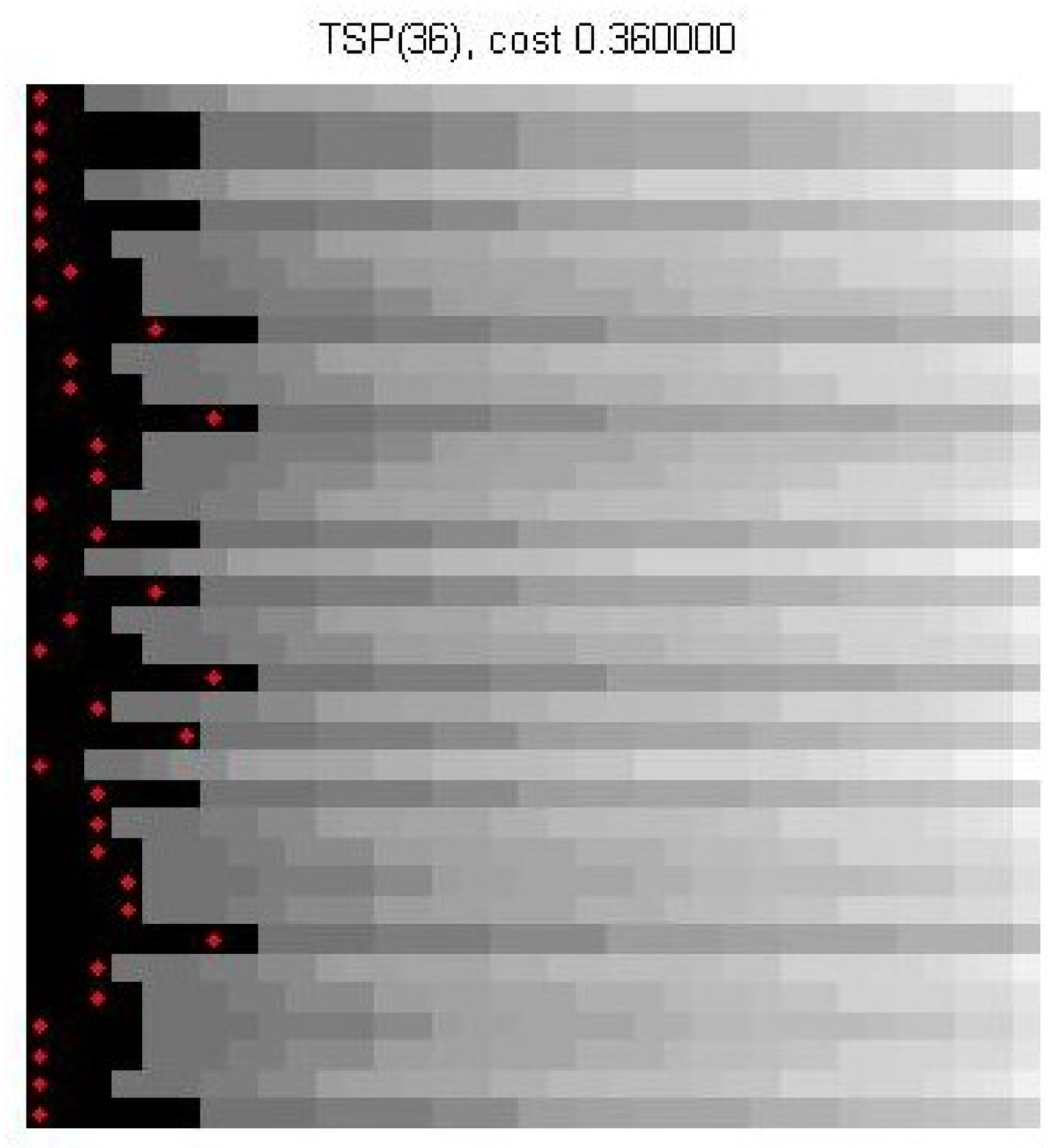,
height=50mm} }
\centerline{ \makebox[2.2in][c]{ a)
}\makebox[2.2in][c]{ b) } } ~
\caption{TKP$_{6\times 6}$ of the final tour. a) Ordered cost matrix, and b)
sorted cost matrix $\mathcal{M}$.}
\label{fig:6x6DgMFT}
\end{figure}

\begin{figure}[tbp]
\centerline{
\psfig{figure=\IMAGESPATH/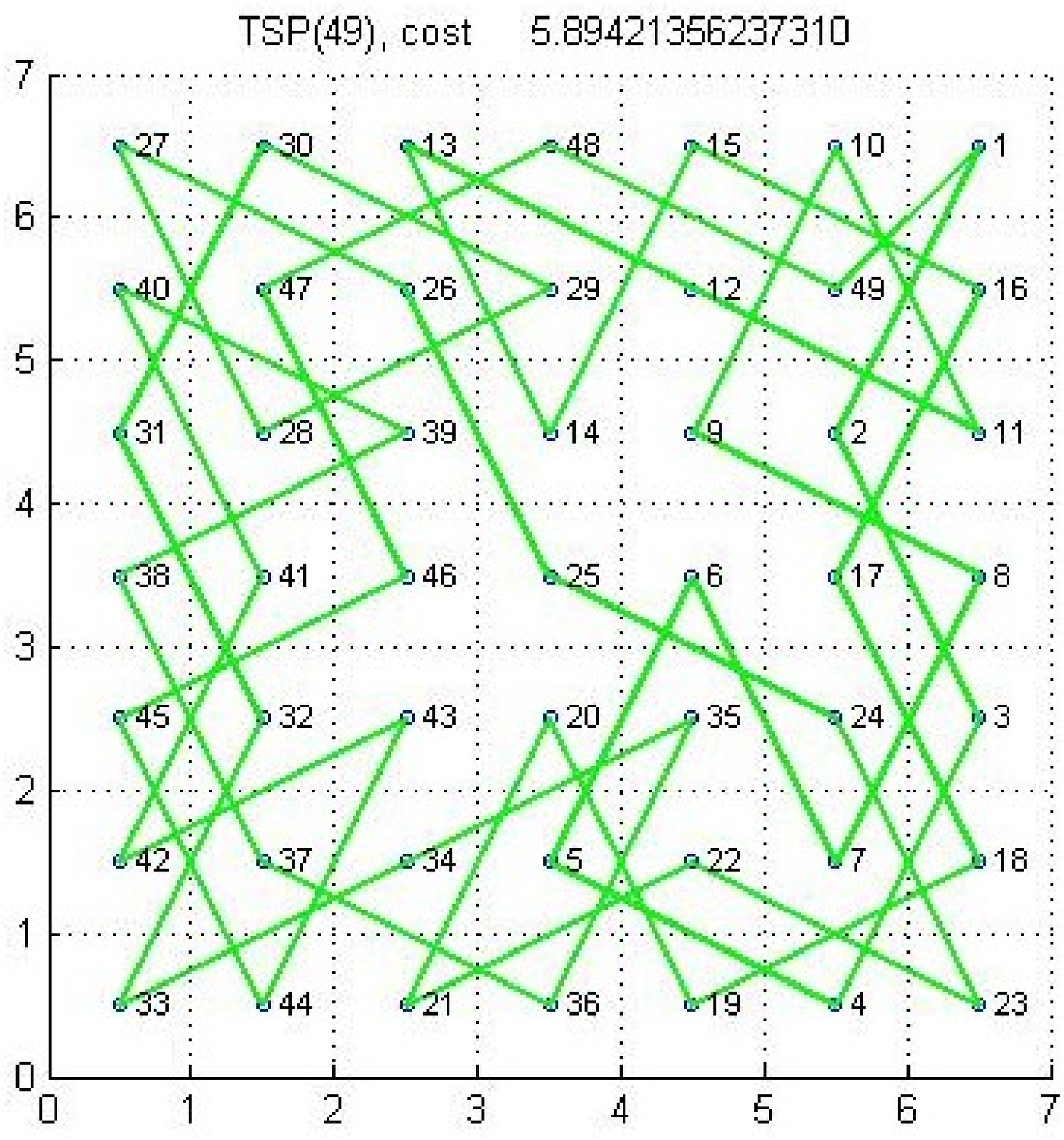,
height=50mm}
\psfig{figure=\IMAGESPATH/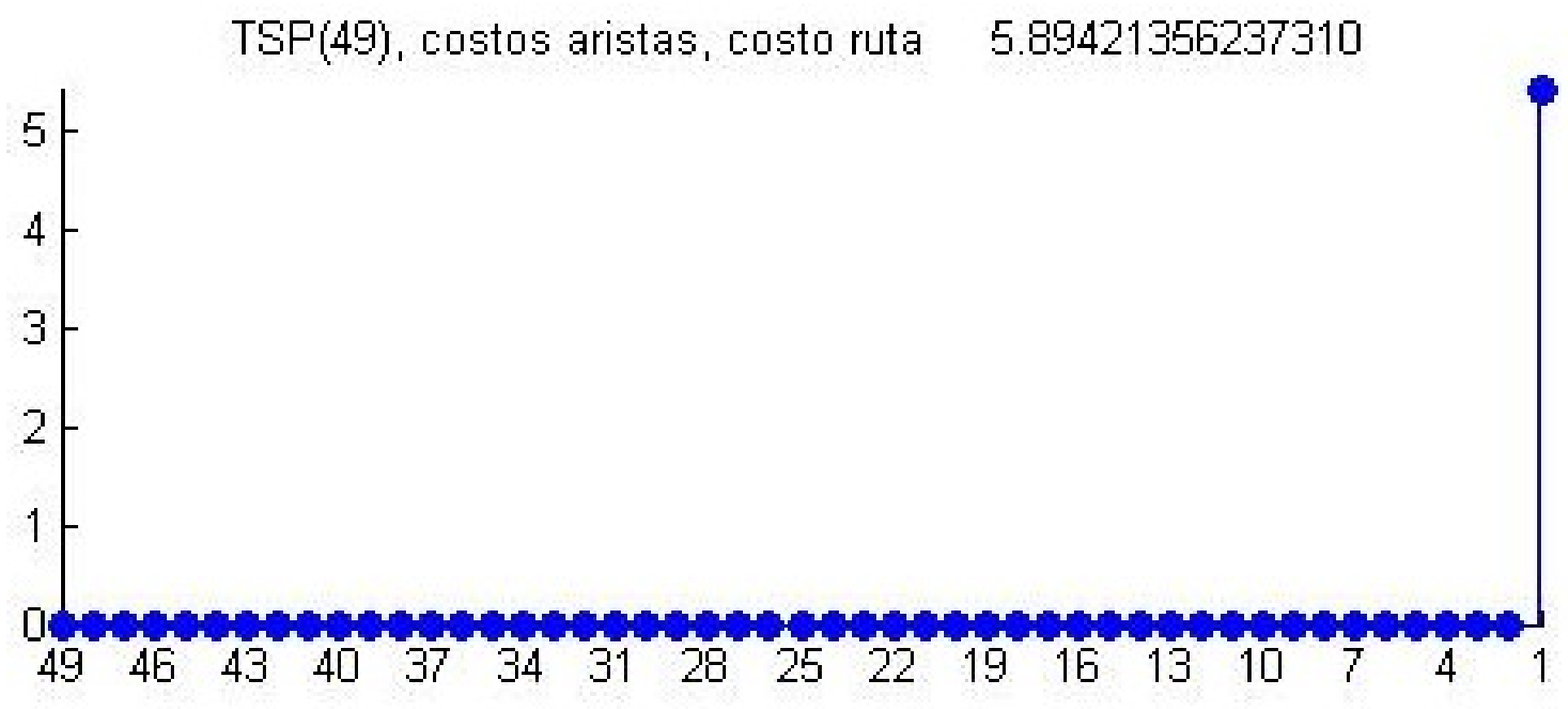,
height=50mm} }
\centerline{ \makebox[2.2in][c]{ a)
}\makebox[2.2in][c]{ b) } } ~
\caption{TKP$_{7\times 7}$. a) final tour, and b) edges'cost.}
\label{fig:7x7FTCT}
\end{figure}

\begin{figure}[tbp]
\centerline{
\psfig{figure=\IMAGESPATH/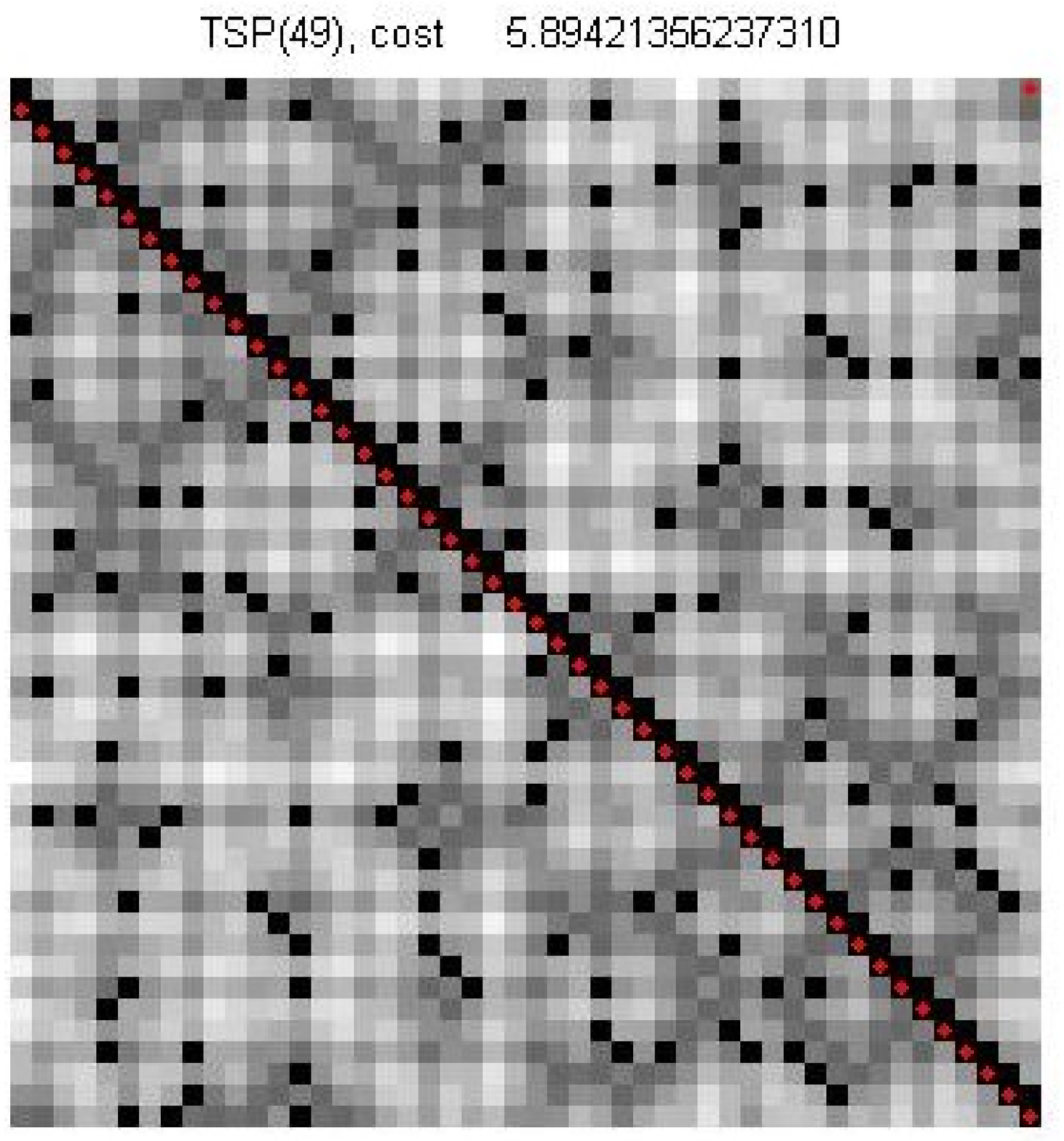,
height=50mm}
\psfig{figure=\IMAGESPATH/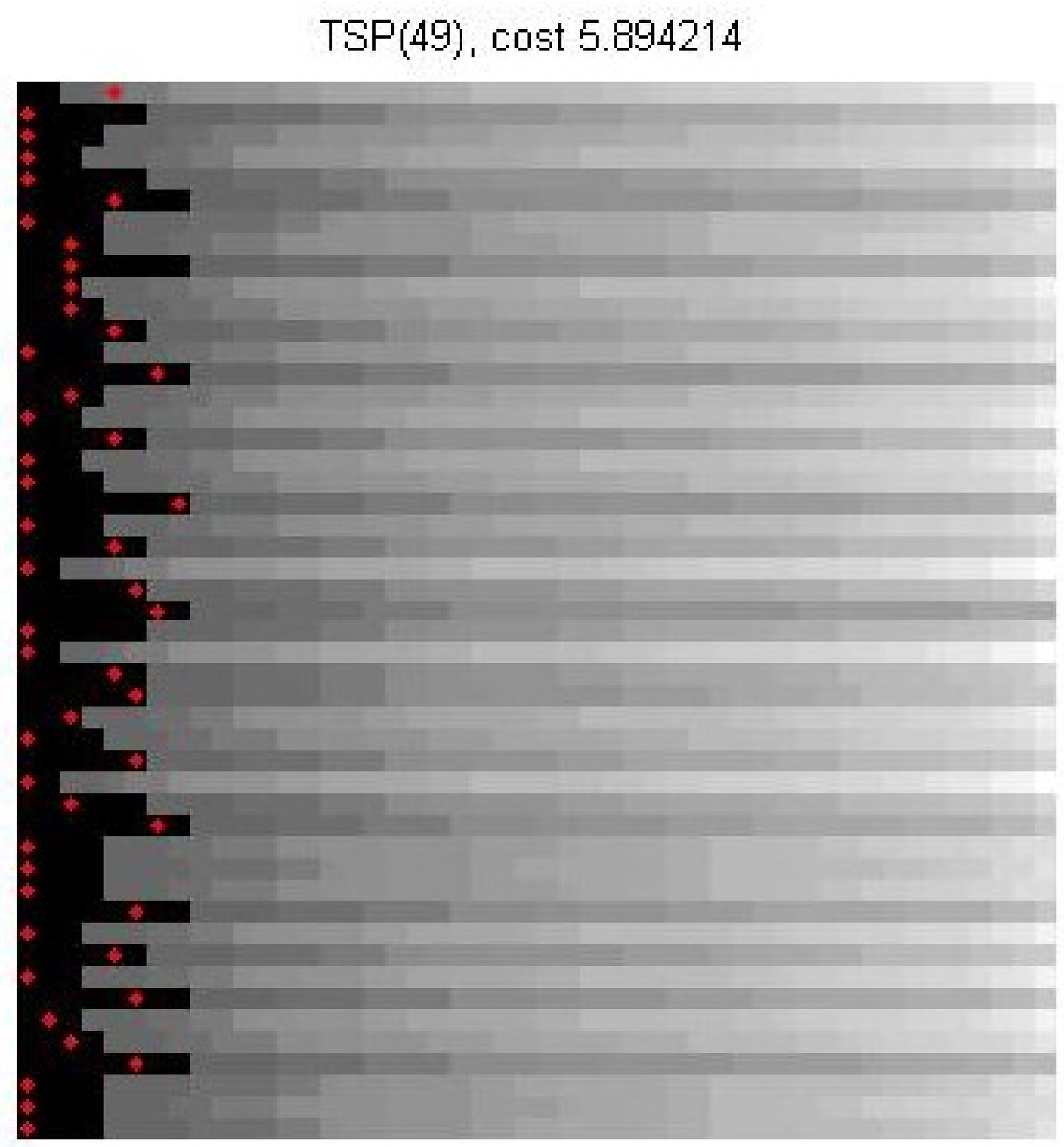,
height=50mm} }
\centerline{ \makebox[2.2in][c]{ a)
}\makebox[2.2in][c]{ b) } } ~
\caption{TKP$_{7\times 7}$ of the final tour. a) Ordered cost matrix, and b)
sorted cost matrix $\mathcal{M}$.}
\label{fig:7x7DgMFT}
\end{figure}

\begin{figure}[tbp]
\centerline{ \psfig{figure=\IMAGESPATH/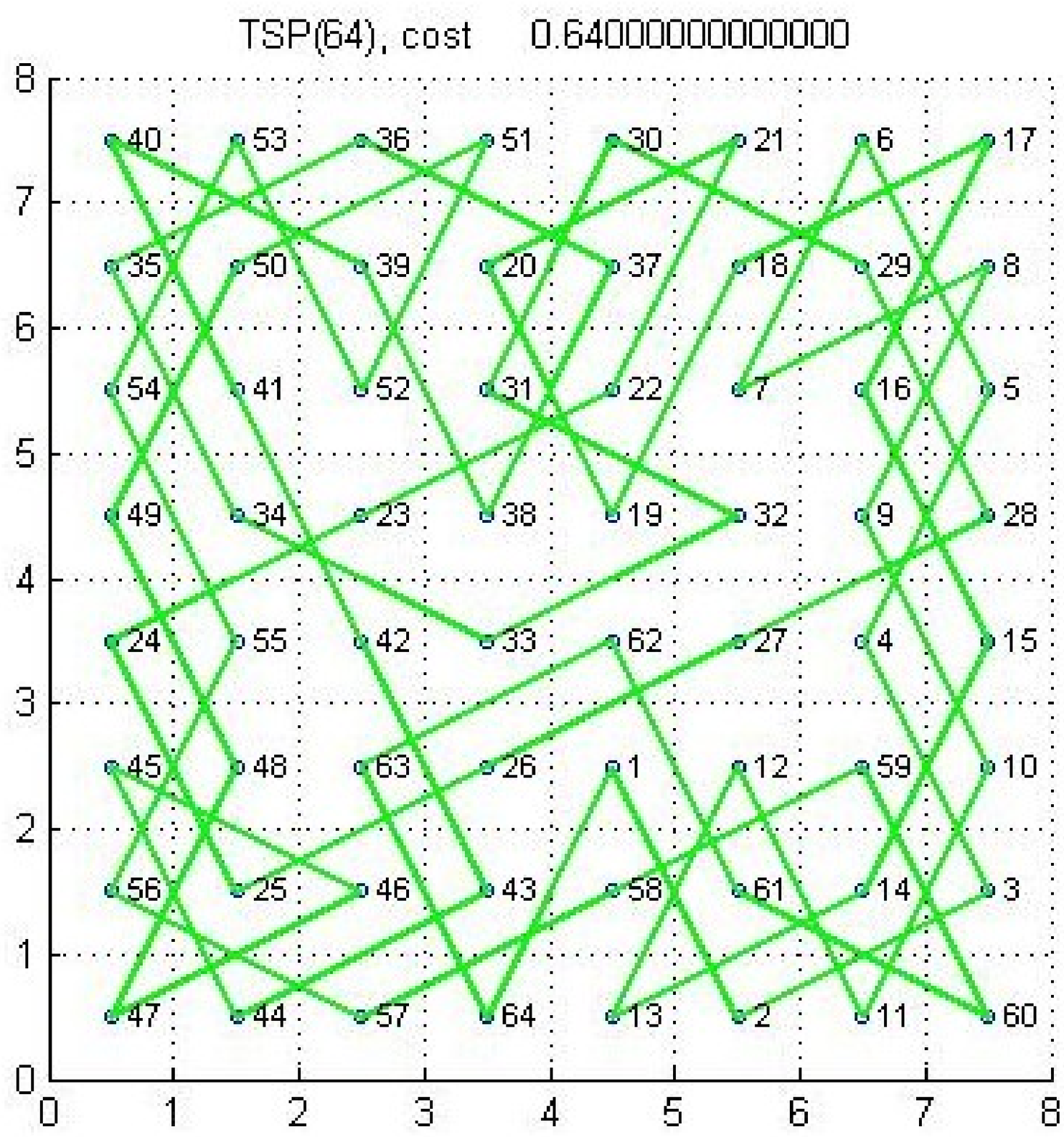,
height=50mm}
\psfig{figure=\IMAGESPATH/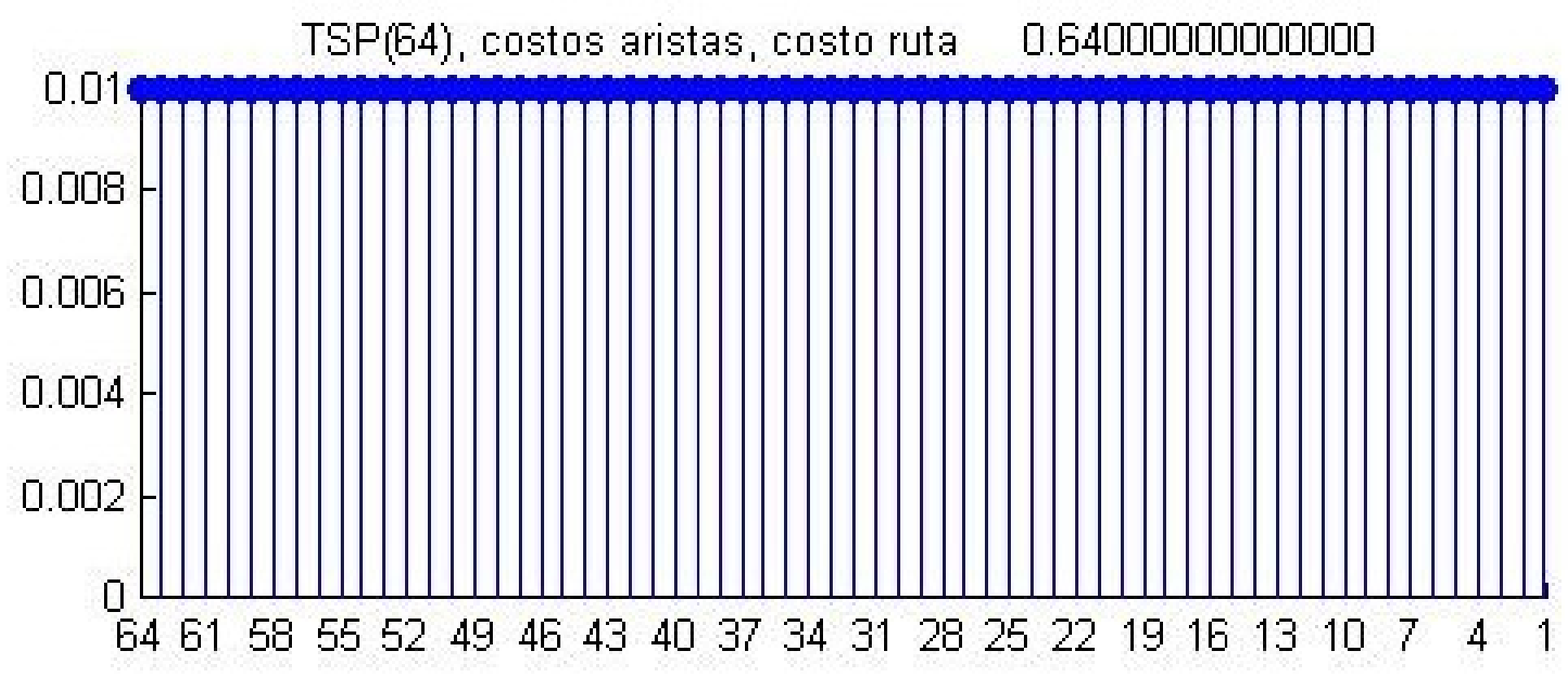,
height=50mm} }
\centerline{ \makebox[2.2in][c]{ a)
}\makebox[2.2in][c]{ b) } } ~
\caption{TKP$_{8\times 8}$. a) final tour, and b) edges'cost.}
\label{fig:8x8NEFTCT}
\end{figure}

\begin{figure}[tbp]
\centerline{
\psfig{figure=\IMAGESPATH/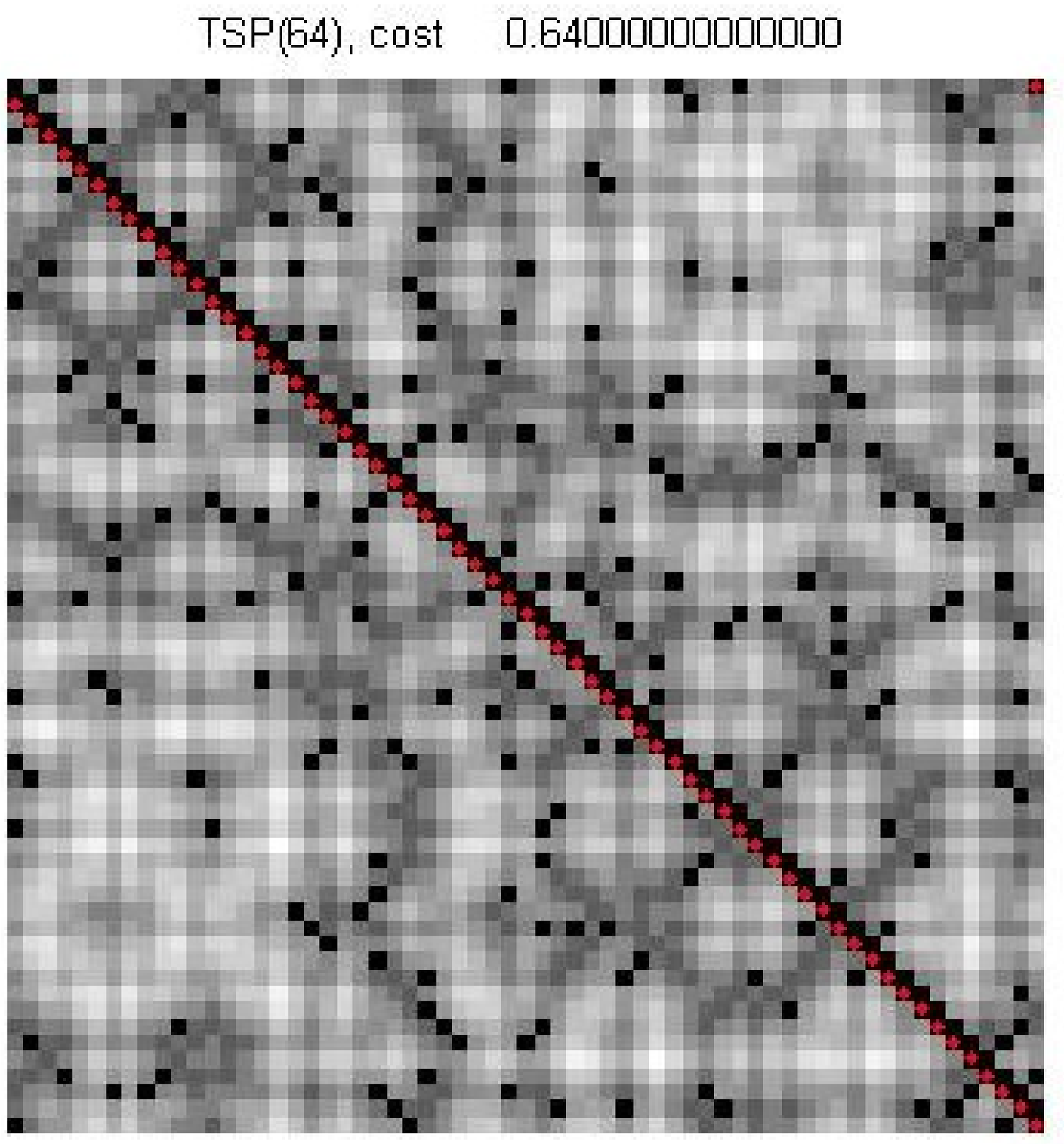,
height=50mm}
\psfig{figure=\IMAGESPATH/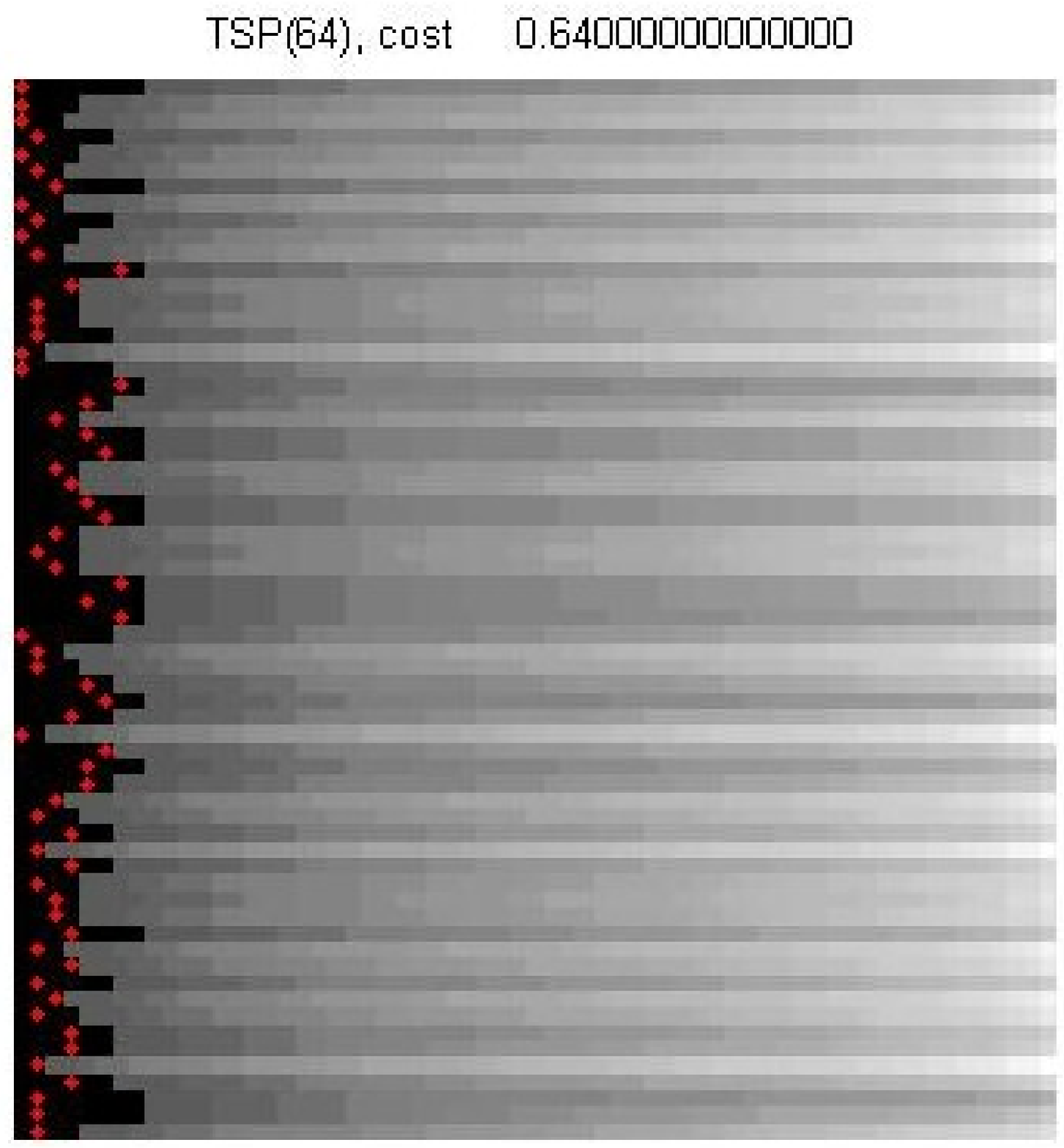,
height=50mm} }
\centerline{ \makebox[2.2in][c]{ a)
}\makebox[2.2in][c]{ b) } } ~
\caption{TKP$_{8\times 8}$ of the final tour. a) Ordered cost matrix, and b)
sorted cost matrix $\mathcal{M}$.}
\label{fig:8x8NEDgMFT}
\end{figure}

\begin{figure}[tbp]
\centerline{
\psfig{figure=\IMAGESPATH/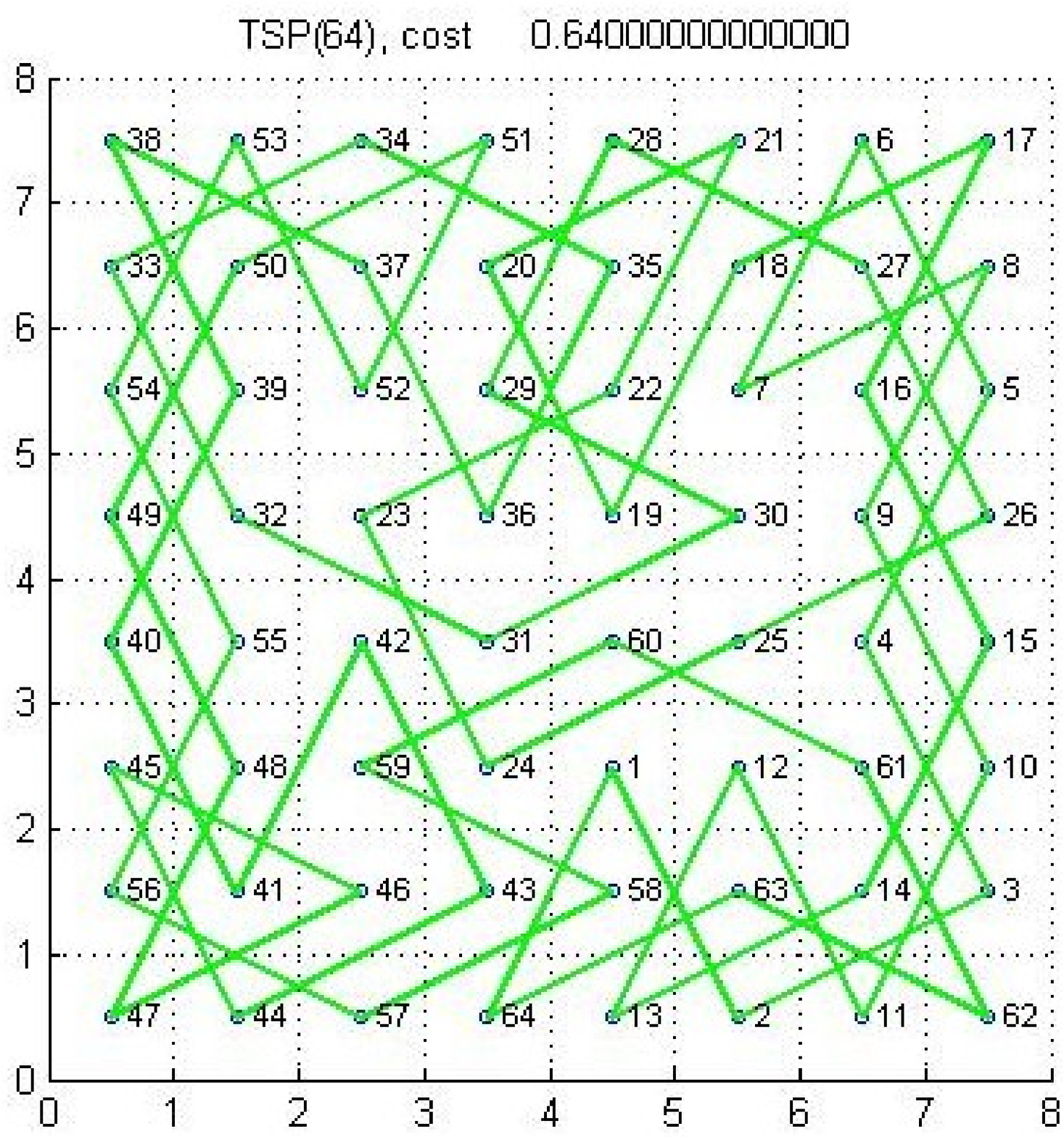,
height=50mm}
\psfig{figure=\IMAGESPATH/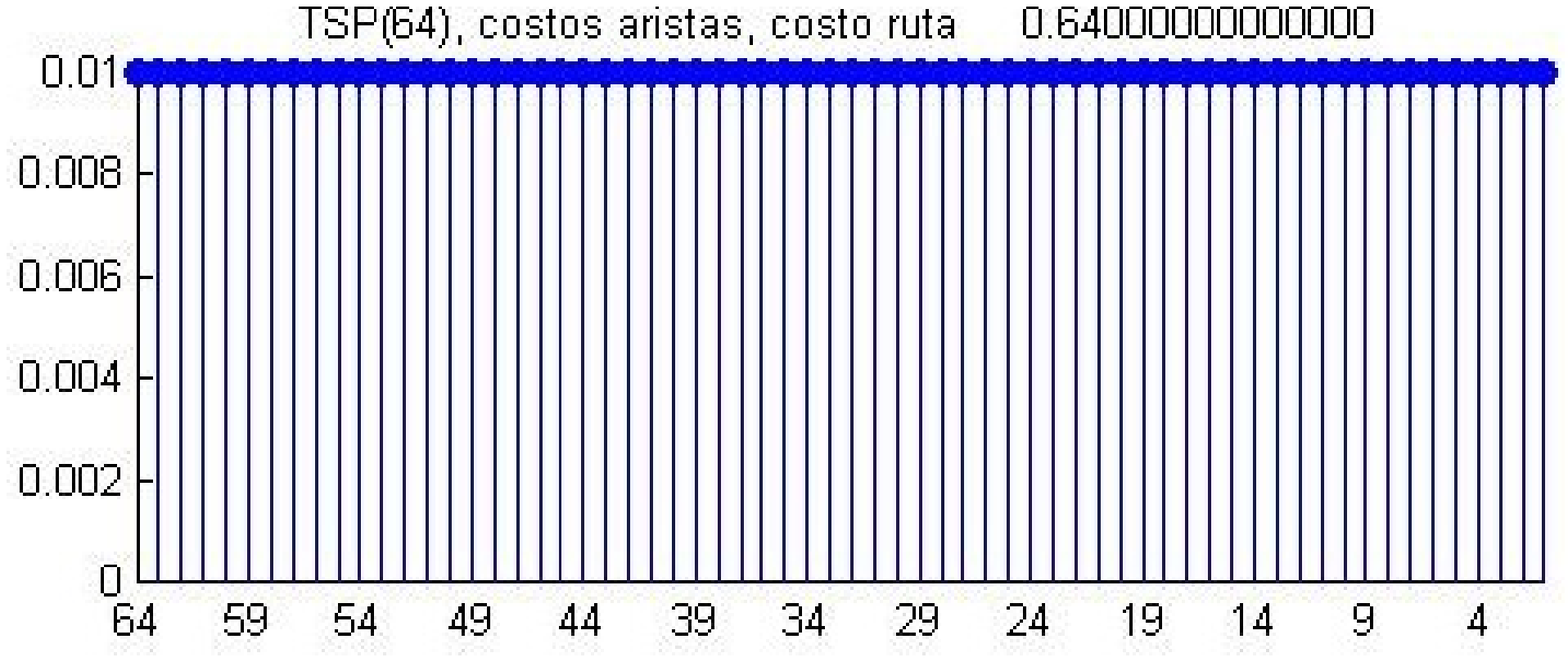,
height=50mm} }
\centerline{ \makebox[2.2in][c]{ a)
}\makebox[2.2in][c]{ b) } } ~
\caption{TKP$_{8\times 8}$. a) final tour 2, and b) edges'cost.}
\label{fig:8x8V2NEFTCT}
\end{figure}

\begin{figure}[tbp]
\centerline{
\psfig{figure=\IMAGESPATH/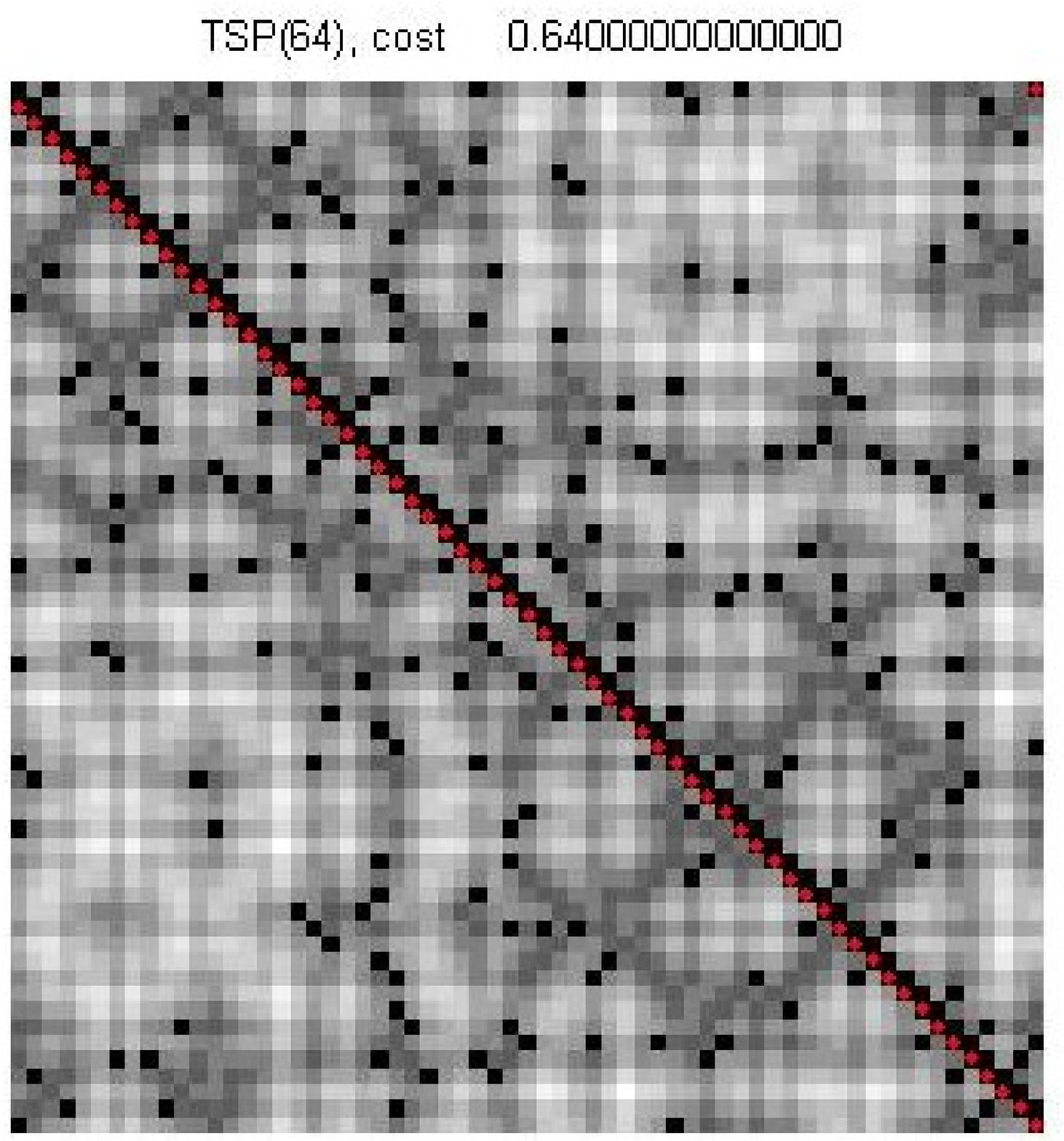,
height=50mm}
\psfig{figure=\IMAGESPATH/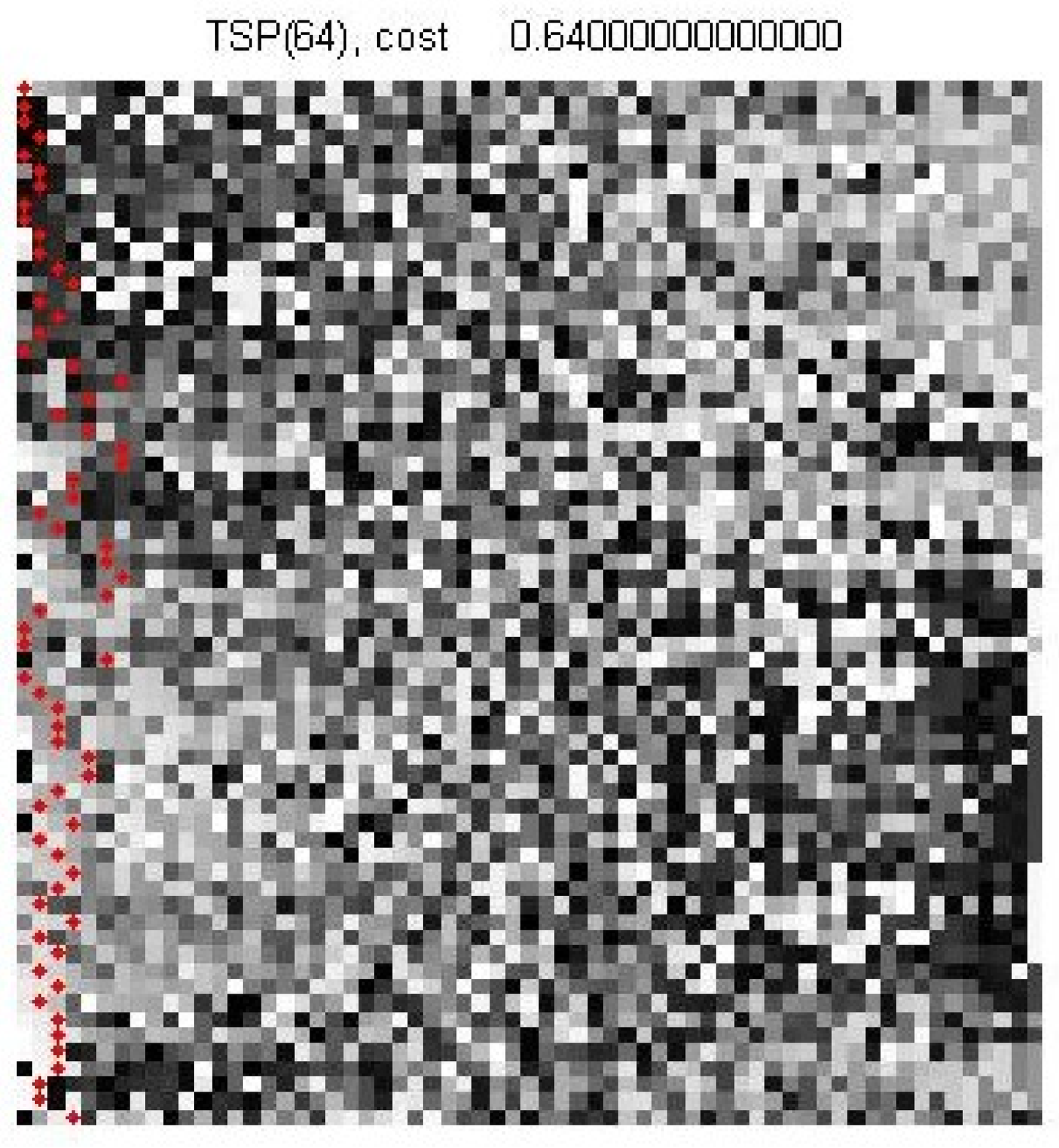,
height=50mm} }
\centerline{ \makebox[2.2in][c]{ a)
}\makebox[2.2in][c]{ b) } } ~
\caption{TKP$_{8\times 8}$ of the final tour 2. a) Ordered cost matrix, and
b) sorted cost matrix $\mathcal{M}$.}
\label{fig:8x8V2NEDgMFT}
\end{figure}


\begin{figure}[tbp]
\centerline{ \psfig{figure=\IMAGESPATH/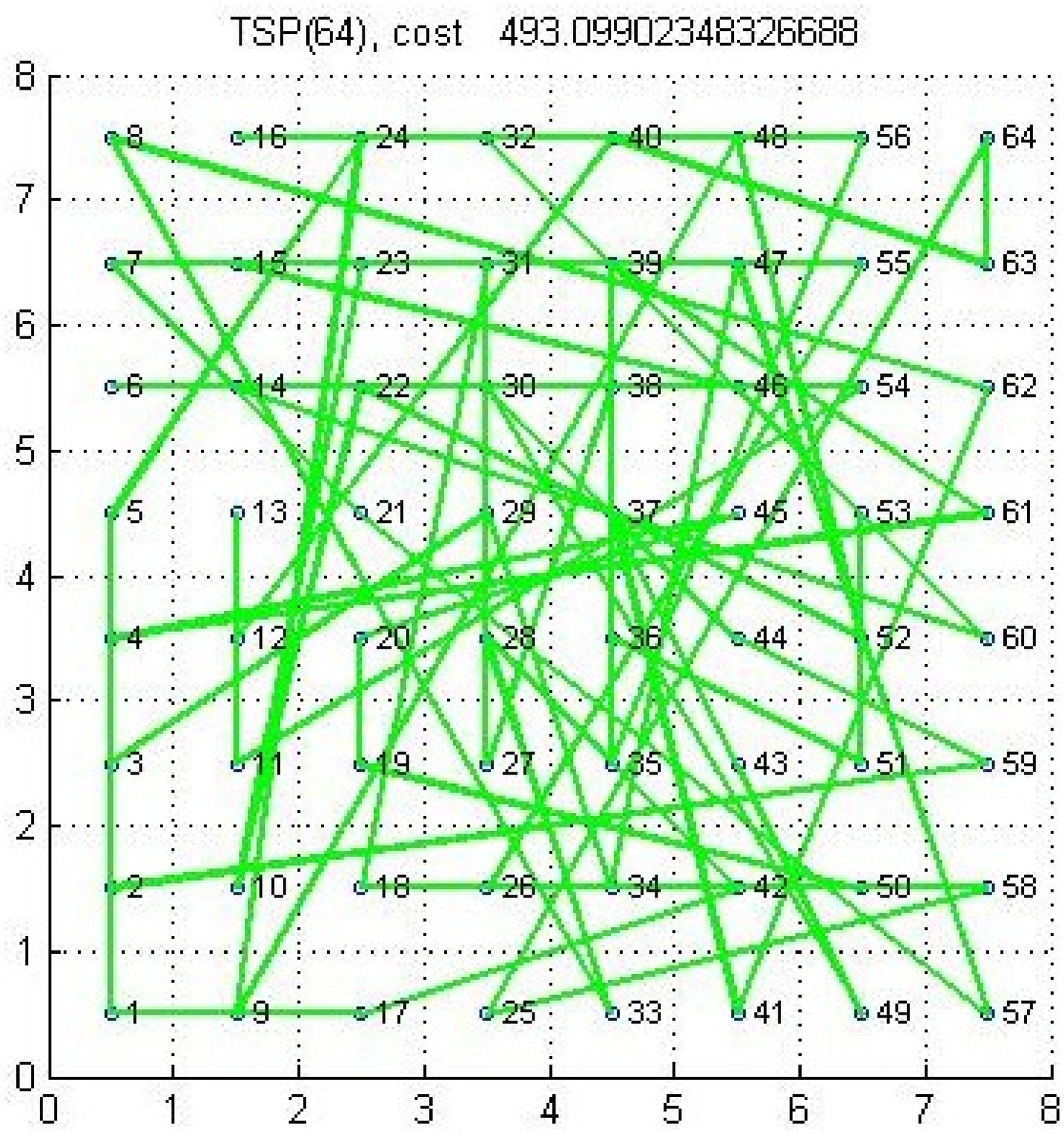, height=50mm}
\psfig{figure=\IMAGESPATH/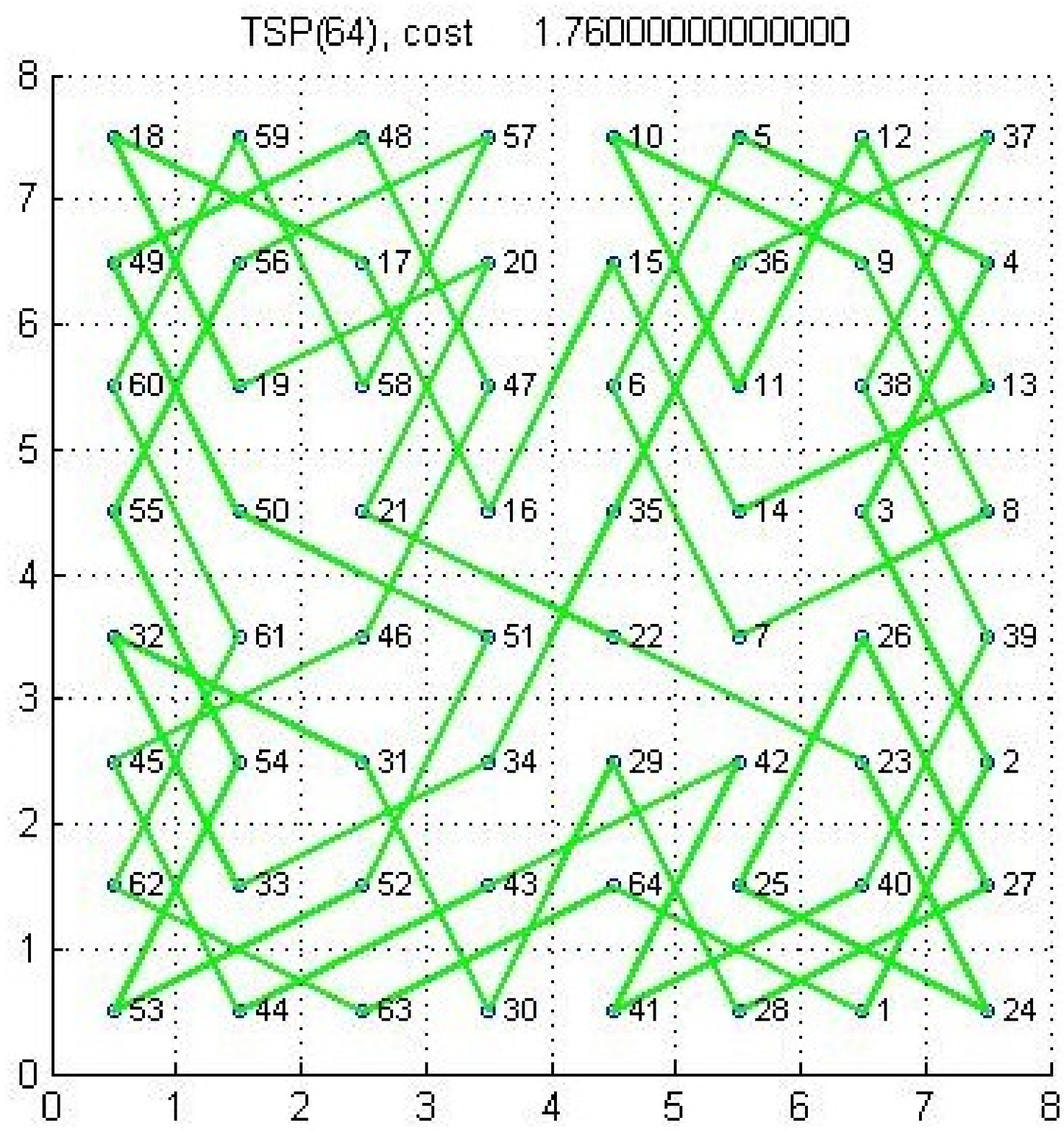, height=50mm} }
\centerline{ \makebox[2.2in][c]{ a) }\makebox[2.2in][c]{ b) } } ~
\caption{TKP$_{8\times 8}$. a) Initial tour, and b) final tour
with cost $<$ 4.} \label{fig:8x8IniFinTour}
\end{figure}

\begin{figure}[tbp]
\centerline{\psfig{figure=\IMAGESPATH/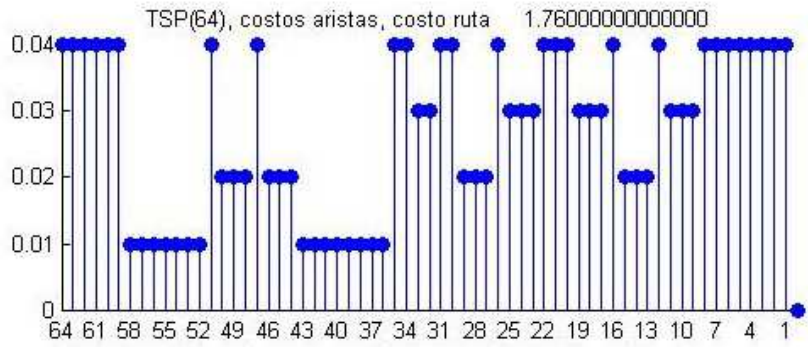,
height=50mm}} ~
\caption{TKP$_{8\times 8}$. Edge's cost depicts that they correspond only to
knight's motions for the final tour.}
\label{fig:8x8FT_cost}
\end{figure}

\begin{figure}[tbp]
\centerline{ \psfig{figure=\IMAGESPATH/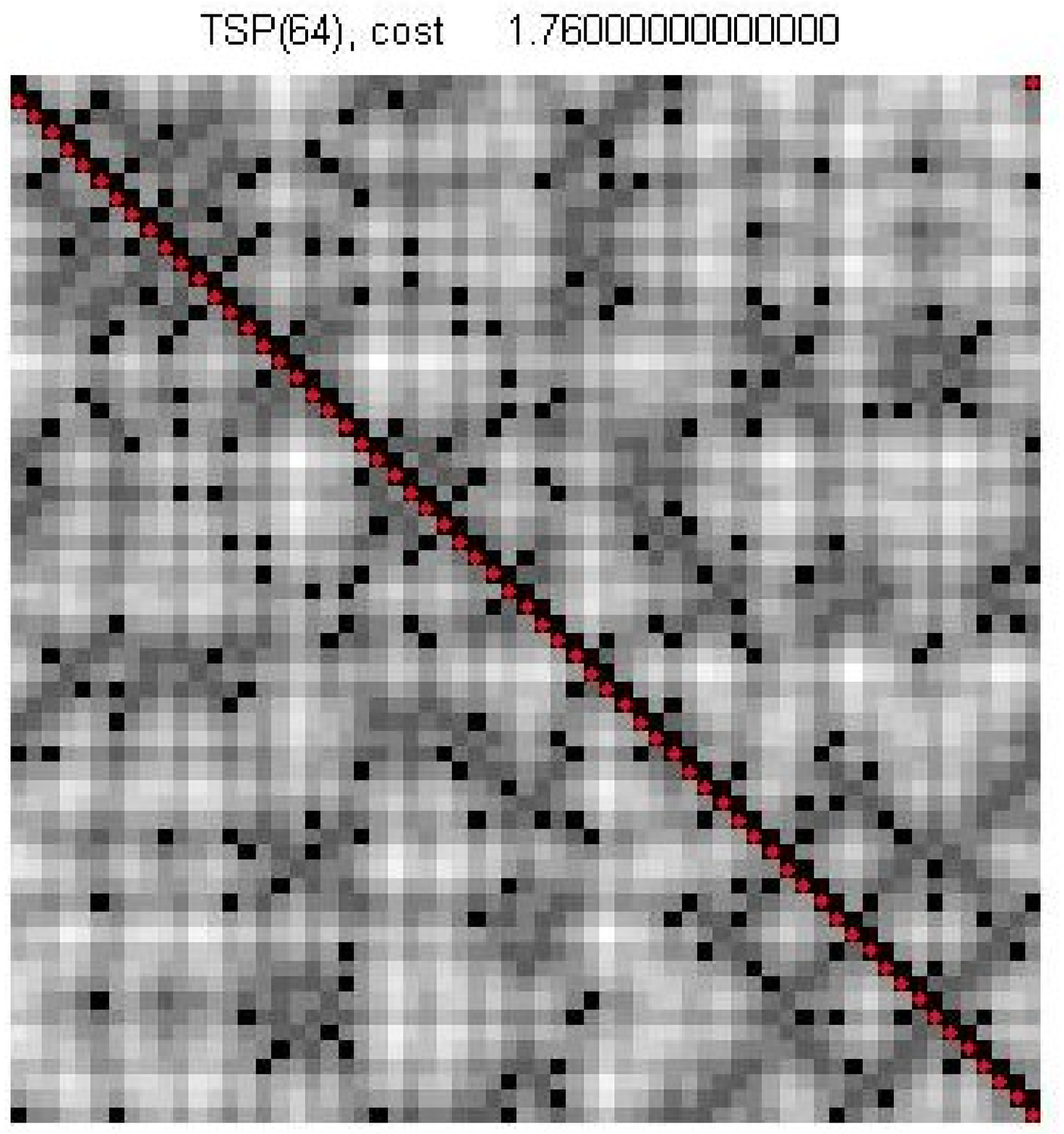,
height=50mm} \psfig{figure=\IMAGESPATH/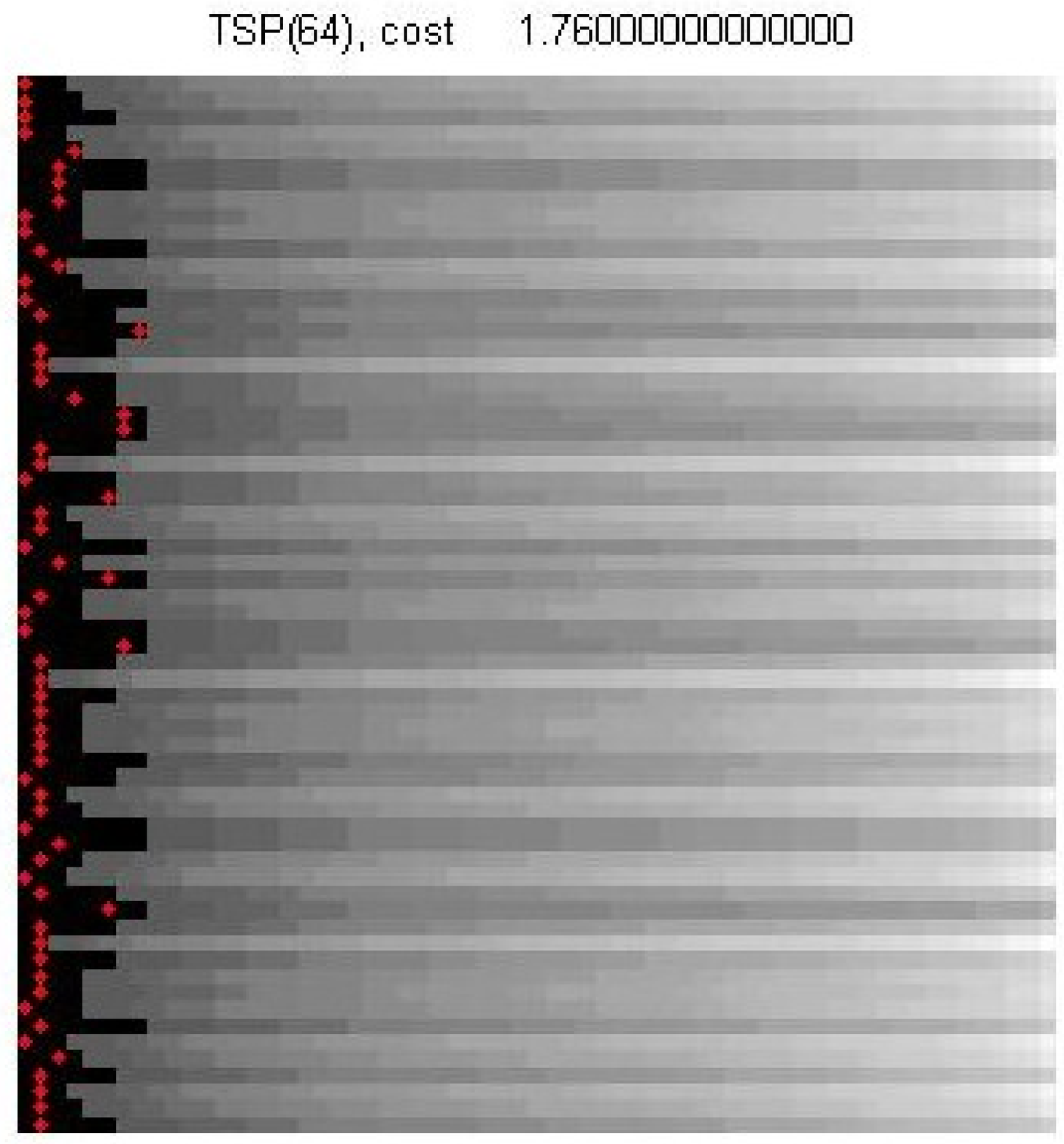,
height=50mm} }
\centerline{ \makebox[2.2in][c]{ a)
}\makebox[2.2in][c]{ b) } } ~
\caption{TKP$_{8\times 8}$ of the final tour. a) Ordered cost matrix, and b)
sorted cost matrix $\mathcal{M}$.}
\label{fig:8x8DgMFT}
\end{figure}

Figure~\ref{fig:8x8IniFinTour} depicts the initial and final tour, and
figure~\ref{fig:8x8FT_cost} depicts that the final tour corresponds only to
knight's paths, i.e., $\left( i-i^{\prime }\right) ^{2}+\left( j-j^{\prime
}\right) ^{2}= 5$ where $(i,j)$ and $(i^{\prime },j^{\prime })$ are the
integer coordinates of the squares of the final hamiltonian cycle. For the
figure~\ref{fig:8x8DgMFT} an heuristical estimation of the posibles
alternatives or a rough estimation of the research space's size is $%
29,030,400$.

\begin{figure}[tbp]
\centerline{
\psfig{figure=\IMAGESPATH/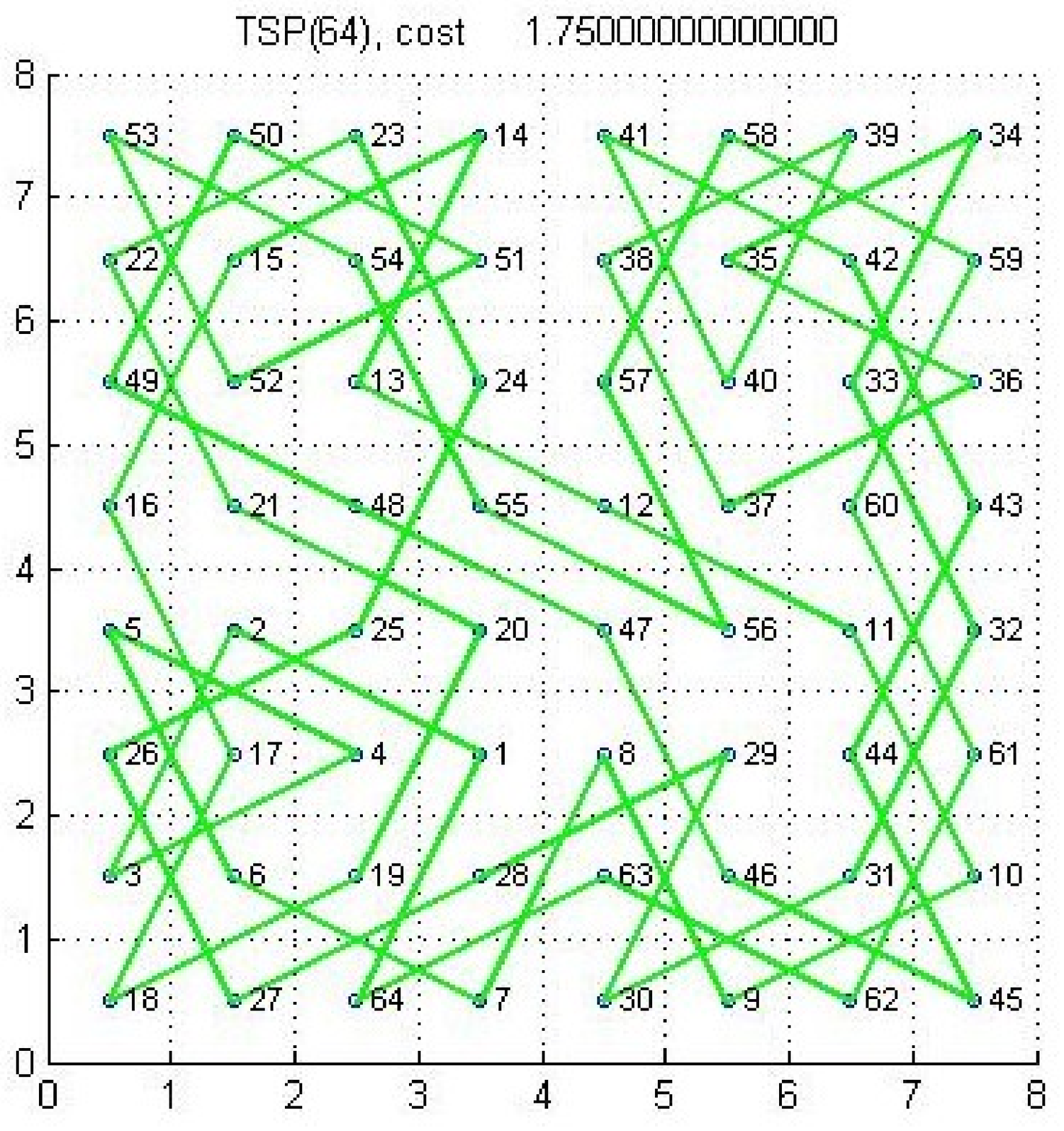,
height=50mm}
\psfig{figure=\IMAGESPATH/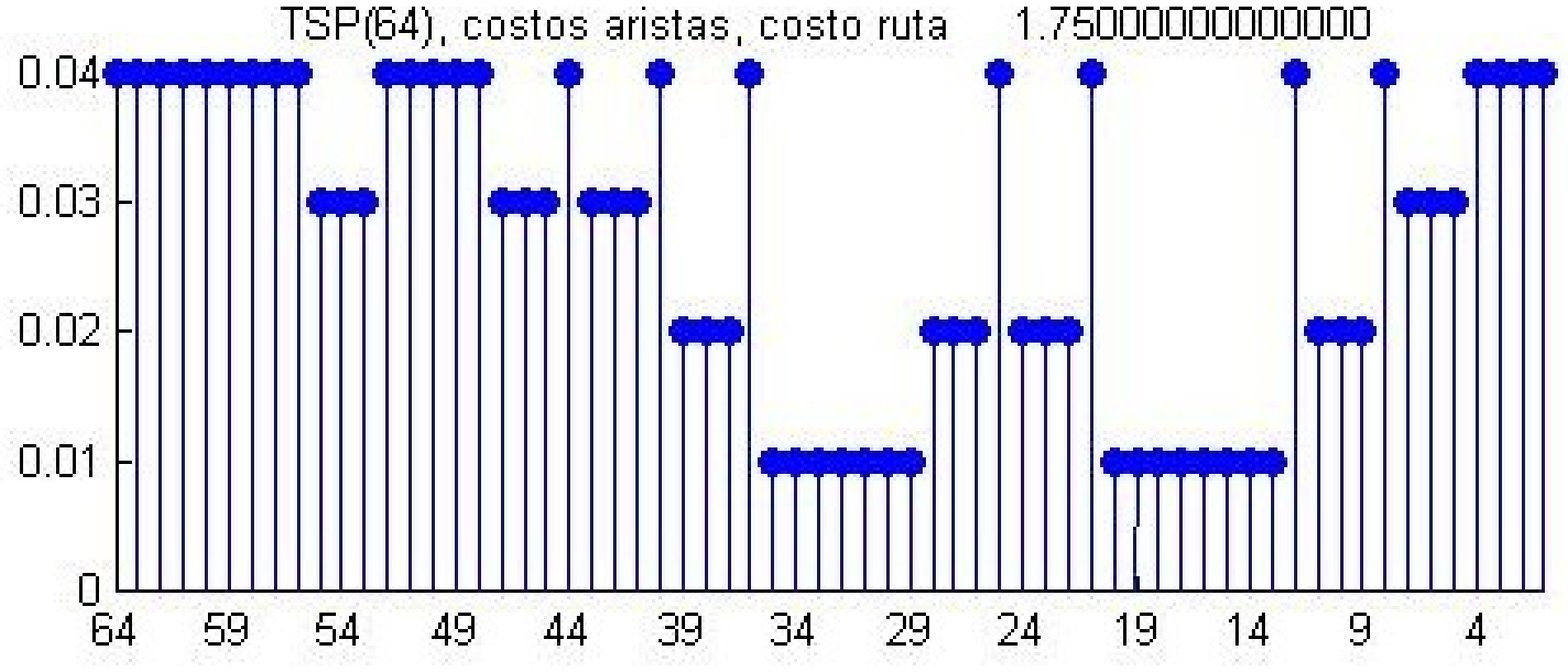,
height=50mm} }
\centerline{ \makebox[2.2in][c]{ a)
}\makebox[2.2in][c]{ b) } } ~
\caption{TKP$_{8\times 8}$. a) final tour 2, and b) edges'cost.}
\label{fig:8x8Eu75PFTCT}
\end{figure}

\begin{figure}[tbp]
\centerline{
\psfig{figure=\IMAGESPATH/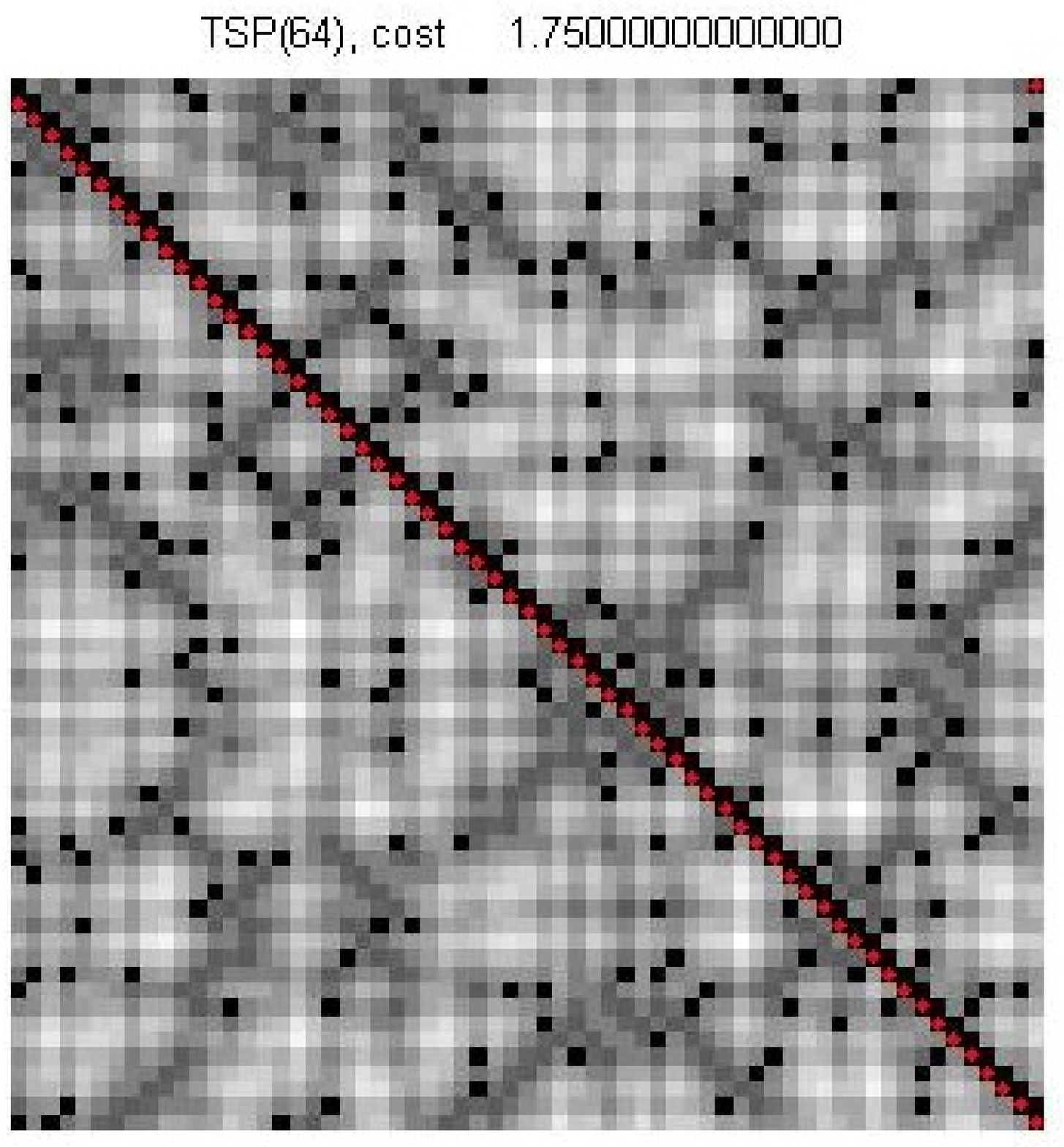,
height=50mm}
\psfig{figure=\IMAGESPATH/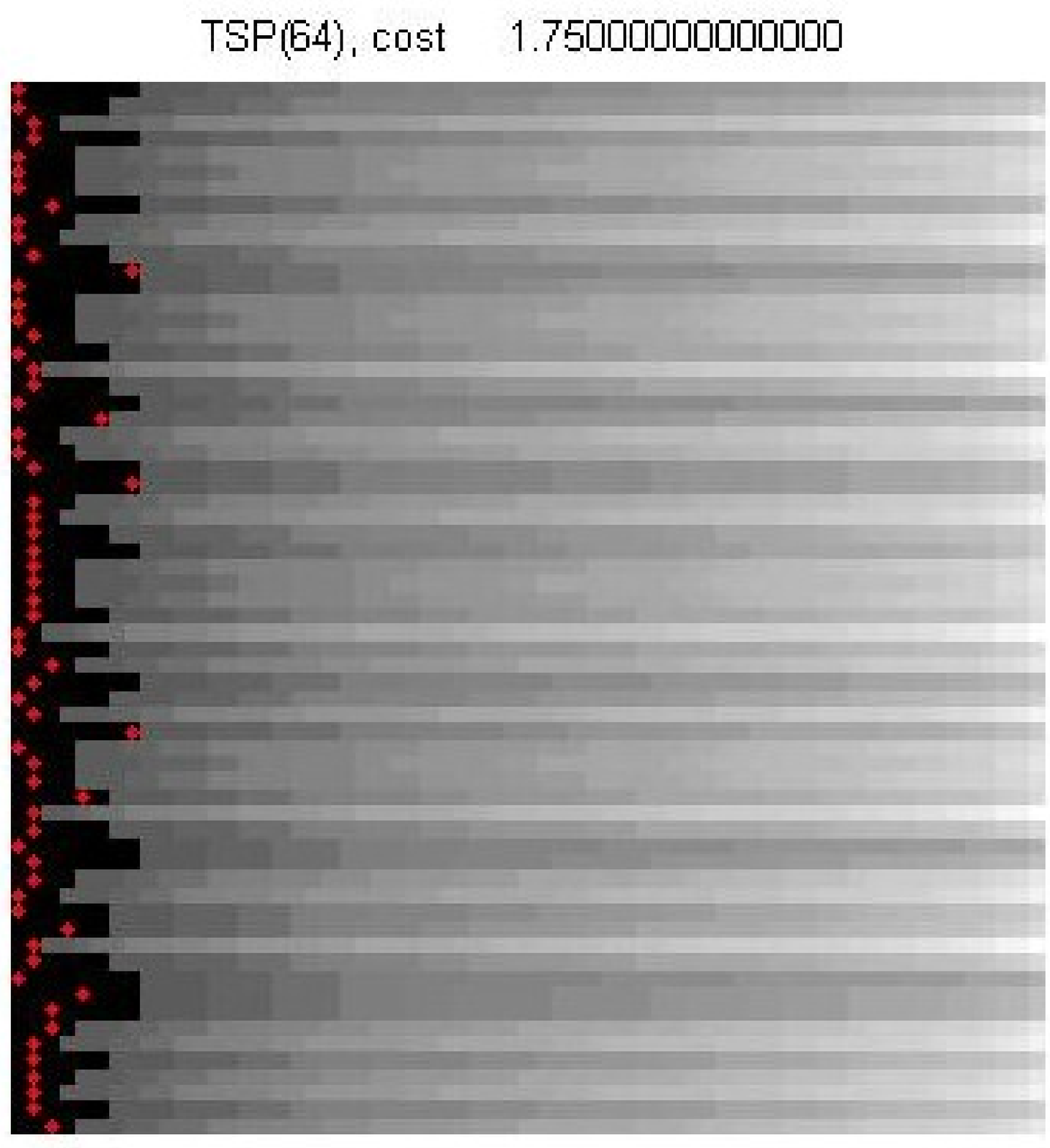,
height=50mm} }
\centerline{ \makebox[2.2in][c]{ a)
}\makebox[2.2in][c]{ b) } } ~
\caption{TKP$_{8\times 8}$ of the final tour 2. a) Ordered cost matrix, and
b) sorted cost matrix $\mathcal{M}$.}
\label{fig:8x8Eu75DgMFT}
\end{figure}

\begin{figure}[tbp]
\centerline{ \psfig{figure=\IMAGESPATH/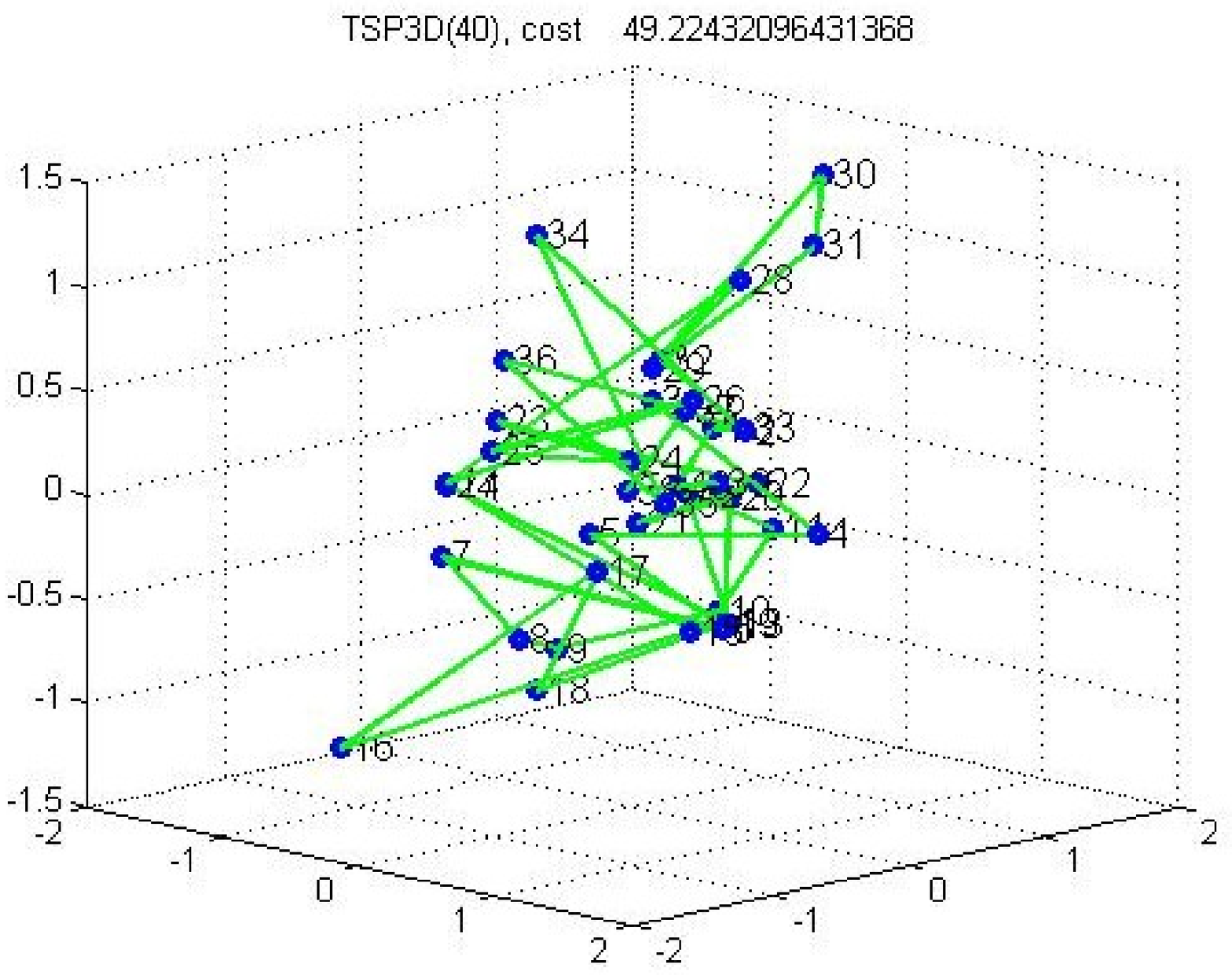,
height=50mm} \psfig{figure=\IMAGESPATH/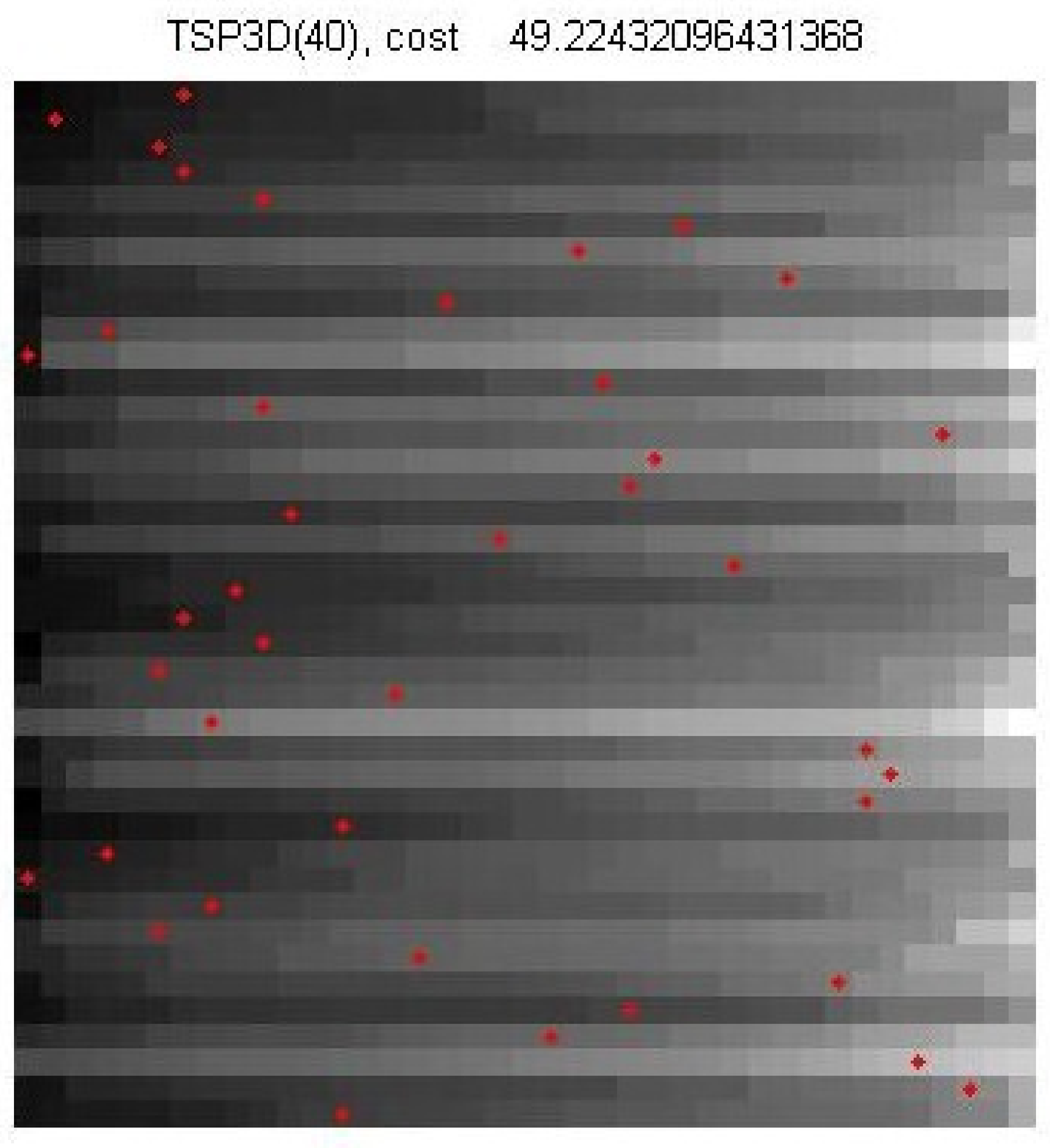,
height=50mm} } \centerline{ \makebox[2.2in][c]{ a)
}\makebox[2.2in][c]{ b) } } ~ ~ \caption{TSP3D$_{40}$. a) Initial
3D Hamiltonian cycle, and b) sorted cost matrix $\mathcal{M}$. The
red dots on the left side implies that there are many alternatives
hamiltonian cycles.} \label{fig:TSP3D_40_ini}
\end{figure}

\begin{figure}[tbp]
\centerline{ \psfig{figure=\IMAGESPATH/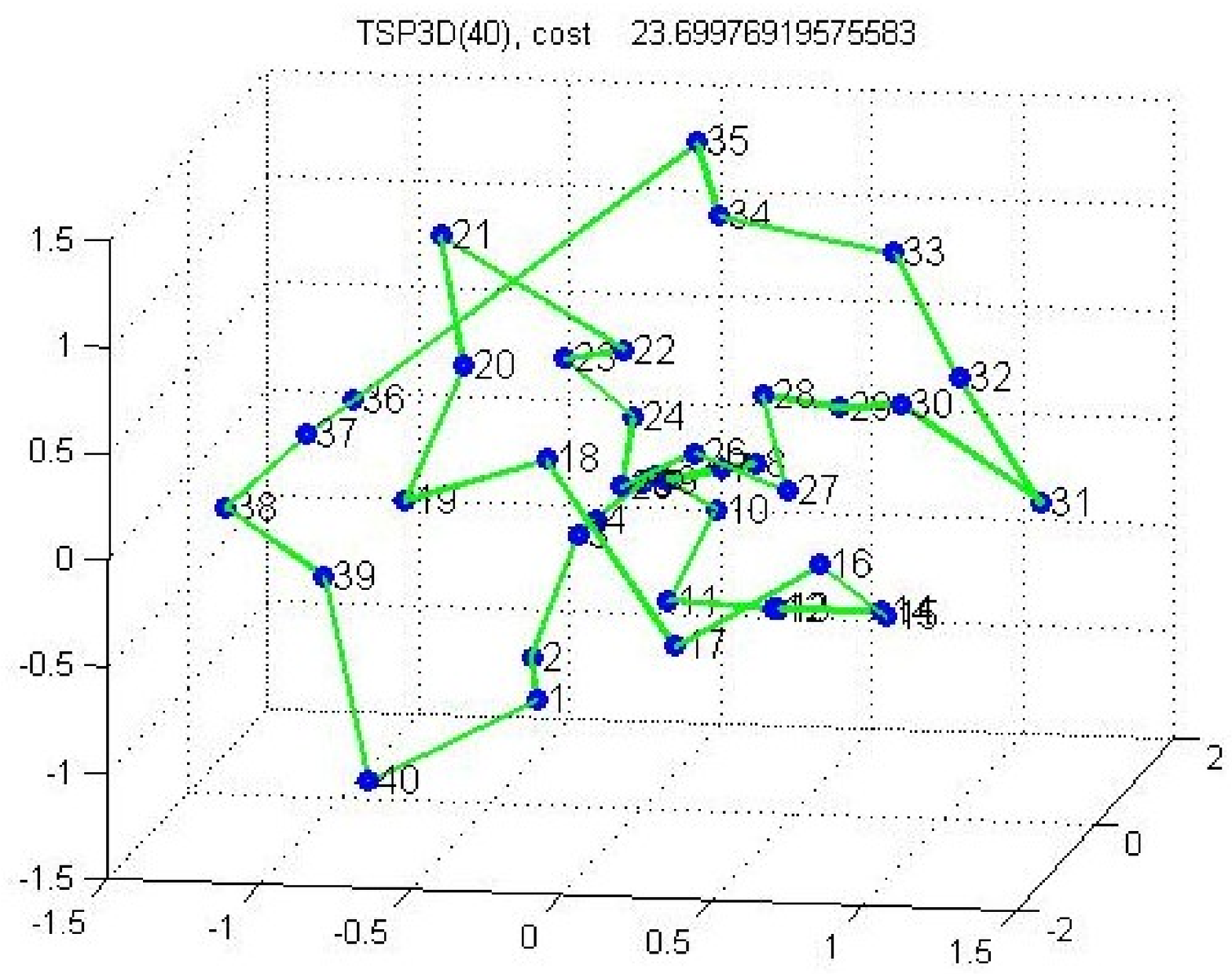,
height=50mm} \psfig{figure=\IMAGESPATH/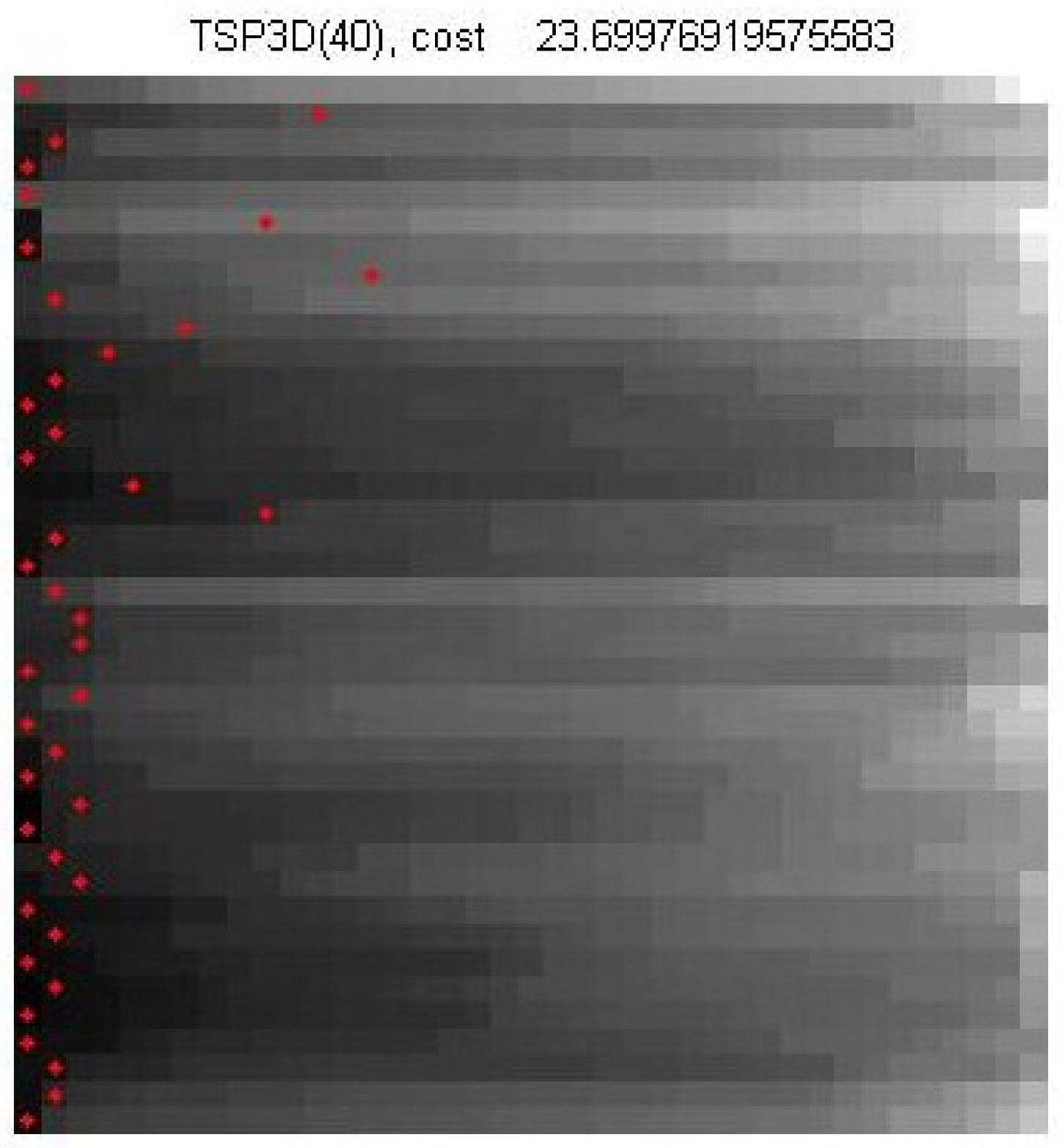,
height=50mm} } \centerline{ \makebox[2.2in][c]{ a)
}\makebox[2.2in][c]{ b) } } ~ \caption{TSP3D$_{40}$. a) Putative
3D Optimal hamiltonian cycle, and b) sorted cost matrix
$\mathcal{M}$. The red dots on the left side implies that there
are still many alternatives hamiltonian cycles.}
\label{fig:TSP3D_40_fin}
\end{figure}


\section{Polygonal shapes and properties}
~\label{sc:PolShPr}

This section depicts efficient algorithms using previous knowledge
to replace the algorithm of section~\ref{sc:greedy}. It is clear
that the population of the searching space for KTP or  TSP
problems are reduced when an algorithm focuses in appropriate
properties. In fact, the next proposition follows trivially from
the prop.~\ref{prop:JscNecSufCond}.
\begin{prop}
Given a TSP$_{n}$ where the cities are the vertex of a regular
polygon in $\mathbb{R}^2$. Then the minimum Hamiltonian cycle is
the polygon using Euclidian, $\max$, and abs distance.
\end{prop}

A regular polygon has sides of the same length. Hereafter, regular
polygon is named polygon.

\begin{figure}[tbp]
\centerline{\psfig{figure=\IMAGESPATH/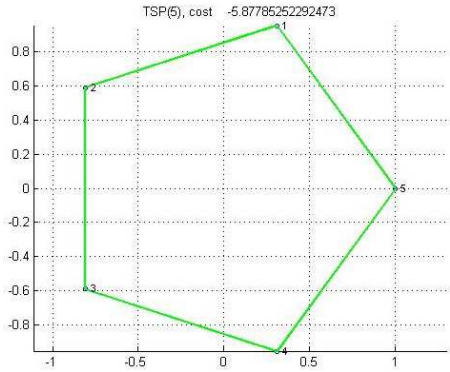,
height=50mm}}\caption{ Pentagon as a TSP$_{5}$.}
~\label{fig:Pol5Min}
\end{figure}

An algorithm to state this obvious solution must be $\bigO (n)$.
Figure~\ref{fig:Pol5Min} depicts an example of the polygon as
TSP$_5$. This is by using the formulas $y_i = r \sin(2\pi i / n )$
and $x_i = r \cos(2\pi i / n ),$ $i=0,\ldots,n-1$. The cost of
minimum Hamiltonian cycle depends of the Euclidian, $\max$, or abs
distance, but not the shape, which is always a Jordan's simple
convex curve.

Polygon$_n$ shape problem have a property to work for searching
the minimum length, but what about the maximum length on cities as
the vertex of a polygon of $n$ vertices in $\mathbb{R}^2$ using
Euclidian, $\max$, and abs distance.

Hereafter, PoMX$_n$ define the problem find the maximum distance
of the vertices of a polygon of $n$ vertices using Euclidian
(POME$_n$), $\max$ (POMm$_n$, or abs ((POMa$_n$) distance where X
denotes E,m or a. Also, an star is a crossing Hamiltonian cycle
over the vertices of a polygon with all angles equals.

\begin{prop}
Given a PoME$_n$ where the cities are the vertex of a polygon in
$\mathbb{R}^2$. Then the maximum Hamiltonian cycle is a star when
$n$ is odd.
\begin{proof}
First, assuming that the Hamiltonian cycle is a star, it has the
maximum length under Euclidian distance because the secants
corresponds to the diagonals of circumscribe triangles.

For $n$ odd, without loss of generality $n=2k+1$ for some $k\in
\Natural$. The sequence $\left\{ \left( 1+\left( i-1\right)
k\right) \text{mod} n \right\},$ $i=1,2,\ldots ,n,n+1,$ form an
star (crossing Hamiltonian cycle)  for the vertices of the given
polygon.

The function $s\left( i,k\right) =\left( 1+\left( i-1\right) k,
\left( 1+\left( i-1\right) k \right) \text{mod}\left( 2k+1\right)
\right)$ computes a positive integer where the second entry is the
residuos of $n$ or the congruent classes $\text{mod}$ $n.$ For a
given $k$, two consecutive numbers $1+\left( i-1\right) k$ and
$1+\left( i\right) k$ are in different class $\text{mod}$ $n.$ For
$i=n+1=2k+2$, corresponds to the class 1 $\text{mod}$ $n.$ And by
construction all sequence numbers $1,1+k,\ldots ,1+\left(
i-1\right) k,$ $i=1,2,\ldots ,n,n+1,$ are in different class
$\text{mod}$ $n$ except the first and the last one.
\end{proof}
\end{prop}

For $n$ even under the Euclidian distance there is not a
proposition as the previous but similar to a star when $n>>0$. In
fact, for any $n$ even never is a star.
Figure~\ref{fig:Pol14MaxCEstEuc} depicts a no star.

\begin{figure}[tbp]
\centerline{\psfig{figure=\IMAGESPATH/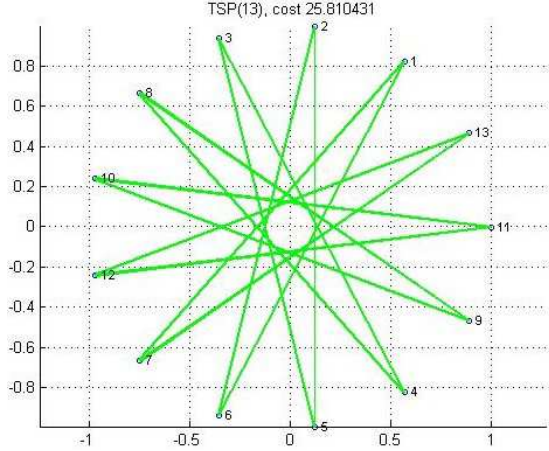,
height=60mm}}\caption{ POME$_{13}$ is a star.}
~\label{fig:pol13MaxEst}
\end{figure}

For $n$ odd the symmetry allow to build an efficient algorithm
under the Euclidian distance. Assuming that the vertices are
numbered using the previous formulas for $(x_i, y_i)$, then $1,
\left( 1+\left( i-1\right)\frac{n-1}{2} \right) \text{%
mod}\left( n\right),\ldots, 1$ defines a star. By example for
$n=13$ the vertices of the star are $1, 7, 13 (\text{or\ } 0), 6,
12, 5, 11, 4, 10, 3, 9, 2, 8,$ and $1$ as is depicted in
Fig.~\ref{fig:pol13MaxEst}. Figure~\ref{fig:Pol99MaxEst} depicts a
star for POME$_{99}$.

\begin{figure}[tbp]
\centerline{ \psfig{figure=\IMAGESPATH/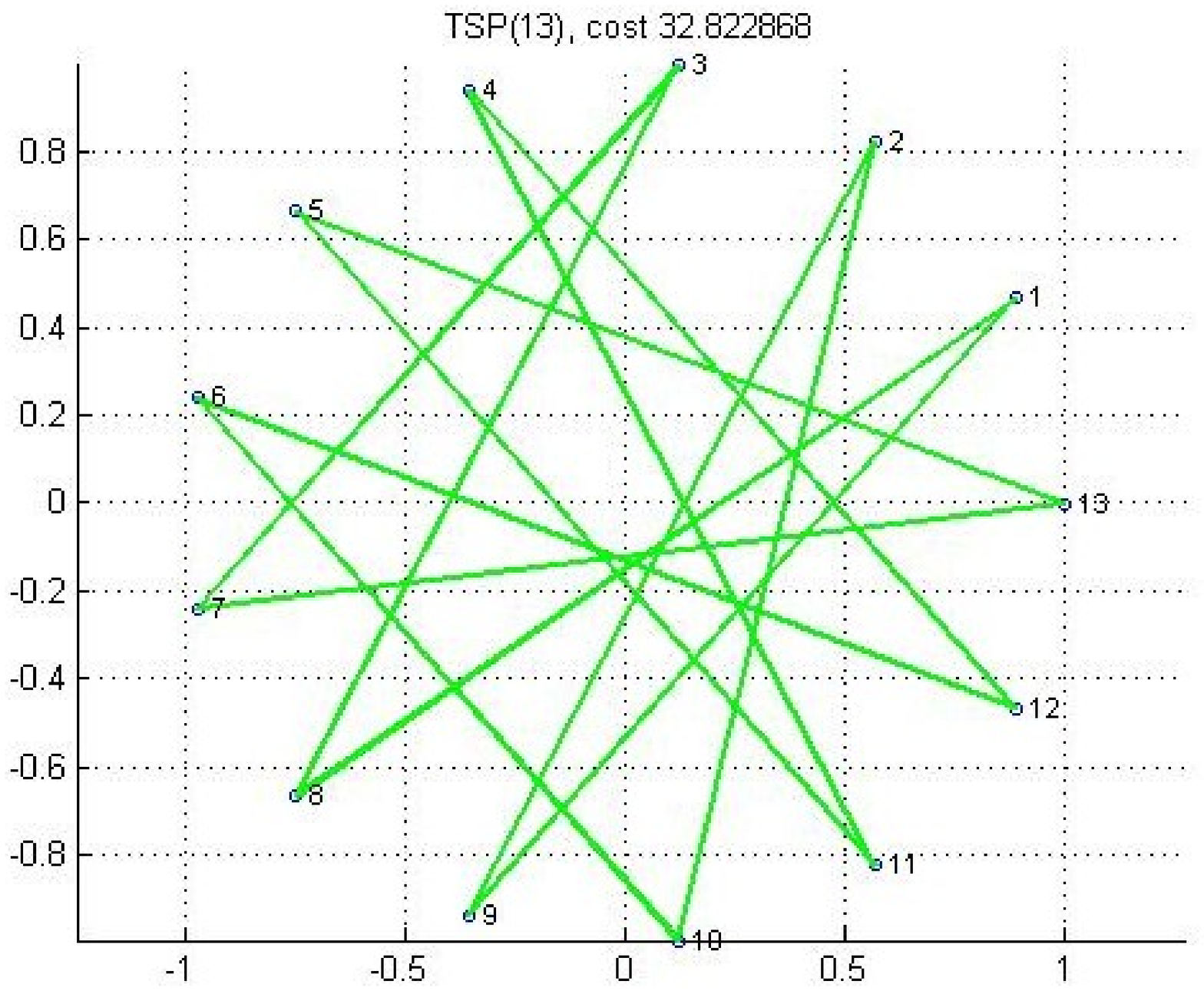,
height=60mm} \psfig{figure=\IMAGESPATH/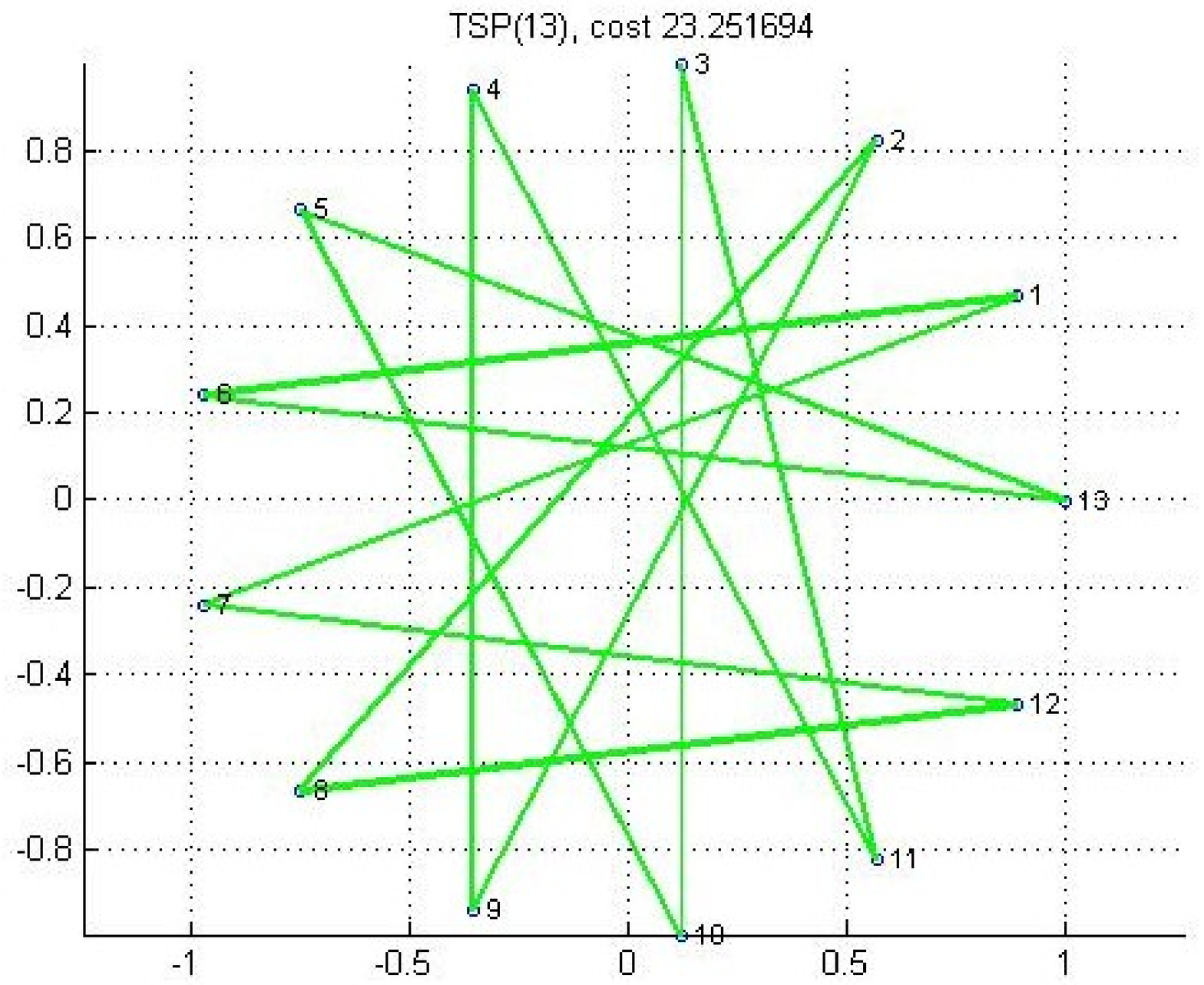,
height=60mm} } \centerline{ \makebox[2.2in][c]{ a)
}\makebox[2.2in][c]{ b)  } } ~ \caption{a) POMa$_{13}$ is no star,
and b) POMm$_{13}$ is no star.} ~\label{fig:pol13MaxAbs}
~\label{fig:pol13MaxMax}

\end{figure}

\begin{figure}[tbp]
\centerline{\psfig{figure=\IMAGESPATH/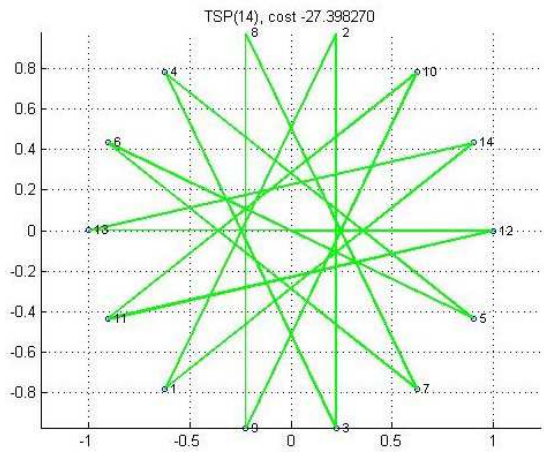,
height=60mm}}\caption{ POME$_{14}$ is no star.}
~\label{fig:Pol14MaxCEstEuc}
\end{figure}

\begin{figure}[tbp]
\centerline{\psfig{figure=\IMAGESPATH/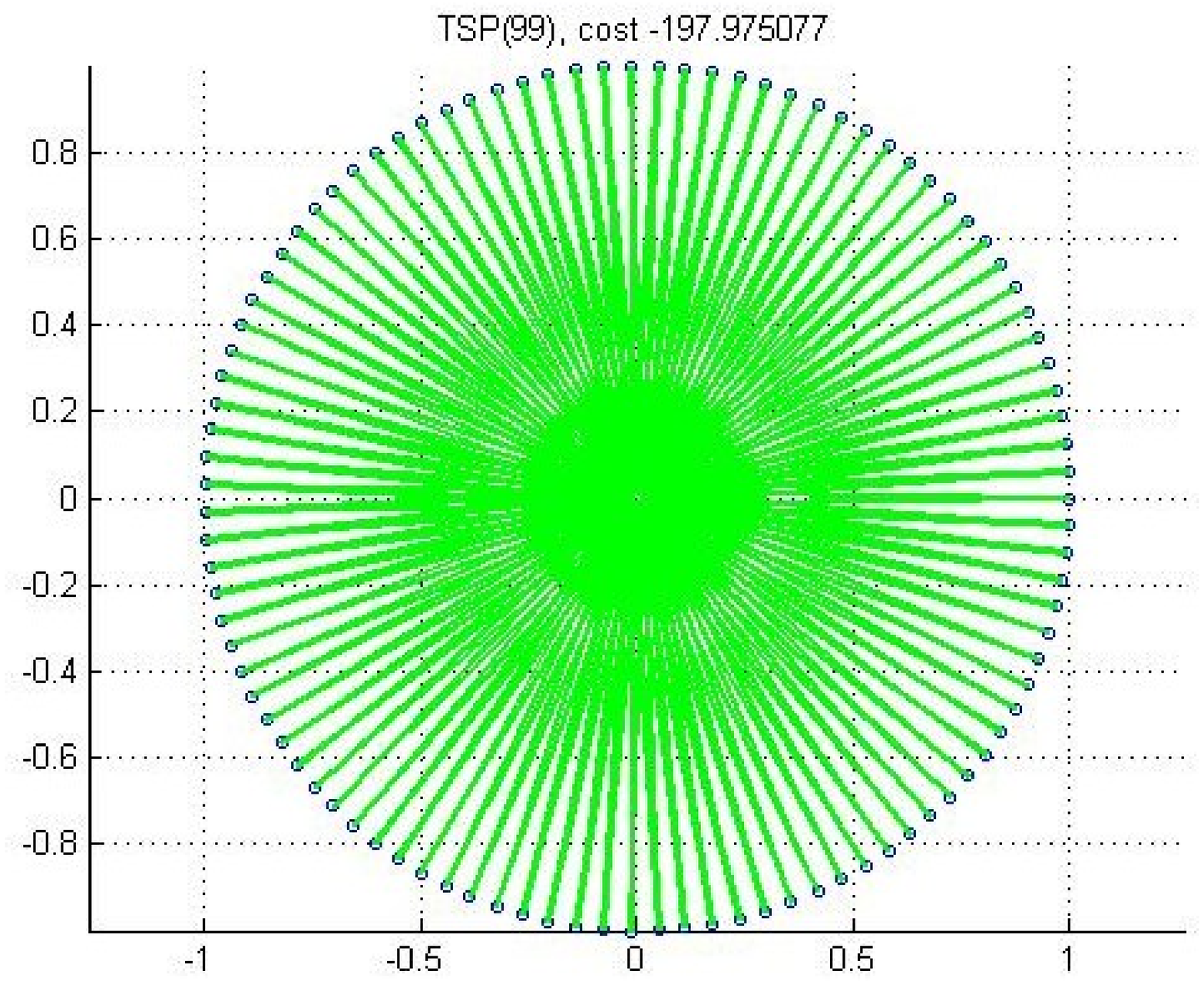,
height=60mm}}\caption{ POME$_{99}$ is a star.}
~\label{fig:Pol99MaxEst}
\end{figure}

On the other hand, POMa$_{13}$ and POMm$_{13}$ are no stars, they
are depicted in Fig.~\ref{fig:pol13MaxAbs} a) and b) respectively.
These differences are caused by using $\max$ distance and abs
distance instead Euclidian distance.

It is well know that an algorithm for minimization can be used for
maximization using $-f$. However, geometry properties, as by
example Simple Jordan's curve is not preserve in maximization.
Moreover, minimum length over a polygon is always a polygon. This
previous knowledge defines a very efficient algorithm to build the
minimum Hamiltonian cycle likes an enumeration of the vertices of
a polygon. Similar situation is with $n$ odd is for PoME$_n$,
where the cities are the vertex of a polygon in $\mathbb{R}^2$.
The efficient algorithm corresponds to an enumeration using the
index formula $\left( 1+\left( i-1\right)\frac{n-1}{2} \right)
\text{mod}\left( n\right),$ for $i=1,\ldots,n+1.$ Complexity is
$\bigO\left(n\right)$.

As in previous sections, the study of a problem allow to discover
properties for building efficient algorithm. The convexity and an
appropriate lattices for Lennard-Jones problems (LJP$_n$) are
important properties used in~\cite{arXiv:Barron2005,
arXiv:Barron2010}. However, these properties can not be
generalized or inherited to other NP problems.


\section{There is not a property for solving GAP}
~\label{sc:Back to GAP}

The algorithm~\ref{alg:GreedyGAP} is blind to the properties of
TSP and KTP. Jordan's simple curve and hamiltonian cycles with
many cross as posible are clearly oppositive properties.

For KTP, the objective function is adapted for solving a decision
problem: is there a hamiltonian cycle for a knight in a chessboard
$8\times8$?

The rippling effects proves that GAP's the objective function
giving by the summation os the edges' cost is not monotonic or
convex. This can see assuming few $c_{ij} > 0$ and their
corresponding $c_{ij} < 0$.

\begin{prop}
~\label{prop:GAPinRm} Given a GAP$_{n}$ there is $\mathbb{R}^K$
where the vertices of GAP can be located as points in
$\mathbb{R}^K$ with the max distance $d_m$.
\begin{proof}
The unknowns are the coordinates of each vertex. Linear equations
can be generated from the edge's cost.

Without loss of generality, I assume that the cost matrix is
asymmetrical. For $K>0, L_j\leq K$, the linear system to solve is:
\begin{eqnarray*}
x^1_1-x^2_1 + x^1_2-x^2_2+ \cdots +x^1_{K-1}-x^2_{K-L_1} + x^1_{K-L_1+1}+\cdots + x^1_{K} &=& c_{1,2} \\
x^1_1-x^2_1 + x^1_2-x^2_2+ \cdots +x^1_{K-1}-x^2_{K-L_2} - x^2_{K-L_2+1}+\cdots- x^2_{K}&=& c_{2,1} \\
\vdots & = &  \vdots, \\
x^n_1-x^{n-1}_1 + x^n_2-x^{n-1}_2+ \cdots +x^n_{K-L_{n^2-1}}-x^{n-1}_{K-L_{n^2-1}}+ x^{n-1}_{K-L{n^2-1}+1}+\cdots + x^{n-1}_{K} &=& c_{n-1,n}, \\
x^n_1-x^{n-1}_1 + x^n_2-x^{n-1}_2+ \cdots +x^n_{K-L_{n^2}}-x^{n-1}_{K-L_{n^2}}- x^{n}_{K-L{n^2}+1}-\cdots - x^{n}_{K} &=& c_{n,n-1}, \\
\end{eqnarray*}
where $x^i=(x^i_1, x^i_2,\ldots,x^i_K)$ $\in$ $\Real^k,$
$i=1,\ldots,n$ are the corresponding coordinates of the vertices.
With sufficient variables, i.e, appropriate values of $K$ and
$L_j$, $j=1,2,\ldots,n^2,$ the previous linear system is always
soluble.
\end{proof}
\end{prop}

\begin{rem}
Any GAP can be mapped in $\mathbb{R}^K$ under $d_m$.

In the core of the algorithms for TSP and KTP is the algorithm~\ref%
{alg:GreedyGAP}. It is designed for exploring edges'cost for
looking a minimum or for doing a random uniform search. The
heuristic used in the algorithm for TSP is well know, but it only
works for a plane where the
distance fulfil the cosine law of the triangles, i.e., where Prop. 6.11 in ~%
\cite{arXiv:Barron2010} holds. Even, in 3D TSP does not comply. Figure~\ref%
{fig:TSP3D_40_ini} depicts an initial hamiltonian cycle for a TSP
with 40 cities in a 3D Space with cost = $29.2243$.
Figure~\ref{fig:TSP3D_40_fin} depicts a putative optimal
hamiltonian cycle with cost = $23.6998$. There is a reduction on
the cost, but, both figures b) imply that there are many
alternatives to explore yet.

On the other hand, KTP could be solved requiring hamiltonian
cycles with length equals to $n \times m \times 5$ with
$dist(x,y)$ $=$ $\left( i-i^{\prime }\right) ^{2}+\left(
j-j^{\prime }\right) ^{2}$, where $x=(i,j),$ and $y=(i^{\prime
},j^{\prime }),$ $1 \leq i,i^\prime \leq n$, and $1 \leq j,
j^\prime \leq m$. In this case the objective function could used
without changes, but algorithm~\ref{alg:GreedyGAP} must be
modified to random uniform search (see ~\ref{rem:AlgGreRnd}).
\end{rem}


\section{SAT$_{n\times m}$ has not properties for an efficient algorithm}
~\label{sc:SAT}

This section is devoted to a former NP problem. I apply my
technique: 1) general problem, 2) simple reduction, and 3) there
is no property for building an efficient algorithm for the simple
reduction problem.

 Here, I use the convention to represent $\overline{x_i}$ to $0$ (false), $x_i$ to $1$ (
true), and 2 when the variable $x_i$ is no present. Let $\Sigma
=\left\{ 0,1\right\},$ $x_{i}:\Sigma \rightarrow \Sigma
,x_{i}\left( v\right) =v,v\in \Sigma ,\overline{x}_{i}:\Sigma
\rightarrow \Sigma ,\overline{x}_{i}\left( v\right) =\rceil v,v\in
\Sigma $ (logical not).

A particular Boolean Satisfiability Problem (SAT) consists in
finding a set of values in $\Sigma$ of $n$ bolean variables and
$m$ formulas to have the following system equal to 1:
\begin{equation*}
\begin{array}{l}
\ \ (x_{1} \vee \overline{x_{2}} \vee \cdots \vee x_l \cdots \vee x_n) \\
\wedge ( x_{1} \vee x_{2} \vee \cdots \vee x_{n-1} ) \\
\vdots \\
\wedge ( \overline{x_{2}}\vee x_3\vee \cdots \vee \overline{
x_{i}} \cdots \vee x_l ).
\end{array}%
\end{equation*}

Then the general problem or complex SAT is where the formulas
could have any subset of boolean variables, where each formula can
be mapped to a ternary number.

For example:
\begin{equation*}
\begin{array}{l}
\ \ (x_{5} \vee \overline{x_{4}} \vee x_1)  \\
\wedge ( x_{3} \vee  x_{2} ) \\
 \wedge ( \overline{x_{4}}\vee x_3\vee \overline{
x_{2}} \vee x_1 ).
\end{array}%
\end{equation*}

It is traduced to:
\begin{equation*}
\begin{array}{l}
 10221  \\
22112 \\
20101.
\end{array}%
\end{equation*}
The assignment $x_1=1$, $x_2=1$ satisfies the previous example. It
is not unique. Moreover, the number 22211 is a representation of
this assignment as a ternary number. Also, the system:

\begin{equation*}
\begin{array}{l}
\ \ (x_{5} \vee \overline{x_{4}} \vee x_1)  \\
\wedge ( x_{3} \vee  x_{2} ) \\
 \wedge ( \overline{x_{4}}\vee x_3\vee \overline{
x_{2}} \vee x_1 )
\\
 \wedge ( x_{2} \vee x_1 ) .
\end{array}%
\end{equation*}

22211 is translated like the last formula and it is a type of a
fixed point system, where 22211 is an appropriate assignment. The
details are given below.

On the other hand, the simple reduction provides a simple version
of SAT, where for convenience all formulas have the same
variables. In any case, this simple versi\'{o}n of SAT is like study
parts of the complex SAT, and it is sufficient to prove that there
is not polynomial time algorithm for the simple SAT, and also, for
general SAT.

 An equivalent functional formulation is:

Let $\overrightarrow{x}=\left( x_{1,}x_{2},\ldots ,x_{n}\right) \in \Sigma
^{n},\overrightarrow{y}\in \Sigma ^{m},S_{j}\subset \{1,\ldots ,n,-1,\ldots
,-n\},j=1,\ldots ,m,$ $F_{j}:\Sigma ^{m}\rightarrow \Sigma ,F_{j}\left(
\overrightarrow{x}\right) =\vee _{k\in S_{j}}z_{k},z_{k}=\left\{
\begin{array}{cc}
x_{k} & k>0 \\
\overline{x}_{k} & k<0%
\end{array}%
\right. ,$ $G:\Sigma ^{m}\rightarrow \Sigma ,G\left( \overrightarrow{y}%
\right) =\wedge _{i=1}^{m}y_{i}$

The SAT\ is

\begin{equation*}
G\circ \left( F_{1},\ldots ,F_{m}\right) \left( \overrightarrow{x}\right).
\end{equation*}

Let be \textsl{U}$=[u_{jk}]$ a matrix of $\Sigma^{m \times 2n},$

$u_{j, k} =\left\{
\begin{array}{cc}
1 & \text{if } x_i \vee \overline{x}_i \text{ are variables of } F_j, k=i,%
\text{if }i>0, k=n+|i|,\text{if }i<0, \\
0 & \text{otherwise.}%
\end{array}%
\right. $

The matrix \textsl{U} let to know how many variables has $F_j$ as the of
summation over the row, $\sum_{i=1}^{2n}(u_{j i})$. Also the summation over
a column, $\sum_{j=1}^m(u_{j k})$ is the number of times that the $x_i$ or $%
\overline{x}_i$ ($k=i,\text{if }i>0, \text{ or } k=n+|i|,\text{if }i<0$) is
used in all $F_j$.

Let be $\blacksquare $ as 0, and $\square $ as 1. The following boards have
not a set of values in $\Sigma$ to satisface them (I called unsatisfactory
boards):

\begin{equation*}
\begin{tabular}{|c|}
\hline
$x_{1}$ \\ \hline
$\blacksquare $ \\ \hline
$\square $ \\ \hline
\end{tabular}%
\ \ \
\begin{tabular}{|l|l|}
\hline
$x_{2}$ & $x_{1}$ \\ \hline
$\blacksquare $ & $\blacksquare $ \\ \hline
$\blacksquare $ & $\square $ \\ \hline
$\square $ & $\blacksquare $ \\ \hline
$\square $ & $\square $ \\ \hline
\end{tabular}%
\end{equation*}

Hereafter, a SAT of $n$ bolean variables and $m$ formulas is
denoted by SAT$n \times m$.

\begin{prop}
~\label{prop:NoSolSAT_binary} Given a SAT$_{n\times 2n}$ where formulas as
the squares correspond to the $0$ to $2^l-1$ binary values, is an
unsatisfactory board.
\begin{proof}
The binary values from $0$ to $2^l-1$ are all posibles assignation
of values for the board. It means that for any possible
assignation, there is the oppositive formula with value 0.
\end{proof}
\end{prop}

\begin{prop}
~\label{prop:NoSolSAT} Given a SAT of $n \times m$, there is not a
satisfactory set of values in $\Sigma$, when there is a subset of $l$ bolean
variables and their set of $Fj$ formulas are isomorphic to an unsatisfactory
board.
\begin{proof}
It is immediately, the subset of $l$ bolean variables can not
satisface their $2^l$, $F_j$ formulas. Therefore, there is not
posible to find satisfactory set of $n$ values for this SAT.
\end{proof}
\end{prop}

When I focusses in the classical techniques for fixed point or finding
roots. The formulation of one can be used in the other. This means $f:%
\mathbb{R}\rightarrow \mathbb{R}$ with a root $x_{1}^{\ast }$ $\in $ $(a,b)$
then the function $g:\mathbb{R}\rightarrow \mathbb{R}$, giving by $%
g(x)=x+f(x)$ has the fixed point $x_{1}^{\ast }$. The reciprocal, let the
function $g:\mathbb{R}\rightarrow \mathbb{R}$ has a fixed point $x^{\ast }$
then the function $f:\mathbb{R}\rightarrow \mathbb{R}$, giving by $%
f(x)=x-g(x)$ has the point $x_{1}^{\ast }$ as a root.

SAT$_{n\times m}$ can be transformed in a fixed point formulation. This
formulation is easy to understand, it consists in:

\begin{enumerate}
\item $\Sigma=\{ 0,1 \}$.

\item giving a boolean variable $x$ is mapped to 1, and $\overline{x}$ to 0;

\item a formula: $\overline{x_{n-1}}\vee \cdots x_{l}\vee \cdots \vee
\overline{x_{1}}\vee x_{0}$ can be transformed in its corresponding binary
number: $0\cdots 1\cdots 01$;

\item if a $y$ is a binary number, $\overline{y}$ is the complement binary
number. this is done bit a bit, 0 is complemented to 1 and 1 is complemented
to 0.

\item a system of SAT$_{n\times m}$ correspond the set $M_{n\times m}$ of
the $m$ binary numbers of its formulas. $M_{n\times m}=\{%
\begin{array}{c}
s_{n-1}^{1}s_{n-2}^{1}\cdots s_{1}^{1}s_{0}^{1}, \\
s_{n-1}^{2}s_{n-2}^{2}\cdots s_{1}^{2}s_{0}^{2}, \\
\ldots , \\
s_{n-1}^{m}s_{n-2}^{m}\cdots s_{1}^{m}s_{0}^{m}%
\end{array}%
\}.$ Note that the number $s_{n-}^{k}s_{n-2}^{k}\cdots s_{1}^{k}s_{0}^{k}$
correspond to formula $k$ of the SAT$_{n\times m}$, $k=1,\ldots ,m$.

\item a system of SAT$_{n\times m}$ as a boolean function is SAT$_{n\times
m}:\Sigma ^{n}\rightarrow \Sigma $. The argument is a binary number of n
bits.
\end{enumerate}

\begin{prop}
~\label{prop:SATFixedPoint1} Evaluation of SAT$_{n\times m}$ is equivalent
to the evaluation of SAT$_{n\times m}(y=y_{n-1}y_{n-2}\cdots y_{1}y_{0})$,
that consists to verify $y_{i}$ match a digit $s_{i}^{k}\in M_{n\times m},$ $%
\forall k=1,\ldots ,m$, SAT$_{n\times m}(y=y_{n-1}y_{n-2}\cdots y_{1}y_{0})$%
=1, otherwise SAT$_{n\times m}(y=y_{n-1}y_{n-2}\cdots y_{1}y_{0})$=0.
\begin{proof}
It is immediately, let $x_i=v_i$, $i=n,\ldots,1$ with $v_i\in
\Sigma$ an assignation for SAT$_{n\times m}$. If the set of values
$v_i$ satisfies SAT$_{n\times m}$ then it means that at least one
bolean variable of each bolean formula is 1, therefore SAT$n\times
m(v_{n} v_{n-1} \cdots v_2 v_1)$ is 1. By other hand, if $x_i=v_i$
does not satisfice SAT$_{n\times m}$, then at least one formula
the SAT$_{n \times m}$ is 0, let assume it is j. This means that
no value $v_i$ match any digit of $s^j_i$ for $i=n,\ldots,1$.
Therefore  SAT$_{n\times m}(v_{n} v_{n-1} \cdots v_2 v_1)=0.$

Reciprocally, SAT$_{n\times m} (y=y_{n} y_{n-1} \cdots y_2
y_1)=1$, means that a digit $y_i$  match a digit $s_i^k \in M,$
$\forall k=1,\ldots,m.$ The k bolean formulas are 1. Therefore,
SAT$_{n\times m}$ is 1 with the assignation $x_i=y_i.$
Otherwise,SAT$_{n\times m} (y=y_{n} y_{n-1} \cdots y_2 y_1)=0$
implies that at least one binary number of $M$ does not coincide
with the digits of $y$. This means a bolean formula of
SAT$_{n\times m}$ is 0, then SAT$_{n\times m}$ is 0.
\end{proof}
\end{prop}

\begin{prop}
An equivalent formulation of SAT$_{n\times m}$ is to look for a binary
number $x^{\ast }$ from $0$ to $2^{n}-1.$

\begin{enumerate}
\item $x^{\ast }\in M_{n\times m}$ and $\overline{x^{\ast }}\notin
M_{n\times m}$ then SAT$_{n\times m}(x^{\ast })=1.$

\item $x^{\ast }\in M_{n\times m}$ and $\overline{x^{\ast }}\in M_{n\times m}
$ then SAT$_{n\times m}(x^{\ast })=0.$ If $m<2^{n}-1$ then $\exists y^{\ast
},0\leq y^{\ast }\leq 2^{n}-1$ with  $\overline{y^{\ast }}\notin M_{n\times
m}$ and SAT$_{n\times m}(y^{\ast })=1.$%
\end{enumerate}
\begin{proof}
\begin{enumerate}
\item How $x^{\ast }\in M_{n\times m}$ and $\overline{x^{\ast
}}\notin M_{n\times m}$, this means that the corresponding formula
of $x^{\ast }$ is not blocked and for each bolean formula of
SAT$_{n\times m}(x^{\ast })$ at least one bolean variable
coincides with one variable of $x^{\ast }.$  Therefore
SAT$_{n\times m}(x^{\ast })=1.$

\item How $m<2^{n}-1$, $\exists y^{\ast },0\leq y^{\ast }\leq
2^{n}-1 $ with $\overline{y^{\ast }}\notin M_{n\times m}.$ Adding
the corresponding formula of $y^{\ast }$ to SAT$_{n\times m}$, a
new SAT$_{n\times m+1}$ is obtained. By the previous case, the
case is proved.
\end{enumerate}
\end{proof}
\end{prop}

This approach is quite forward to verify and get a solution for any SAT$%
_{n\times m}$. By example given  SAT$_{6\times 4}$, its correspond set $%
M_{6\times 4}$:

\begin{equation*}
\begin{tabular}{|l|l|l|l|l|l|l|}
\hline
& $x_{5}=0$ & $x_{4}=0$ & $x_{3}=0$ & $x_{2}=0$ & $x_{1}=0$ & $x_{0}=0$ \\
\hline
& $\overline{x_{5}}\vee $ & $\overline{x_{4}}\vee $ & $\overline{x_{3}}\vee $
& $\overline{x_{2}}\vee $ & $\overline{x_{1}}\vee $ & $\overline{x_{0}})$ \\
\hline
$\wedge ($ & $\overline{x_{5}}\vee $ & $\overline{x_{4}}\vee $ & $\overline{%
x_{3}}\vee $ & $\overline{x_{2}}\vee $ & $\overline{x_{1}}\vee $ & $x_{0})$
\\ \hline
$\wedge ($ & $x_{5}\vee $ & $x_{4}\vee $ & $x_{3}\vee $ & $x_{2}\vee $ & $%
x_{1}\vee $ & $\overline{x_{0}})$ \\ \hline
$\wedge ($ & $\overline{x_{5}}\vee $ & $x_{4}\vee $ & $x_{3}\vee $ & $%
\overline{x_{2}}\vee $ & $x_{1}\vee $ & $x_{0})$ \\ \hline
\end{tabular}%
\text{ \ }%
\begin{array}{c}
\\
\left[ \overline{x_{0}}\right] \equiv 1 \\
\left[ \overline{x_{1}}\right] \equiv 1 \\
\left[ \overline{x_{0}}\right] \equiv 1 \\
\left[ \overline{x_{2}}\right] \equiv 1%
\end{array}%
\text{ \ }%
\begin{tabular}{|l|l|l|l|l|l|}
\hline
$x_{5}$ & $x_{4}$ & $x_{3}$ & $x_{2}$ & $x_{1}$ & $x_{0}$ \\ \hline
$0$ & $0$ & $0$ & $0$ & $0$ & $0$ \\ \hline
$0$ & $0$ & $0$ & $0$ & $0$ & $1$ \\ \hline
$1$ & $1$ & $1$ & $1$ & $1$ & $0$ \\ \hline
$0$ & $1$ & $1$ & $0$ & $1$ & $1$ \\ \hline
\end{tabular}%
\text{.}
\end{equation*}

The left side depicts that SAT$_{6\times 4}(y=000000)=1$. The
right side depicts the set $M_{6\times 4}$ as an array of binary
numbers , where $000000\in M_{6\times 4}.$ The middle column
depicts at least one variable that satisfies the boolean formulas of SAT$%
_{6\times 4}.$ Also, $y$ $=$ $000000$ can be interpreted as the
satisfied assignment $x_{5}=0,$  $x_{4}=0,$  $x_{3}=0,$ $x_{2}=0,$
$x_{1}=0,$ and $x_{0}=0.$

In fact, the previous proposition gives an interpretation of the
SAT as a type fixed point problem.

\begin{prop}
~\label{prop:SATFixedPoint2} Given a SAT$_{n\times m}$, there is a
binary number $y\in M_{n\times m}$ such that $\overline{y}\notin
M_{n\times m}$ then $y$ is fixed point for SAT$_{n\times m}$ or
SAT$_{n\times m+1}(y)$, where SAT$_{n\times m+1}$ is SAT$_{n\times
m}$ with adding the corresponding formula of $y$ to SAT$_{n\times
m}.$
\begin{proof}
This result follows from the previous proposition.
\end{proof}
\end{prop}

Using the propositions and properties for SAT$_{n\times m},$ a
computable algorithm for solving SAT$_{n\times m}$ is:

\begin{algorithm}
~\label{alg:SAT} \textbf{Input:} SAT$_{n\times m}$.

\textbf{Output:} $x^{\ast }$ such that SAT$_{n\times m}(x^{\ast
})=1$ or SAT$_{n\times m+1}(x^{\ast })=1$ or SAT$_{n\times m}$ is
not satisfied.

\textbf{Variables in memory}: T[0:$2^{n-1}$]=0 of boolean.
$mi$:=$2^n$ : Integer, $mx$:=-1: Integer; $ct$:=0 : Integer;

\begin{enumerate}

\item \textbf{for} i:=1 to m

\item \hspace{0.5cm} Translate formula $i$ of SAT$_{n\times m}$ to
a binary number $k$.

\item \hspace{0.5cm} \textbf{if} T[$k$] not equal 1 \textbf{then}

\item \hspace{0.5cm} \hspace{0.5cm} \textbf{if} SAT$_{n\times
m}$($k$) equal 1 \textbf{then}

\item \hspace{0.5cm} \hspace{0.5cm} \hspace{0.5cm}
\textbf{output:} $k$ is the solution for SAT$_{n\times m}$.

\item \hspace{0.5cm} \hspace{0.5cm} \hspace{0.5cm} \textbf{stop}

\item \hspace{0.5cm} \hspace{0.5cm} \textbf{end if}

 \item \hspace{0.5cm} T[$k$]  := 1;

\item \hspace{0.5cm} $ct$  := $ct$ + 1;

\item \hspace{0.5cm}  $mi$ := Min($k$,$mi$);

\item \hspace{0.5cm} $mx$ := Max($k$, $mx$);

\item \hspace{0.5cm} \textbf{end if}

\item \textbf{end for}

\item \textbf{if} mi == 0 \textbf{and} mx == $2^{n-1}$
\textbf{and} ct $\geq$ $2^n$ \textbf{then}

\item \hspace{0.5cm} \textbf{output:} There is not solution for
SAT$_{n\times m}$.

\item \hspace{0.5cm} \textbf{stop}

\item \textbf{end if}

\item  \textbf{if} $mi$  $>$ $1$ \textbf{and} SAT$_{n\times
m}$($mi-1$) == 1 \textbf{then}

\item \hspace{0.5cm} \textbf{output:} $mi-1$ is the solution for
SAT$_{n\times m+1}$ with the corresponding formula for $mi-1$.

\item  \hspace{0.5cm} \textbf{stop}

\item \textbf{end if}

\item  \textbf{if} mx  $<$ $2^n-1$ \textbf{and} SAT$_{n\times
m}$($mx+1$) == 1 \textbf{then}

\item \hspace{0.5cm} \textbf{output:} $mx+1$ is the solution for
SAT$_{n\times m+1}$ with the corresponding formula for $mx+1$.

\item  \hspace{0.5cm} \textbf{stop}

\item \textbf{end if}

\item \textbf{for} i:=0 to $2^n-1$

\item \hspace{0.5cm} \textbf{if} T[$i$] not equal 1 \textbf{then}

\item \hspace{0.5cm} \hspace{0.5cm} \textbf{output:} $i$ is the
solution for SAT$_{n\times m+1}$ with the corresponding formula
for $i$.

\item \hspace{0.5cm} \hspace{0.5cm} \textbf{stop}

\item \hspace{0.5cm} \textbf{end if}

\item \textbf{end for}

\end{enumerate}
\end{algorithm}

At this point, SAT$_{n\times m}$ is equivalent to set of binary
numbers M$_{n\times m}$. The knowledge for a given SAT$_{n\times
m}$ depends at least from exploring M$_{n\times m}$. Before of
exploring SAT$_{n\times m}$ no binary numbers associated with it
are know. Now, let $\mathbb{K}_{n\times m}$ be knowledge of a
given SAT$_{n\times m}$. $\mathbb{K}_{n\times m}$ $=$
$\left\{y^\ast\right\}$ $\cup$ M$_{n\times m}$ $\cup$
$\mathbb{S}_{n\times m}$, where $y^\ast$ is the satisfying number
for SAT$_{n\times m}$, and $\mathbb{S}_{n\times m}$ $=$
M$_{n\times m}^c$, it  is the set of binary numbers that satisfies
SAT$_{n\times m}$. Using the previous proposition, without lost of
generality, $y^\ast\in$ M$_{n\times m}$.

\begin{prop}
~\label{prop:SATKnow} Given a SAT$_{n\times m}$, and
$\mathbb{K}_{n\times m}$:
\begin{enumerate}
    \item It is trivial to solve SAT$_{n\times m}.$
    \item It is trivial to solve SAT$_{n\times m+1}$ from M$_{n\times
    m}$ $\cup$ $\{\overline{y}^\ast\}$.
\end{enumerate}
\begin{proof} \

\begin{enumerate}
    \item $y^\ast$ solves SAT$_{n\times m}.$
    \item Any $s\in \mathbb{S}_{n\times m}\setminus \{\overline{y}^\ast \}$ solves SAT$_{n\times m+1}$.
\end{enumerate}\end{proof}
\end{prop}

The previous proposition depicts the condition needed to solve
efficiently SAT$_{n\times m}$, which is $\mathbb{K}_{n\times m}$
the knowledge associated with the specific given SAT$_{n\times
m}$.

It is not complicate and complex to create an object data
structure combined index array and double links, the drawback is
the amount of memory needed to do the insertion and erase with
cost $\bigO(k)$ ($k$ very small integer) by updating only the
links.

Assuming that the initial states of the structures for
$\mathbb{S}$ and M are giving, it cost for building them can be
neglected or considered a constant. Of course the amount of memory
needed is exponential $(2^{n}),$ where $n$ is the numbers of
boolean variables for SAT$_{n\times m}$

The next tables depict that inserting and erasing binary numbers
into the structures  $\mathbb{S}$ and M only consists on updating
links with fixed complexity $\bigO(k)$ ($k$ very small integer
comparing with $2^n$).

\begin{center}
$%
\begin{tabular}{|c|c|c|}
\hline Est & prev & next \\ \hline $\mathbb{S}$ & 8 & 1 \\ \hline
M & 0 & 0 \\ \hline
\end{tabular} \
\begin{tabular}{|c|c|c|c|}
\hline \multicolumn{4}{|c|}{$\mathbb{S}_{n\times m}$} \\ \hline
$i$ & $n$ & prev$_{i}$ & next$_{i}$ \\ \hline 1 & 000 & 8 & 2 \\
\hline 2 & 001 & 1 & 3 \\ \hline 3 & 010 & 2 & 5 \\ \hline 4 & 011
& 0 & 0 \\ \hline 5 & 100 & 3 & 6 \\ \hline 6 & 101 & 5 & 7 \\
\hline 7 & 110 & 6 & 8 \\ \hline 8 & 111 & 7 & 1 \\ \hline
\end{tabular} \
\begin{tabular}{|c|c|c|c|}
\hline \multicolumn{4}{|c|}{M$_{n\times m}$} \\ \hline $i$ & $n$ &
prev$_{i}$ & next$_{i}$ \\ \hline 1 & - - - & 0 & 0 \\ \hline 2 &
- - - & 0 & 0 \\ \hline 3 & - - - & 0 & 0 \\ \hline 4 & - - - & 0
& 0 \\ \hline 5 & - - - & 0 & 0 \\ \hline 6 & - - - & 0 & 0 \\
\hline 7 & - - - & 0 & 0 \\ \hline 8 & - - - & 0 & 0 \\ \hline
\end{tabular}
$
\end{center}

Erasing $011$ from $\mathbb{S}_{n\times m}$ and inserting it in
M$_{n\times m}$

\begin{center}
$%
\begin{tabular}{|c|c|c|}
\hline Est & prev & next \\ \hline $\mathbb{S}$ & 8 & 1 \\ \hline
M & 4 & 4 \\ \hline
\end{tabular} \
\begin{tabular}{|c|c|c|c|}
\hline \multicolumn{4}{|c|}{$\mathbb{S}_{n\times m}$} \\ \hline
$i$ & $n$ & prev$_{i}$ & next$_{i}$ \\ \hline 1 & 000 & 8 & 2 \\
\hline 2 & 001 & 1 & 3 \\ \hline 3 & 010 & 2 & 5 \\ \hline 4 & - -
- & 0 & 0 \\ \hline 5 & 100 & 3 & 6 \\ \hline 6 & 101 & 5 & 7 \\
\hline 7 & 110 & 6 & 8 \\ \hline 8 & 111 & 7 & 1 \\ \hline
\end{tabular} \
\begin{tabular}{|c|c|c|c|}
\hline \multicolumn{4}{|c|}{M$_{n\times m}$} \\ \hline $i$ & $n$ &
prev$_{i}$ & next$_{i}$ \\ \hline 1 & - - - & 0 & 0 \\ \hline 2 &
- - - & 0 & 0 \\ \hline 3 & - - - & 0 & 0 \\ \hline 4 & 011 & 4 &
4 \\ \hline 5 & - - - & 0 & 0 \\ \hline 6 & - - - & 0 & 0 \\
\hline 7 & - - - & 0 & 0 \\ \hline 8 & - - - & 0 & 0 \\ \hline
\end{tabular}%
$
\end{center}

Erasing  $001$ from $\mathbb{S}_{n\times m}$ and inserting it in
M$_{n\times m}$

\begin{center}
$%
\begin{tabular}{|c|c|c|}
\hline Est & prev & next \\ \hline $\mathbb{S}$ & 8 & 1 \\ \hline
M & 4 & 2 \\ \hline
\end{tabular} \
\begin{tabular}{|c|c|c|c|}
\hline \multicolumn{4}{|c|}{$\mathbb{S}_{n\times m}$} \\ \hline
$i$ & $n$ & prev$_{i}$ & next$_{i}$ \\ \hline 1 & 000 & 8 & 3 \\
\hline 2 & - - - & 0 & 0 \\ \hline 3 & 010 & 1 & 5 \\ \hline 4 & -
- - & 0 & 0 \\ \hline 5 & 100 & 3 & 6 \\ \hline 6 & 101 & 5 & 7 \\
\hline 7 & 110 & 6 & 8 \\ \hline 8 & 111 & 7 & 1 \\ \hline
\end{tabular} \
\begin{tabular}{|c|c|c|c|}
\hline \multicolumn{4}{|c|}{M$_{n\times m}$} \\ \hline $i$ & $n$ &
prev$_{i}$ & next$_{i}$ \\ \hline 1 & - - - & 0 & 0 \\ \hline 2 &
001 & 4 & 4 \\ \hline 3 & - - - & 0 & 0 \\ \hline 4 & 011 & 2 & 2
\\ \hline 5 & - - - & 0 & 0 \\ \hline 6 & - - - & 0 & 0 \\ \hline
7 & - - - & 0 & 0 \\ \hline 8 & - - - & 0 & 0 \\ \hline
\end{tabular}%
$
\end{center}

Erasing  $000$ from $\mathbb{S}_{n\times m}$ and inserting it in
M$_{n\times m}$

\begin{center}
$%
\begin{tabular}{|c|c|c|}
\hline Est & prev & next \\ \hline $\mathbb{S}$ & 8 & 3 \\ \hline
M & 4 & 1 \\ \hline
\end{tabular} \ %
\begin{tabular}{|c|c|c|c|}
\hline \multicolumn{4}{|c|}{$\mathbb{S}_{n\times m}$} \\ \hline
$i$ & $n$ & prev$_{i}$ & next$_{i}$ \\ \hline 1 & - - - & 0 & 0 \\
\hline 2 & - - - & 0 & 0 \\ \hline 3 & 010 & 8 & 5 \\ \hline 4 & -
- - & 0 & 0 \\ \hline 5 & 100 & 3 & 6 \\ \hline 6 & 101 & 5 & 7 \\
\hline 7 & 110 & 6 & 8 \\ \hline 8 & 111 & 7 & 3 \\ \hline
\end{tabular} \ %
\begin{tabular}{|c|c|c|c|}
\hline \multicolumn{4}{|c|}{M$_{n\times m}$} \\ \hline $i$ & $n$ &
prev$_{i}$ & next$_{i}$ \\ \hline 1 & 000 & 4 & 2 \\ \hline 2 &
001 & 1 & 4 \\ \hline 3 & - - - & 0 & 0 \\ \hline 4 & 011 & 2 & 1
\\ \hline 5 & - - - & 0 & 0 \\ \hline 6 & - - - & 0 & 0 \\ \hline
7 & - - - & 0 & 0 \\ \hline 8 & - - - & 0 & 0 \\ \hline
\end{tabular}%
$
\end{center}

Using the structures, propositions and properties for
SAT$_{n\times m},$ a computable algorithm for solving
SAT$_{n\times m}$ and building $\mathbb{K}_{n\times m}$ is:

\begin{algorithm}
~\label{alg:SATKnow} \textbf{Input:} SAT$_{n\times m}$.

\textbf{Output:} Y set of binary numbers, such that when Y$\neq \{
\},$ any $y\ast\in$ Y, SAT$_{n\times m}(y^{\ast })=1$. The
structure $\mathbb{S}_{n\times m}$, such that any $x^\ast \in
\mathbb{S}_{n\times m}$,
 SAT$_{n\times m+1}(x^{\ast })=1$. $\mathbb{K}_{n\times m}$ (object structures: M,
$\mathbb{S}_{n\times m}$ and Y).

\textbf{Variables in memory}: Y $= \{ \}$: set of binary numbers,
initialized to empty;  Initialized object structures: M,
$\mathbb{S}_{n\times m}$.

\begin{enumerate}

\item \textbf{for} i:=1 to m

\item \hspace{0.5cm} Translate formula $i$ of SAT$_{n\times m}$ to
a binary number $k$.

\item \hspace{0.5cm} \textbf{if} $\mathbb{S}$.$\mathbb{S}_{n\times
m}$[$k$] not equal "- - -" \textbf{then}

\item \hspace{0.5cm} \hspace{0.5cm} \textbf{if} SAT$_{n\times
m}$($k$) equal 1 \textbf{then}

\item \hspace{0.5cm} \hspace{0.5cm} \hspace{0.5cm}
\textbf{output:} $y^\ast=k$ is a solution for SAT$_{n\times m}$.

\item \hspace{0.5cm} \hspace{0.5cm} \hspace{0.5cm}
\textbf{insert}($k$, Y).

\item \hspace{0.5cm} \hspace{0.5cm} \hspace{0.5cm}
$\textbf{delete}( k, \mathbb{S}_{n\times m})$

\item \hspace{0.5cm} \hspace{0.5cm} \hspace{0.5cm}
$\textbf{insert}(k, \text{M}_{n\times m})$

\item \hspace{0.5cm} \hspace{0.5cm} \textbf{else}

\item \hspace{0.5cm} \hspace{0.5cm} \hspace{0.5cm}
$\textbf{delete}( k, \mathbb{S}_{n\times m})$

\item \hspace{0.5cm} \hspace{0.5cm} \hspace{0.5cm}
$\textbf{insert}(k, \text{M}_{n\times m})$

\item \hspace{0.5cm} \hspace{0.5cm} \hspace{0.5cm}
$\textbf{delete}(\overline{k}, \mathbb{S}_{n\times m})$

\item \hspace{0.5cm} \hspace{0.5cm} \hspace{0.5cm}
$\textbf{insert}(\overline{k}, \text{M}_{n\times m})$

\item \hspace{0.5cm} \hspace{0.5cm} \textbf{end if}

\item \hspace{0.5cm} \textbf{end if}

\item \textbf{end for}

\end{enumerate}
\end{algorithm}

At this point, SAT$_{n\times m}$ is equivalent to set of binary
numbers M$_{n\times m}$. And $\mathbb{K}_{n\times m}$ has
information to solve SAT$_{n\times m}$ immediately.

The knowledge for a given SAT$_{n\times m}$ depends of exploring
M$_{n\times m}$. Before of exploring SAT$_{n\times m}$ no binary
numbers associated with $\mathbb{K}_{n\times m}$.

Now, let $\mathbb{K}_{n\times m}$ be knowledge of SAT$_{n\times
m}$. $\mathbb{K}_{n\times m}$ $=$ Y $\cup$ M$_{n\times m}$ $\cup$
$\mathbb{S}_{n\times m}$, where Y is the set of satisfying number
for SAT$_{n\times m}$, and $\mathbb{S}_{n\times m}$ $=$ (Y $\cup$
M$_{n\times m})^c$, it  is the set of binary numbers that
satisfies SAT$_{n\times m}$.

\begin{remark}

Having $\mathbb{K}_{n\times m}$, the knowledge of SAT$_{n\times
m}$, it is easy to solve any modified SAT$_{n\times m}$. It is
need to keep tracking of changes in M and $\mathbb{S}$, then
modified SAT can be solved by the proposition~\ref{prop:SATKnow}.

\end{remark}

So far, SAT$_{n\times m}$ is now a fixed point type problem over a
set of binary numbers. It consists in look for a number in
M$_{n\times m}$ which is not blocked or to pick up any element $s$
$\in \mathbb{S}_{n \times m}$ when $\mathbb{K}$ is given.

If all number of M$_{n\times m}$ are blocked then the first binary
number $s$ from 0 to $2^n-1$ which is not in M$_{n\times m}$ is
the solution for M$_{n\times m}$, i.e. any element $s$ $\in
\mathbb{S}_{n \times m}$. This is trivial when knowledge is given
or can be created in an efficient way. This results of this
section are related to my article ~\cite{arXiv:Barron2010}, where
I stated that NP problems need to look for its solution in an
search space using a Turing Machine: "It is a TM the appropriate
computational model for a simple algorithm to explore at full the
GAPn's research space or a reduced research space of it". As
trivial as it sound, pickup a solution for SAT$_{n \times m}$
depends of exploring its $m$ boolean equations.

The question if there exists an efficient algorithm for any
SAT$_{n\times m}$ now can be answered. The
algorithm~\ref{alg:SATKnow} is technologically implausible because
the amount of memory needed. However, building $\mathbb{K}_{n
\times m}$ provides a very efficient telephone algorithm (see
proposition 8, and remarks 5 in~\cite{arXiv:Barron2005}) where
succeed is guaranteed for any $s$ $\in$ $\mathbb{S}_{n \times m}$.
On the other hands, following ~\cite{arXiv:Barron2005}) there are
tree possibilities exhaustive algorithm (exploring all the
searching space), scout algorithm (previous knowledge or heuristic
facilities the search in the searching space), and wizard
algorithm (using necessary and sufficient properties of the
problem).

The study of NP problems depicts that only for an special type of
problem exists properties such as Jordan's simple curve or $n$ odd
for polygon for TSP2D, or convexity, IF lattice, or CB lattice for
Lennard Jones structural and potential minimization problems for
building an ad-hoc efficient algorithms, but these properties can
not be generalizad for any GAP or any member of class NP.

The boolean formulas of SAT$_{n\times m}$ correspond a binary
numbers in disorder (assuming an order of the boolean variables as
binary digits).  The algorithm~\ref{alg:SAT} has a complexity of
the size of $m$ (the numbers of formulas of SAT).It is not worth
to considered sorting algorithm because complexity increase by a
factor $\bigO(n(2^n))$ when $m \approx 2^n.$

Without exploring, given SAT$_{n \times m}$ then build M$_{n
\times m}$. M can be consider an arbitrary set of numbers. The
numbers in M have not relation or property to point out the
 satisfied binary number, in fact one or many numbers
could be solution or not one in M, but in the range from 0 to
$2^n-1$ there is a solution or not depending of M. Of course, this
is true for an arbitrary set of numbers.

On the other hand, if the formulas of a SAT problem are formulated
assuming, by example, no boolean formula is complement of other
one, then any translation to binary number is a satisfied
assignment.

Hereafter, for the sake of my argumentation the boolean formulas
of SAT$_{n\times m}$ are considered a translation to binary
numbers in disorder and without any correlation in M$_{n \times
m}$.

\begin{prop} With algorithm~\ref{alg:SAT}:

\begin{enumerate}

\item A solution for SAT$_{n\times m}$  is efficient when $m << $
$2^{n}.$

\item A solution for SAT$_{n\times m}$  is not efficient when $m
\approx $ $2^{n}.$

\end{enumerate}
\begin{proof}
It follows from algorithm~\ref{alg:SAT}.
\end{proof}
\end{prop}

In order to find a solution for SAT, there are a probabilistic
 algorithm rather than algorithm~\ref{alg:SAT}. By example a
 simple
probabilistic algorithm for SAT$_{n\times m}$ is:

\begin{algorithm}
~\label{alg_prob:SAT} \textbf{Input:} SAT$_{n\times m}$.

\textbf{Output:} $x^{\ast }$ such that SAT$_{n\times m}(x^{\ast
})=1$ or SAT$_{n\times m+1}(x^{\ast })=1$ or SAT$_{n\times m}$ is
not satisfied.

\textbf{Variables in memory}: T[0:$2^{n-1}$]=0 of boolean. $ct$:=0
: Integer;

\begin{enumerate}

\item \textbf{while (1)}

\item \hspace{0.5cm} \textbf{Select randomly} $k$ \textbf{in}
$[0,2^n-1]$ \textbf{minus marked} T[i]==1;

\item \hspace{0.5cm} \textbf{if} SAT$_{n\times m}$($k$) == 1
\textbf{then}

\item \hspace{0.5cm} \hspace{0.5cm} \textbf{output:} $k$ is the
solution for SAT$_{n\times m}$.

\item \hspace{0.5cm} \hspace{0.5cm} \textbf{stop}

\item \hspace{0.5cm} \textbf{end if}

\item \hspace{0.5cm} \textbf{if} T[$k$] $<> 1$ \textbf{then} \item
\hspace{0.5cm} \hspace{0.5cm} T[$k$]:=1;

\item \hspace{0.5cm} \hspace{0.5cm} $ct$ := $ct$ +1;

\item \hspace{0.5cm} \hspace{0.5cm} \textbf{if} $ct \geq 2^n$
\textbf{then}

\item \hspace{0.5cm} \hspace{0.5cm} \hspace{0.5cm}
\textbf{output:} There is not solution for SAT$_{n\times m}$.

\item \hspace{0.5cm} \hspace{0.5cm} \hspace{0.5cm} \textbf{stop}

\item \hspace{0.5cm} \hspace{0.5cm} \textbf{end if}

\item \textbf{end while}

\end{enumerate}

\end{algorithm}

\begin{prop} With algorithm~\ref{alg_prob:SAT}:

\begin{enumerate}

\item A solution for SAT$_{n\times m}$  is efficient when $m << $
$2^{n}.$

\item A solution for SAT$_{n\times m}$  is not efficient when $m
\approx $ $2^{n}.$

\end{enumerate}

\begin{proof} \

\begin{enumerate}

\item It follows from algorithm~\ref{alg_prob:SAT} for finding $k$
and $m << $ $2^{n}$. It implies, for many $k$ that P(SAT$_{n\times
m}$($k$)==1) $\approx 1$.

\item It follows from algorithm~\ref{alg_prob:SAT} for finding $k$
and $m \approx $ $2^{n}$. Here, it is possible that many selected
randomly number in $[0,2^n-1]$ are blocked. Therefore
P(SAT$_{n\times m}$($k$)==0) $\approx 1$. This could cause that
algorithm~\ref{alg_prob:SAT} does not solve SAT$_{n\times m}$ in
an appropriate amount of time.
\end{enumerate}
\end{proof}
\end{prop}

Finally,
\begin{prop}~\label{prop:SATnxm}
 SAT$_{n\times m}$ has not property or heuristic to build an
efficient algorithm.

\begin{proof}
Given SAT$_{n\times m}$, its translation is a set of binary
numbers, in disorder and without any any previous knowledge, nor
correlation in M$_{n \times m}$. If such property or heuristic
exist then any arbitrary subset of number has it. Such
characteristic or property must imply that any natural number is
related to each other. This means it is in the intersection of all
properties for all natural numbers. Also it is no related to the
value, because by example, the intersection of natural clases
under the modulo of a prime number is empty. Moreover, it is
something a priory, otherwise, it will imply to revise all number
at least in M$_{n\times m}$. Using it in a random algorithm, to
point out efficiently to the number, which is solution, means that
this number is inherently not equally probable. Also, this binary
number correspond to an arbitrary arrangement of the boolean
variables of SAT$_{n\times m}$, so in a different arrangement of
the positions for the boolean variables, any number is inherently
not equally probable!
\end{proof}
\end{prop}

\begin{prop}~\label{prop:NPnP} NP has not property or heuristic to build an
efficient algorithm.

\begin{proof}
Let be X a problem in NP. PH${_\text{X}}$ is the set of properties
or heuristics for building an efficient algorithm for problem X.
$$\bigcap_\text{X} \text{PH}{_\text{X}}=\{\}$$
by the previous proposition.
\end{proof}
\end{prop}


\section*{Conclusions and future work}

~\label{sc:conclusions and future work}

The Jordan's simple curve is an example of a property to create an
efficient algorithm for a necessary type of solution of a NP hard
problem. It is for solving approximately Euclidian Travelling
Salesman problem in 2D planes but not in $\Real^m$ spaces, $m>2$.
Here the Euclidian metric implies that the quadrilateral'sides are
less than the quadrilateral's diagonals. Polygon or convex
distribution of the cities, Jordan is a sufficient and necessary
property, this previous knowledge allows to build efficient
algorithms. For $\Real^m$,  even with fast triangulations, this
means more alternatives to explore, without a property to reduce
global complexity in the corresponding searching space.

Heuristic techniques, using previous knowledge of a problem in
$\mathbb{R}^m$ do not provide reducibility (see 6 in
~\cite{arXiv:Barron2010}) for any NP problems.

There are nice properties Jordan's simple curve and star. Any
POME$_n$, $n>3$ odd has a solution type star. Any TSP2D has a
solution as a Jordan's simple curve. Crossing lines in a solution
is the opposite of not crossing lines in other type of problem.

For LJ problems or for the Searching of the Optimal Geometrical
Structures of clusters of $n$ particles remains out of the
existence of an efficient algorithm, the selection of the $n$
particles is the bottle's neck (see ~\cite{arXiv:Barron2005}).

Here, the classical SAT (a NP decision problem) provides a general
case, where formulas have any number of boolean variables, then it
is is reduced to a simple version SAT${n\times m}$, all boolean
formulas have the same number of boolean variables. This allows to
see that does not exist a property for solving SAT with
reducibility of the searching space. This shows that there is not
a general property that allows to solve SAT problems in polynomial
time. The classical SAT depicted in section~\ref{sc:SAT} has
clearly no properties for solving efficiently it when $m \approx
2^n$.

For the future, I believe that the implementation of the
algorithm~\ref{alg:Detcross} for TSP brings an efficient time's
reduction for finding the optimal hamiltonian cycle under
euclidian distance versus looking to solve a global optimization
problem without a necessary stop condition.

Quantum Computation is on the future road.  To my knowledge the
algorithm~\ref{alg:SAT} can be adapted for using quantum variables
for exploring $[0,2^n-1]$ states at the same time, instead of the
cycle \textbf{for} i:=0 to $2^n-1$. This implies that the
complexity for solving NP problems can be reduced to one cycle
over quantum variables.

Finally, this reformulation of my previous works, using Jordan's
simple curve versus hamiltonian cycles with many crossings as
possible, and the research of the properties of SAT$_{n\times m}$
supports  that there is not a general property to reduce the
complexity of the worst case NP problem.


%
%
%
%
%
%
%

\bibliographystyle{abbrv}

\end{document}